\documentclass[12,draftcls,onecolumn]{IEEEtran}
\usepackage[T1]{fontenc}
\usepackage[latin9]{inputenc}
\usepackage{color}
\usepackage{verbatim}
\usepackage{mathrsfs}
\usepackage{amsmath}
\usepackage{amsthm}
\usepackage{amssymb}
\usepackage{graphicx}
\usepackage[unicode=true,
 bookmarks=true,bookmarksnumbered=true,bookmarksopen=true,bookmarksopenlevel=1,
 breaklinks=false,pdfborder={0 0 1},backref=false,colorlinks=false]
 {hyperref}
\hypersetup{pdftitle={Your Title},
 pdfauthor={Your Name},
 pdfborderstyle=,pdfpagelayout=OneColumn,pdfnewwindow=true,pdfstartview=XYZ,plainpages=false}
\usepackage{breakurl}

\makeatletter
\let\oldforeign@language\foreign@language
\DeclareRobustCommand{\foreign@language}[1]{%
  \lowercase{\oldforeign@language{#1}}}
\theoremstyle{plain}
\newtheorem{thm}{\protect\theoremname}
\theoremstyle{plain}
\newtheorem{fact}[thm]{\protect\factname}
\theoremstyle{plain}
\newtheorem{cor}[thm]{\protect\corollaryname}
\theoremstyle{plain}
\newtheorem{prop}[thm]{\protect\propositionname}
\theoremstyle{plain}
\newtheorem{lem}[thm]{\protect\lemmaname}

\providecommand{\corollaryname}{Corollary}
\providecommand{\factname}{Fact}
\providecommand{\lemmaname}{Lemma}
\providecommand{\theoremname}{Theorem}

\makeatother

\providecommand{\corollaryname}{Corollary}
\providecommand{\factname}{Fact}
\providecommand{\lemmaname}{Lemma}
\providecommand{\propositionname}{Proposition}
\providecommand{\theoremname}{Theorem}

\begin{document}

\title{Capacity and Algorithms for a Cognitive Network with Primary-Secondary
User Cooperation}

\author{Athanasios Papadopoulos,~\IEEEmembership{Student~Member,~IEEE,}
Nestor~D.~Chatzidiamantis,~\IEEEmembership{Member,~IEEE,} and~Leonidas~Georgiadis,~\IEEEmembership{Senior~Member,~IEEE}\thanks{Athanasios Papadopoulos, Nestor D. Chatzidiamantis and Leonidas Georgiadis
are with Department of Electrical and Computer Engineering, Aristotle
University of Thessaloniki, Thessaloniki, Greece, Emails: \{athanapg,nestoras,leonid\}@auth.gr}}

\markboth{}{Athanasios Papadopoulos \MakeLowercase{\emph{et al.}}: Capacity
and Algorithms for a Cognitive Network with Primary-Secondary User
Cooperation}

\IEEEpubid{}
\maketitle
\begin{abstract}
In this work, we examine cognitive radio networks, where secondary
users may act as relays for messages sent by the primary user, hence
offering performance improvement of primary transmissions, while at
the same time obtaining more transmission opportunities for their
own data. In particular, assuming the broadcast packet erasure model
with feedback, we investigate the capacity of the fundamental cooperative
cognitive radio network which consists of one primary and one secondary
transmitter-receiver pairs. The primary transmitter is the owner of
the channel and as such, we intend to keep its operations simple and
to avoid increasing its storage requirements. Specifically, the primary
transmitter does not receive data sent by the secondary transmitter
and does not perform any coding operations. The only requirement on
the primary transmitter is to listen to public feedback and take appropriate
scheduling actions. \textcolor{black}{On the other hand, }the secondary
transmitter can overhear primary transmissions and is allowed to perform
any coding operations\textcolor{black}{.} We develop an outer bound
to the capacity of the fundamental cooperative cognitive radio network
under consideration. Then, we propose a coding-scheduling algorithm
suitable for this type of networks, which involves only XOR network
coding operations. The complexity of the scheduling decisions of the
proposed algorithm depends on the channel statistical parameters and
three cases, depending on the relations between channel erasure probabilities,
are distinguished. For the first two cases the rate region of the
proposed algorithm coincides with the developed capacity outer bound,
hence the algorithm is capacity achieving. For the third case, the
rate region of the proposed algorithm is not identical to the outer
bound; however, numerical results show that it is fairly close to
the derived outer bound for a wide range of the statistical parameters
of the system. 
\end{abstract}


\section{Introduction}

Cognitive networks attracted a lot of attention in recent years due
to their potential for improving spectral efficiency \cite{A:Haykin}.
In this type of networks, unlicensed users, also known as \emph{secondary
users,} are allowed to communicate with each other utilizing the licensed
spectrum, thus taking advantage of the underutilized shared spectrum,
while maintaining limited or no interference to the licensed users,
also known as \emph{primary users}.

Initial designs of cognitive radio networks assumed that there is
no interaction between primary and secondary users (see \cite{DSA2}
and the references therein). However, it was soon realized that by
allowing secondary users to cooperate with primary users, several
benefits for both types of users arise. These benefits stem from the
fact that by allowing secondary users to relay primary transmissions,
the channel between the secondary transmitter and primary receiver
can be exploited, thus, increasing the primary user's effective transmission
rate, as well as offering more transmission opportunities to secondary
user. This type of cognitive radio networks are referred to in the
literature as cooperative cognitive radio networks.

Due to their advantages, cooperative cognitive radio networks have
gained a lot of attention in recent years. Physical layer cooperation
between primary and secondary users was examined in \cite{Goldsmith-2009-ID359},
while non-orthogonal multiple access techniques based on successive
interference cancellation were proposed in \cite{lv2017application}.
Queuing theoretic analysis and transmission protocol design for cooperative
cognitive radio networks were presented in \cite{Simeone-2007-ID29,Krikidis-2009-ID233,Neely_cooper,Nestor_ours}.
Specifically, a cooperative transmission protocol for cognitive radio
networks where the secondary transmitter acts as a relay for primary
user's transmissions was initially presented in \cite{Simeone-2007-ID29}
and the benefits of such cooperation for both types of users were
investigated. In \cite{Krikidis-2009-ID233}, cooperative cognitive
radio networks with multiple secondary users were investigated and
advanced relaying techniques which involved physical layer coding
between primary and secondary transmissions were suggested. Cooperative
transmission policies which take into account the available power
resources at the secondary transmitter in order for the latter to
decide whether to cooperate or not, have been presented in \cite{Neely_cooper},
\cite{Nestor_ours}.

Network coding has been applied in cooperative cognitive radio networks
as a means to increase capacity for both type of users (see \cite{network_survey}
and the references therein). However, in most of these works, the
network coding operations that were performed by secondary users (acting
as relays) involved only primary user's packets. Relatively recently,
network coding schemes which involved both primary and secondary users
packets have been suggested as an effective means of cooperation in
cooperative cognitive radio networks \cite{li2014cooperation,li2018performance,papadopoulos_conf1,URL:arxiv_version}.
Specifically, in \cite{li2014cooperation} and \cite{li2018performance}
a first attempt was made to design transmission algorithms where secondary
users employ network coding between their data and the overheard primary
transmissions; however, the presented algorithms leave room for improvement
by exploiting more opportunities for transmitting network coded packets.
More efficient similar network coding based transmission algorithms
for cooperative cognitive radio networks were presented in our previous
works, \cite{papadopoulos_conf1,URL:arxiv_version}, whose performance
was investigated based on queuing theory. While the presented algorithms
offered an enhancement of the primary-secondary user throughput region
compared to previously proposed cooperation schemes, this approach
did not address the problem of optimality or near optimality of the
proposed algorithms in terms of achievable throughput region. This
issue is addressed in the current work using an information theoretic
approach.

The current work focuses on investigating the capacity region of the
fundamental cooperative cognitive radio network when the channel is
modeled as broadcast erasure with feedback, that models well the network
at the MAC layer, and aims on designing efficient coding-scheduling
algorithms - transmission algorithms for short. In the past, the capacity
of several wireless communications systems setups has been investigated
under the assumption of erasure channel model \cite{wang2014capacity,gatzianas2013multiuser,papadopoulos_journal,Wang:2016,wang2013two,wang2007beyond}.
Specifically, the capacity of broadcast erasure networks was investigated
in \cite{wang2014capacity} and \cite{gatzianas2013multiuser}, while
the capacity with side information available to the receivers has
been characterized in \cite{papadopoulos_journal}. Moreover, the
capacity region for the fully-connected 3-node packet erasure network
is investigated in \cite{Wang:2016}, while a simple and a more complicated
butterfly erasure network is analyzed in \cite{wang2013two} and \cite{wang2007beyond},
respectively. A related channel model is investigated in \cite{Wang_conf_2016},
where a single source broadcast erasure channel with two receivers
and a relay is examined; the source has two independent messages,
one for each destination and the messages may be delivered to the
receivers either directly or through the relay, using Linear Network
Coding. 

The major difference between the previous setups and the setup considered
in this work stems from the requirements imposed by the fact that
the primary transmitter, as owner of the channel, has certain privileges.
Specifically, motivated by our previous works, \cite{papadopoulos_conf1,URL:arxiv_version},
we require that the primary transmitter does not receive any data
sent by the secondary transmitter and, in order to avoid increasing
its complexity and memory requirements, does not perform any coding
operations; in contrast the secondary transmitter may perform arbitrary
coding operations. The only requirement on the primary transmitter
is to listen to public feedback and take appropriate scheduling actions.
Based on the above, the contribution of the paper is summarized as
follows:
\begin{enumerate}
\item We\textcolor{black}{{} consider a basic cognitive radio network setup
which is composed by one primary and one secondary transmitter-receiver
pairs. }All the underlying channels are considered to be broadcast
packet erasure channels with public feedback.\textcolor{black}{{} }The
primary transmitter does not receive any data transmitted by the secondary
transmitter and does not perform coding operations\textcolor{black}{;
it only listens to the feedback and takes scheduling actions. On the
other hand, }the secondary transmitter can overhear primary transmissions
and is allowed to perform network coding operations based on its own
packets as well as the overheard packets during primary transmissions.\textcolor{black}{{}
The objective of the presented analysis is to maximize the secondary
user's transmission rate without reducing primary user's channel capacity.}
\item We develop an outer bound to the capacity region of the fundamental
cooperative cognitive radio network under consideration. 
\item We propose a transmission algorithm suitable for the cooperative cognitive
radio system under consideration. The proposed algorithm involves
only XOR network coding operations, while the complexity of scheduling
decisions depends on channel statistical parameters. Specifically
we consider three cases depending on relations between channel erasure
probabilities. For the first two cases the rate region of the proposed
algorithm coincides with the developed capacity outer bound, hence
the algorithm is capacity achieving. For the third case, involving
more complex scheduling decisions, the rate region of the proposed
algorithm is not identical to the outer bound, but in general it is
fairly close to it.
\end{enumerate}
The remainder of the paper is organized as follows. In Section \ref{sec:Notation,-System-model}
we provide the notation that is used in the analysis that follows
along with the system model studied in this work. In Section \ref{sec:Main-Results}
we present the main results of this paper which include the derived
outer bound, the description of the cases where this outer bound is
in fact the capacity region of the system and an inner bound for the
case where the system capacity is not known. Section \ref{sec:Coding-Algorithm}
describes the proposed transmission algorithm and investigates its
performance in terms of achievable rate region. Section \ref{sec:Conclusion}
provides concluding remarks and suggestions for future research. Proofs
of the main results are provided in the Appendix \ref{sec:Proof-of-Theorem}.

\section{\label{sec:Notation,-System-model}Notation, System model and Channel
Codes}

\subsection{Notation}

We use the following notation. 
\begin{itemize}
\item Sets are denoted by calligraphic letters e.g., ${\cal F}$. 
\item Random variables are denoted by capital letters and their values by
small letters. 
\item Vectors are denoted by bold letters. 
\item For a sequence $Y(t),\ t=1,\cdots$, we denote $\boldsymbol{Y}^{t}=\left(Y(1),\cdots,Y(t)\right).$
Also, $X\in\boldsymbol{Y}^{t}$ means that $X=Y(s)$ for some $s\in\left\{ 1,\cdots,t\right\} $. 
\item If $\boldsymbol{Y}=\left(Y_{1},\cdots,Y_{n}\right)$ and $\mathcal{S}\subseteq\left\{ 1,2,\cdots,n\right\} ,$
we denote by $\boldsymbol{Y}_{\mathcal{S}}=\left(Y_{i}:i\in\mathcal{S}\right)$
the vector of coordinates of $\boldsymbol{Y}$with index in the set
$\mathcal{S}$. If $\mathcal{S=\emptyset}$ by convection we set $\boldsymbol{Y}_{\mathcal{S}}=c,$
a constant.
\item For $i\in\{1,2\}$ we denote by $i^{c}$ the element in the set $\left\{ 1,2\right\} -\{i\}.$ 
\item For random variables $X,Y,$ the notation $X\perp Y$ means that the
random variables are independent. 
\item A sentence between brackets next to a formula provides explanation
of the relations involved in the formula, e.g., $f(x)=y\ \ \text{"since ... "}.$ 
\end{itemize}

\subsection{System Model}

\begin{figure}[tbh]
\includegraphics[scale=0.4]{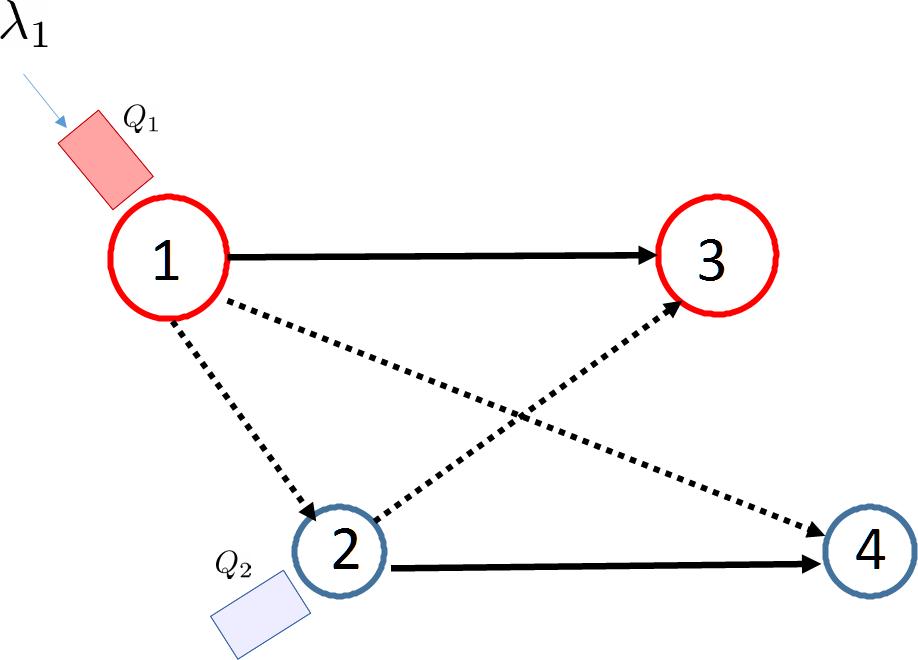}\caption{System Model\label{fig:System_Model}}
\end{figure}

We consider the four-node cognitive radio system model depicted in
Fig. \ref{fig:System_Model}. The system consists of two (transmitter,
receiver) pairs (1,3), (2,4). Pair (1,3) - odd numbers- represents
the primary channel. Node 1 is the primary transmitter who is the
licensed owner of the channel . Node 2 is the secondary transmitter;
this node does not have any licensed spectrum and seeks transmission
opportunities on the primary channel in order to deliver data to secondary
receiver, node 4.

Let 
\[
\mathcal{N}_{1}=\left\{ 2,3,4\right\} ,\ \mathcal{N}_{2}=\left\{ 3,4\right\} .
\]

\textbf{Erasure events}

Erasure events are characterized by a sequence of tuples of 0-1 random
variables, 
\[
\boldsymbol{Z}\left(t\right)=\left(\boldsymbol{Z}_{1}\left(t\right),\boldsymbol{Z}_{2}\left(t\right)\right)\triangleq\left(\left[Z_{1j}(t),\ j\in\mathcal{N}_{1}\right],\ \left[Z_{2j}(t),\ j\in\mathcal{N}_{2}\right]\right),\ t=1,\cdots,
\]
with the following interpretation. A symbol transmitted by node $i$
at time $t$ is received correctly by node $j$ if $Z_{ij}\left(t\right)=1$,
and erased at node $j$ if $Z_{ij}(t)=0$.

We assume that the tuples $\boldsymbol{Z}\left(t\right)=\left(\boldsymbol{Z}_{1}\left(t\right),\boldsymbol{Z}_{2}\left(t\right)\right),\ t=1,\cdots,$
are independent, however, for given $t,$ the random variables $Z_{ij}\left(t\right)$
can be arbitrarily dependent. We denote by $\epsilon_{\mathcal{S}}^{i},\ i\in\left\{ 1,2\right\} ,\ \mathcal{S}\subseteq\mathcal{N}_{i}$,
the probability that a message transmitted by node $i$ is erased
at all nodes in the set $\mathcal{S}$, i.e., $Z_{1j}(t)=0$ for all
$j\in\mathcal{S}$. Let $\mathcal{Z}_{i}=\left\{ \boldsymbol{z}=\left(z_{l}\right)_{l\in\mathcal{N}_{i}}:z_{l}\in\left\{ 0,1\right\} \right\} ,$
be the set of possible erasure events when node $i$ transmits, and
for $\mathcal{S}\subseteq\mathcal{N}_{i},\ \mathcal{S}\neq\emptyset$,
\[
\mathcal{Z}_{\mathcal{S}}^{i}=\left\{ \boldsymbol{z}\in\mathcal{Z}_{i}:\ \exists l\in\mathcal{S},\ z_{l}=1\right\} ,
\]
be the set of all vectors in $\mathcal{Z}_{i}$ for which at least
one component with index in set $\mathcal{S}$ is equal with 1.

\textbf{Feedback and Scheduling}

We assume that after a transmission by node $1\ (2)$ a 1-0 feedback
is sent by each node in $\mathcal{N}_{1}\ \left(\mathcal{N}_{2}\right)$
to the rest of the nodes, indicating correct reception-1 or erasure-0.
Hence, if node $i$ transmits at time $t,$ at the end of transmission
all nodes know $\boldsymbol{Z}_{i}\left(t\right).$

At each time $t=0,\cdots,$ only one of the nodes in $\left\{ 1,2\right\} $
is scheduled to transmit. This scheduling depends only on node feedback.
Specifically, denoting by $\sigma(t)\in\left\{ 1,2\right\} $ the
index of the node scheduled for transmission at time $t$, we set
$\sigma(1)=1$ (or $\sigma(1)=2)$ and 
\begin{equation}
\sigma(t)=\sigma\left(\boldsymbol{Z}^{t-1}\right),\ t=2,\cdots.\label{eq:feedback}
\end{equation}
where with a slight abuse of notation we denote $\boldsymbol{Z}^{t}\triangleq\left(\boldsymbol{Z}_{\sigma(s)}(s)\right)_{s=1}^{t}$.

\textbf{Transmission and reception symbol alphabets }

The transmitted symbols, called ``packets'', belong to a finite
field $\mathcal{\mathcal{F}}.$

\textcolor{black}{We denote by $X_{i}\left(t\right)\in\mathcal{\mathcal{F}}$
the symbol transmitted by node $i$ at time $t$ - if node $i$ is
not scheduled for transmission at time $t,$ i.e., $\sigma\left(t\right)\neq i$
we set $X_{i}\left(t\right)=\eta$ (null). }

We denote by $Y_{ij}\left(t\right)$ the symbol received by node $j\in\mathcal{N}_{i}$
if node $i$ transmits at time $t$, where erasure is indicated by
the symbol $\varepsilon$ - if node $i$ is not scheduled for transmission
at time $t,$ we set $Y_{ij}\left(t\right)=\eta.$ %
\begin{comment}
Hence, the output alphabet at each receiving node consists of the
set $\mathcal{\mathcal{F}}\cup\left\{ \varepsilon\right\} \cup\left\{ \eta\right\} $. 
\end{comment}
{} 

The following facts follow directly from the definitions. 
\begin{fact}
\label{fact:Def}Let $i\in\left\{ 1,2\right\} ,\ \mathcal{S}_{1}\subseteq\mathcal{N}_{1},\ \mathcal{S}_{2}\subseteq\mathcal{N}_{2},\ \mathcal{S}_{i}\neq\emptyset.$
For any $t,$ 
\begin{enumerate}
\item $\boldsymbol{Z}\left(t\right)$ is independent of $\sigma(t)=\sigma\left(\boldsymbol{Z}^{t-1}\right),\ t\geq2.$ 
\item \label{enu:LemDef1}Let $\sigma\left(t\right)=i$ and $\boldsymbol{Z}_{i\mathcal{S}_{i}}\left(t\right)=$$\boldsymbol{z}\notin{\cal Z}_{{\cal S}_{i}}^{i}$.
Then, $\boldsymbol{Y}_{i{\cal S}_{i}}\left(t\right)=\boldsymbol{\varepsilon},\ \boldsymbol{Y}_{i^{c}\mathcal{S}_{i^{c}}}\left(t\right)=\boldsymbol{\eta}.$ 
\item \label{enu:one-to-one}Let $\sigma\left(t\right)=i$ and $\boldsymbol{Z}_{i\mathcal{S}_{i}}\left(t\right)=$$\boldsymbol{z}\in{\cal Z}_{{\cal S}_{i}}^{i}$.
Then, $X_{i}\left(t\right)=f\left(\boldsymbol{Y}_{i\mathcal{S}_{i}}\right)$
(specifically, $X_{i}\left(t\right)=Y_{il}\left(t\right),$ where
$l$ is any coordinate of $\boldsymbol{z}$ with $z_{l}=1$) and $\boldsymbol{Y}_{i^{c}\mathcal{S}_{i^{c}}}\left(t\right)=\boldsymbol{\eta}$.
Moreover, the function $f\left(\cdot\right)$ is one-to-one. 
\end{enumerate}
\end{fact}

\subsection{\textcolor{black}{Channel Codes and Channel Capacity}}

\textcolor{black}{A channel code $C_{n}$ of rate vector $\boldsymbol{R}=\left(R_{1},R_{2}\right),\ R_{i}\geq0,$
consists of the following: } 
\begin{itemize}
\item $n$ symbol transmissions. 
\item \textcolor{black}{Messages, $\left(\boldsymbol{W}_{1,n},\boldsymbol{W}_{2,n}\right)$.
Message $\boldsymbol{W}_{1,n}$ consists o}f $k_{i,n}=k_{i,n}=\left\lceil nR_{i}\right\rceil $
packets, i.e. $\boldsymbol{W}_{i,n}=\left(W_{i,1},\cdots,W_{i,k_{i,n}}\right)$,
$W_{i,l}\in\mathcal{F},$ that need to be delivered to node $3$ if
$i=1$, and node $4$ if $i=2$\textcolor{black}{.} Messages are independent
of feedback variables $\left(\boldsymbol{Z}_{1}\left(t\right),\boldsymbol{Z}_{2}\left(t\right)\right),\ t=1,\cdots$.
We assume that each packet is a uniformly selected element from the
finite field $\mathcal{F}$ and that packets are independent. 
\item \textcolor{black}{Encoders that specify the symbol to be transmitted
by one of the nodes $1$, 2, as follows.} 
\begin{itemize}
\item If $\sigma(t)=2,$ then 
\[
X_{2}\left(t\right)=f_{2,n}\left(\boldsymbol{W}_{2,n},\boldsymbol{Y}_{1\left\{ 2\right\} }^{t-1},\boldsymbol{Z}^{t-1}\right),
\]
where $f_{2}$ is an arbitrary function. Thus the secondary transmitter
can perform any coding operation that depends on its own packets,
the packets received by primary node $1$ and the channel feedback. 
\item If $\sigma(t)=1,$ then node $1$ transmits one of the packets in
$\boldsymbol{W}_{1,n}$, where the index $J_{n}\left(t\right)$ of
the packet to be transmitted depends only on channel feedback, i.e.,
$J_{n}\left(1\right)$ is selected arbitrarily, and for $t\geq2$,
$J_{n}\left(t\right)=J_{n}\left(\boldsymbol{Z}^{t-1}\right)$. Hence,
\[
X_{1}\left(t\right)=W_{1,J_{n}\left(t\right)}.
\]
Thus the primary node 1 does not perform coding operations, and only
schedules packets according to received feedback. For convenience
in the description below, whenever $\sigma(t)=2$, we define $J_{n}\left(t\right)=\eta.$ 
\end{itemize}
\item \textcolor{black}{Decoders $g_{j,n}(\boldsymbol{Y}_{1\left\{ j\right\} }^{n},\boldsymbol{Y}_{2\left\{ j\right\} }^{n},\boldsymbol{Z}^{n})$,
for receivers $j\in\left\{ 3,4\right\} .$ Within $n$ channel uses,
receiver $j$ estimates the message transmitted by its intended transmitter
($j-2),$ 
\begin{equation}
\hat{\boldsymbol{W}}_{j-2,n}=g_{j,n}(\boldsymbol{Y}_{1\left\{ j\right\} }^{n},\boldsymbol{Y}_{2\left\{ j\right\} }^{n},\boldsymbol{Z}^{n}).\label{eq:decode}
\end{equation}
} 
\end{itemize}
\textcolor{black}{Thus the channel code $C_{n}$ is fully specified
by the tuple $(n,\ \left\lceil nR_{1}\right\rceil ,\left\lceil nR_{2}\right\rceil ,\sigma,J_{n},f_{2,n},g_{3,n},g_{4,n})$.
The probability of erroneous decoding of code $C_{n}$ is $\lambda_{n}=\Pr(\underset{i\in\left\{ 1,2\right\} }{\cup}\{\hat{W}_{i,n}\neq W_{i,n}\})$.
A vector rate $\boldsymbol{R}$ is called achievable under the sequence
of codes $C_{n}$ if for this rate vector, $\lim_{n\rightarrow\infty}\lambda_{n}=0$.
In this case, we also say that the sequence of code $C_{n}$ achieves
rate $\boldsymbol{R}.$ A rate vector $\boldsymbol{R}$ is achievable
under a class of codes $\mathscr{C}$ if there is a sequence of codes
in $\mathscr{C}$ that achieves $\boldsymbol{R}.$ The closure of
the set of rate vectors $\boldsymbol{R}$ that are achievable under
$\mathscr{C}$ constitutes the rate region of $\mathscr{C}$. The
capacity region of the channel, $\mathbb{\mathcal{C}}$, is the closure
of the set of all achievable rates under the class of all codes.}

\section{\label{sec:Main-Results}Main Results}

In this section we present the main results of the paper. Since the
Primary transmitter is the owner of the channel and our intention
is not to degrade its performance, we concentrate on cases where cooperation
has the potential of increasing the Primary rate, i.e., $\epsilon_{3}^{1}\geq\epsilon_{3}^{2}$.
The next theorem provides an outer bound to system capacity. 
\begin{thm}
\label{thm:final}Let $\epsilon_{3}^{1}\geq\epsilon_{3}^{2}$. If
$(R_{1},R_{2})$ is achievable, then $\left(R_{1},R_{2}\right)\in\mathcal{R}$
where $\mathcal{R}$ is the region defined by, 
\begin{align}
\frac{R_{1}}{1-\epsilon_{23}^{1}}+\frac{R_{2}}{1-\epsilon_{4}^{2}} & \leq1-G-S-U,\label{eq:finfin1}\\
\left(\frac{\epsilon_{3}^{1}-\epsilon_{23}^{1}}{\left(1-\epsilon_{3}^{2}\right)\left(1-\epsilon_{23}^{1}\right)}+\frac{1}{1-\epsilon_{23}^{1}}\right)R_{1}+\frac{R_{2}}{1-\epsilon_{34}^{2}} & \leq1-G-S-U+\left(\frac{1-\epsilon_{3}^{1}}{1-\epsilon_{3}^{2}}\right)\left(G+S\right),\label{eq:finfin2}\\
\left(\frac{\epsilon_{34}^{1}-\epsilon_{234}^{1}}{\left(1-\epsilon_{234}^{1}\right)\left(1-\epsilon_{34}^{2}\right)}+\frac{1}{1-\epsilon_{23}^{1}}\right)R_{1}+\frac{1}{1-\epsilon_{4}^{2}}R_{2} & \leq1-G-S-U+\frac{1-\epsilon_{34}^{1}}{1-\epsilon_{34}^{2}}G+\frac{1-\epsilon_{4}^{1}}{1-\epsilon_{4}^{2}}\left(S+U\right)\label{eq:finfin3}
\end{align}
\[
G\geq0,S\geq0,U\geq0,R_{i}\geq0,i\in\left\{ 1,2\right\} .
\]
\end{thm}
\begin{IEEEproof}
The proof can be found in Appendix \ref{sec:Proof-of-Theorem}. 
\end{IEEEproof}
The next corollary provides a more concise description of the outer
bound in Theorem \ref{thm:final}. 
\begin{cor}
\label{cor:FinCor}If $\epsilon_{3}^{1}\geq\epsilon_{3}^{2}$, the
region $\mathcal{R}$ can be described as follows depending on system
erasure probabilities: 
\begin{enumerate}
\item \label{enu:cor1}If 
\begin{equation}
\max\left\{ \frac{1-\epsilon_{34}^{1}}{1-\epsilon_{34}^{2}},\frac{1-\epsilon_{4}^{1}}{1-\epsilon_{4}^{2}}\right\} \leq1,\label{eq:conda}
\end{equation}
then $\mathcal{R}=\mathcal{R}_{1}$ where $\mathcal{R}_{1}$ is defined
by the following inequalities 
\begin{align}
\left(\frac{\epsilon_{3}^{1}-\epsilon_{23}^{1}}{\left(1-\epsilon_{3}^{2}\right)\left(1-\epsilon_{23}^{1}\right)}+\frac{1}{1-\epsilon_{23}^{1}}\right)R_{1}+\frac{R_{2}}{1-\epsilon_{34}^{2}} & \leq1,\label{eq:newfin2a}\\
\left(\frac{\epsilon_{34}^{1}-\epsilon_{234}^{1}}{\left(1-\epsilon_{234}^{1}\right)\left(1-\epsilon_{34}^{2}\right)}+\frac{1}{1-\epsilon_{23}^{1}}\right)R_{1}+\frac{1}{1-\epsilon_{4}^{2}}R_{2} & \leq1,\label{eq:newfin3a}
\end{align}
\[
R_{i}\geq0,i\in\left\{ 1,2\right\} .
\]
\item \label{enu:cor2}If 
\begin{equation}
\max\left\{ \frac{1-\epsilon_{34}^{1}}{1-\epsilon_{34}^{2}},\frac{1-\epsilon_{4}^{1}}{1-\epsilon_{4}^{2}}\right\} =\frac{1-\epsilon_{34}^{1}}{1-\epsilon_{34}^{2}}>1,\label{eq:condb}
\end{equation}
then $\mathcal{R}=\mathcal{R}_{2}$ where $\mathcal{R}_{2}$ is defined
by the following inequalities, 
\begin{align}
G & \leq\frac{\epsilon_{34}^{1}-\epsilon_{234}^{1}}{\left(1-\epsilon_{234}^{1}\right)\left(1-\epsilon_{34}^{1}\right)}R_{1},\label{eq:r2a}\\
\left(\frac{\epsilon_{3}^{1}-\epsilon_{23}^{1}}{\left(1-\epsilon_{3}^{2}\right)\left(1-\epsilon_{23}^{1}\right)}+\frac{1}{1-\epsilon_{23}^{1}}\right)R_{1}+\frac{R_{2}}{1-\epsilon_{34}^{2}} & \leq1-G+\left(\frac{1-\epsilon_{3}^{1}}{1-\epsilon_{3}^{2}}\right)G,\label{eq:r2b}\\
\left(\frac{\epsilon_{34}^{1}-\epsilon_{234}^{1}}{\left(1-\epsilon_{234}^{1}\right)\left(1-\epsilon_{34}^{2}\right)}+\frac{1}{1-\epsilon_{23}^{1}}\right)R_{1}+\frac{1}{1-\epsilon_{4}^{2}}R_{2} & \leq1-G+\frac{1-\epsilon_{34}^{1}}{1-\epsilon_{34}^{2}}G,\label{eq:r2c}
\end{align}
\[
Q\geq0,\ R_{i}\geq0,i\in\left\{ 1,2\right\} .
\]
\item \label{enu:cor3}If 
\begin{equation}
\max\left\{ \frac{1-\epsilon_{34}^{1}}{1-\epsilon_{34}^{2}},\frac{1-\epsilon_{4}^{1}}{1-\epsilon_{4}^{2}}\right\} =\frac{1-\epsilon_{4}^{1}}{1-\epsilon_{4}^{2}}>1,\label{eq:condc}
\end{equation}
then $\mathcal{R}=\mathcal{R}_{3}$ where $\mathcal{R}_{3}$ is defined
by the following inequalities 
\begin{align}
S & \leq\frac{\epsilon_{34}^{1}-\epsilon_{234}^{1}}{\left(1-\epsilon_{234}^{1}\right)\left(1-\epsilon_{34}^{1}\right)}R_{1},\label{eq:3-first}\\
\left(\frac{\epsilon_{3}^{1}-\epsilon_{23}^{1}}{\left(1-\epsilon_{3}^{2}\right)\left(1-\epsilon_{23}^{1}\right)}+\frac{1}{1-\epsilon_{23}^{1}}\right)R_{1}+\frac{R_{2}}{1-\epsilon_{34}^{2}} & \leq1-S+\left(\frac{1-\epsilon_{3}^{1}}{1-\epsilon_{3}^{2}}\right)S,\label{eq:3-second}\\
\left(\frac{\epsilon_{34}^{1}-\epsilon_{234}^{1}}{\left(1-\epsilon_{234}^{1}\right)\left(1-\epsilon_{34}^{2}\right)}+\frac{1}{1-\epsilon_{23}^{1}}\right)R_{1}+\frac{1}{1-\epsilon_{4}^{2}}R_{2} & \leq1-S+\frac{1-\epsilon_{4}^{1}}{1-\epsilon_{4}^{2}}S,\label{eq:3-third}
\end{align}
\begin{equation}
S\geq0,R_{i}\geq0,i\in\left\{ 1,2\right\} .\label{eq:3-fifth}
\end{equation}
\end{enumerate}
\end{cor}
\begin{IEEEproof}
The proof can be found in Appendix \ref{sec:ProofOfFinCor}. 
\end{IEEEproof}
The next theorem expresses either the system capacity or an inner
bound to system capacity region, depending on system erasure probabilities. 
\begin{thm}
\label{thm:4}Let $\epsilon_{3}^{1}\geq\epsilon_{3}^{2}.$ 
\begin{enumerate}
\item \label{enu:Th4-part1}If 
\[
\max\left\{ \frac{1-\epsilon_{34}^{1}}{1-\epsilon_{34}^{2}},\frac{1-\epsilon_{4}^{1}}{1-\epsilon_{4}^{2}}\right\} \leq1,
\]
the system capacity region is $\mathcal{R}_{1}.$ 
\item \label{enu:Th4-part2}If 
\[
\max\left\{ \frac{1-\epsilon_{34}^{1}}{1-\epsilon_{34}^{2}},\frac{1-\epsilon_{4}^{1}}{1-\epsilon_{4}^{2}}\right\} =\frac{1-\epsilon_{34}^{1}}{1-\epsilon_{34}^{2}}>1,
\]
the system capacity region is $\mathcal{R}_{2}.$ 
\item \label{enu:Th4-part3}If 
\[
\max\left\{ \frac{1-\epsilon_{34}^{1}}{1-\epsilon_{34}^{2}},\frac{1-\epsilon_{4}^{1}}{1-\epsilon_{4}^{2}}\right\} =\frac{1-\epsilon_{4}^{1}}{1-\epsilon_{4}^{2}}>1,
\]
then an inner bound to system capacity region is the region described
by the equations below. 
\[
\left(\frac{\epsilon_{3}^{1}-\epsilon_{23}^{1}}{\left(1-\epsilon_{3}^{2}\right)\left(1-\epsilon_{23}^{1}\right)}+\frac{1}{1-\epsilon_{23}^{1}}\right)R_{1}+\frac{R_{2}}{1-\epsilon_{34}^{2}}\leq1-G-S-U+\left(\frac{1-\epsilon_{3}^{1}}{1-\epsilon_{3}^{2}}\right)\left(G+S\right),
\]
\[
\left(\frac{\epsilon_{34}^{1}-\epsilon_{234}^{1}}{\left(1-\epsilon_{234}^{1}\right)\left(1-\epsilon_{34}^{2}\right)}+\frac{1}{1-\epsilon_{23}^{1}}\right)R_{1}+\frac{1}{1-\epsilon_{4}^{2}}R_{2}\leq1-G-S-U+\frac{1-\epsilon_{34}^{1}}{1-\epsilon_{34}^{2}}G+\frac{1-\epsilon_{4}^{1}}{1-\epsilon_{4}^{2}}\left(S+U\right),
\]
\[
\left(\epsilon_{3}^{2}-\epsilon_{34}^{2}\right)G+\left(1-\epsilon_{34}^{2}\right)S\leq\frac{\left(\epsilon_{3}^{2}-\epsilon_{34}^{2}\right)\left(\epsilon_{34}^{1}-\epsilon_{234}^{1}\right)}{\left(1-\epsilon_{34}^{1}\right)\left(1-\epsilon_{234}^{1}\right)}R_{1},
\]
\[
\left(\epsilon_{3}^{2}-\epsilon_{34}^{2}\right)U\leq\left(\frac{\left(1-\epsilon_{4}^{2}\right)\left(1-\epsilon_{34}^{1}\right)}{\left(1-\epsilon_{4}^{1}\right)}-\left(\epsilon_{3}^{2}-\epsilon_{34}^{2}\right)\right)S,
\]
\[
G+\frac{\left(1-\epsilon_{4}^{1}\right)\left(1-\epsilon_{34}^{2}\right)}{\left(1-\epsilon_{4}^{2}\right)\left(1-\epsilon_{34}^{1}\right)}\left(U+S\right)\leq\frac{\left(\epsilon_{34}^{1}-\epsilon_{234}^{1}\right)}{\left(1-\epsilon_{34}^{1}\right)\left(1-\epsilon_{234}^{1}\right)}R_{1},
\]
\[
S\leq\frac{\left(\epsilon_{3}^{2}-\epsilon_{34}^{2}\right)\left(\epsilon_{4}^{2}-\epsilon_{34}^{2}\right)}{\left(1-\epsilon_{34}^{1}\right)\left(1-\epsilon_{4}^{2}\right)\left(1-\epsilon_{34}^{2}\right)}R_{2},
\]
\[
G\geq0,S\geq0,U\geq0,R_{i}\geq0,i\in\left\{ 1,2\right\} .
\]
\end{enumerate}
\end{thm}
\begin{IEEEproof}
The proof follows from the performance analysis of the transmission
algorithm that is proposed in Section \ref{sec:Coding-Algorithm}. 
\end{IEEEproof}
To examine the proximity of the inner and outer bound in part \ref{enu:Th4-part3}
of Theorem \ref{thm:4}, we conducted the following numerical investigation.
Assuming that erasure events are independent, all statistical parameters
of the system are determined by the erasure probabilities, $\epsilon_{j}^{1},\ j\in\left\{ 2,3,4\right\} ,\ \epsilon_{j}^{2},\ j\in\left\{ 3,4\right\} .$
We varied these probabilities from 0.1 to 0.9 in step 0.1 and kept
the values satisfying condition (\ref{eq:condc}). For these values
we varied the rate $R_{1}$ from $0.1B$ to $0.9B$ in steps of 0.05,
where
\[
B=\left(\frac{\epsilon_{3}^{1}-\epsilon_{23}^{1}}{\left(1-\epsilon_{3}^{2}\right)\left(1-\epsilon_{23}^{1}\right)}+\frac{1}{1-\epsilon_{23}^{1}}\right)^{-1},
\]
is the upper bound on $R_{1}$ determined by (\ref{eq:3-first})-(\ref{eq:3-fifth}).
Next, for a given rate $R_{1}$, based on the inequalities determining
the inner bound in part \ref{enu:Th4-part3} of Theorem \ref{thm:4}
we calculated the maximum rate $R_{2}$ as well as the rate $\hat{R}_{2}$
obtained using the inequalities of the outer bound, and registered
the relative deviation, 
\[
D=\frac{\hat{R}_{2}-R_{2}}{\hat{R}_{2}}.
\]

\begin{figure}
\includegraphics[scale=0.5]{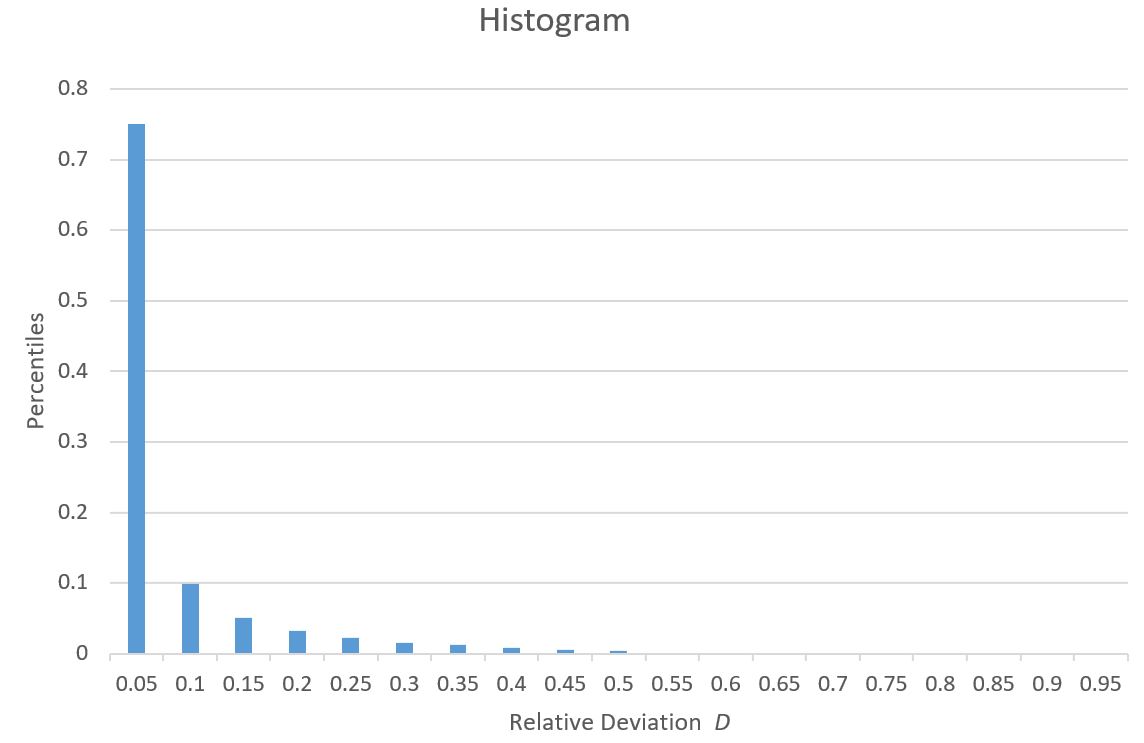} \caption{\label{fig:histogram}Histogram of Deviation Proportion $D$}
\end{figure}

In Figure \ref{fig:histogram} we present the histogram of this relative
deviation. We see that deviation smaller that 0.05 is achieved for
75\% of the cases. We note that most of larger deviations occur for
large values of $\epsilon_{4}^{1},\ \epsilon_{4}^{2}.$ For example,
if we restrict these values to be below 0.6, deviation of at most
$0.05$ occurs for $99.9\%$ of the cases, while the rest of the cases
have deviation between $0.05$ and $0.087$. It is worth noting that
the regions $R_{1}$ and $R_{2}$ described above are the same as
the throughput regions of the algorithms presented in \cite{URL:arxiv_version}.

\section{Transmission Algorithms\label{sec:Coding-Algorithm}}

In this section we present transmission algorithms that achieve the
rates described by Theorem \ref{thm:4}. For the reader's convenience,
we initially present an algorithm, Algorithm 1, whose description
is simple to follow and which achieves capacity under the condition
of part \ref{enu:Th4-part1} of Theorem \ref{thm:4}. Next, we describe
the general transmission algorithm, Algorithm 2, which achieves the
rates described in every part of Theorem \ref{thm:4}. 

In the description of the transmissions algorithms that follow, for
a given number of packets $k_{i}=\left\lceil nR_{i}\right\rceil $,
$n\in\left\lceil n\right\rceil $, instead of stopping after $n$
transmissions, packets are transmitted until all receivers receive
correctly all the packets destined to them - in general this requires
a random number of transmissions. By stopping the algorithm after
$n$ transmissions, and declaring an error if at least one receiver
does not receive all packets destined to it, we obtain an algorithm
that performs only $n$ transmission steps (please see the proof of
Proposition \ref{prop:Achievability-1}).

Furthermore, to simplify the description of the algorithms, we use
the following notation. Queue $Q_{i},\ i=1,2$, contains $\left\lceil nR_{i}\right\rceil $
packets, initially located at node $i$, that form message $\boldsymbol{W}_{i,n}$.
These packets must be delivered to node $i+2$; we refer to them as
``packets destined to node $i+2$'' or ``packets with origin node
$i$''. A generic symbol $X_{j,k\bar{l}}$ denotes a queue that is
located at node $j$, contains packets that were initially in queue
$Q_{j}$ (destined to node $j+2$), have been received by node $k$
and has not been received by node $l.$ Queue $X_{j,\bar{k}l}^{i}$
is located at node $j$, contains packets that were initially in queue
$Q_{i}$ (i.e. these packets have been received by node $j$ through
earlier transmissions of node $i$), have not been received by node
$k$ and have been received by node $l.$ A similar notation, with
small letters instead of capital, is used for packets. For example,
$x_{2,\bar{3}4}^{1}$ is a packet from queue $X_{2,\bar{3}4}^{1}$,
hence the packet was originally in $Q_{1}$, is located at node 2,
has been received by node 4 and has not been received by node 3. Note
that by definition all packets in $X_{2,\bar{3}4}^{1}$ are also in
$X_{4,2\bar{3}}^{1}$; $X_{2,\bar{3}4}^{1}$ is located at node 2,
while $X_{4,2\bar{3}}^{1}$ is located at node 4. Similarly, $X_{2,3\bar{4}}$
and $X_{3,\bar{4}}^{2}$ contain the same packets.

At some steps of the algorithms described below, XOR combinations
of packets from different queues may be sent. If the sent packet is
of the form $q=q_{1}\oplus q_{2}$, we say that packets $q_{1},\ q_{2}$
``constitute'' packet $q.$ We note that movements and insertion
of packets in queues can be done distributively by the nodes at which
the queues are located, by following the channel feedback.

\subsection{Description of Algorithm 1}

The full description of Algorithm 1 is given in detail in subsection
\ref{subsec:Detailed-description} and is summarized as follows. In
Step \ref{enu:Alg1-st1}, transmitter 1 sends packets from $Q_{1}$
until they are received by at least one of the nodes 2, 3; during
this process, packets that are received by node 4 are ``marked''
by node 1 and placed in buffer $B_{4,\bar{2}\bar{3}}^{1}$. At the
end of this step, $Q_{1}$ is empty, and queues $Q_{2,\bar{3}\bar{4}}^{1}$
, $Q_{2,\bar{3}4}^{1}$ and $Q_{4,2\bar{3}}^{1}$ may be nonempty.
In Step \ref{enu:Alg1-st2}, transmitter 2 sends packets from $Q_{2,\bar{3}\bar{4}}^{1}$
until they are received by at least one of the nodes 3 or 4, placing
packets that are erased at node 3 and received by node 4 in $Q_{2,\bar{3}4}^{1}$
and $Q_{4,2\bar{3}}^{1}$. In Step \ref{enu:Alg1-st3}, transmitter
2 sends packets from $Q_{2}$ until they are received by at least
one of the nodes 3, 4, placing packets that are erased at node 4 and
received by node 3 in $Q_{2,3\bar{4}}$ and $Q_{3,\bar{4}}^{2}$ .
At the end of this step, only queues $Q_{2,\bar{3}4}^{1}$ and $Q_{2,3\bar{4}}$
(and the corresponding queues $Q_{4,2\bar{3}}^{1},\ $$Q_{3,\bar{4}}^{2}$)
may be nonempty. Finally, in Step \ref{enu:Alg1-st4}, transmitter
2 sends XOR combinations of packets from $Q_{2,\bar{3}4}^{1}$ and
$Q_{2,3\bar{4}}$, i.e., packets of the form $q=q_{2,\bar{3}4}^{1}\oplus q_{2,3\bar{4}}$.
Since each of the nodes 3, 4 has already received one of the packets
that constitute the XOR combination of $q$, upon reception of $q$
the node can extract the packet that is destined to it. For example,
if node 4 receives $q,$ then since $q_{2,\bar{3}4}^{1}\in Q_{4,\bar{23}}^{1},$
node 4 can extract $q_{2,3\bar{4}}=q\oplus q_{2,\bar{3}4}^{1}$; packet
$q_{2,3\bar{4}}$ can therefore be removed from $Q_{2,3\bar{4}}$
and $Q_{3,\bar{4}}^{2}$. This process continues until one of the
queues $Q_{2,\bar{3}4}^{1}$, $Q_{2,3\bar{4}}$ empties; the packets
remaining in the nonempty queue (if any) are transmitted by node 2
until received by the corresponding destination.

\subsubsection{Detailed description\label{subsec:Detailed-description}}

$ $

\textbf{Algorithm 1}
\begin{enumerate}
\item \label{enu:Alg1-st1}If $Q_{1}$ is nonempty, transmitter 1 sends
packet $q$ from the head of line of $Q_{1}$ until it is received
by at least one of the nodes 2,3. 
\begin{enumerate}
\item \label{enu:Alg1St1a}If $q$ is received by node 3, it is removed
from $Q_{1}$. 
\item \label{enu:Alg1St1b}If $q$ is received by node $4$ and erased at
nodes 2, 3, it is ``marked'' by node 1 and placed in $B_{4,\bar{2}\bar{3}}^{1}$
(this queue is a buffer containing at most one packet). The packet
is re-transmitted by transmitter 1. 
\item \label{enu:Alg1St1c}If $q$ is received by nodes 2 and 4 and erased
at node 3, it is removed from $Q_{1}$ and placed in $Q_{2,\bar{3}4}^{1}$
and $Q_{4,2\bar{3}}^{1}.$ If $q$ is also ``marked'' by node 1,
hence it is in $B_{4,\bar{2}\bar{3}}^{1}$ (i.e., the $p$ has been
received earlier by node 4), $q$ is removed from this buffer. 
\item \label{enu:Alg1St1d}If $q$ is received by node 2, erased at nodes
3, 4 and and $q$ is ``marked'' (i.e. it has been received earlier
by node 4), it is removed from $Q_{1}$ and $B_{4,\bar{2}\bar{3}}^{1}$,
and placed in $Q_{2,\bar{3}4}^{1}$ and $Q_{4,2\bar{3}}^{1}$. 
\item If $q$ is received by node 2, erased at nodes 3, 4 and $q$ is not
``marked'' (i.e., the packet has not been received earlier by node
4), $q$ is removed from $Q_{1}$ and placed in $Q_{2,\bar{3}\bar{4}}^{1}.$ 
\end{enumerate}
\item \label{enu:Alg1-st2}If $Q_{2,\bar{3}\bar{4}}^{1}$ is nonempty, transmitter
2 sends packet $q$ from from the head of line of $Q_{2,\bar{3}\bar{4}}^{1}$
until it is received by at least one of the nodes 3, 4. 
\begin{enumerate}
\item If $q$ is received by node 3, it is removed removed from $Q_{2,\bar{3}\bar{4}}^{1}$. 
\item If $q$ is erased at node 3 and received by node 4, it is removed
from $Q_{2,\bar{3}\bar{4}}^{1}$ and placed in queue $Q_{2,\bar{3}4}^{1}$.
Also, $q$ is placed in $Q_{4,2\bar{3}}^{1}$. 
\end{enumerate}
\item \label{enu:Alg1-st3}If $Q_{2}$ is nonempty, transmitter 2 sends
packet $q$ from the head of line of $Q_{2}$ until it is received
by at least one of the nodes 3, 4. 
\begin{enumerate}
\item If $q$ is received by node 4, it is removed from $Q_{2}$. 
\item If $q$ is erased at node 4 and received by node 3, it is removed
from $Q_{2}$ and placed in queue $Q_{2,3\bar{4}}$. Also, $p$ is
placed in $Q_{3,\bar{4}}^{2}.$ 
\end{enumerate}
\item \label{enu:Alg1-st4}Transmitter 2 sends packet $q=q_{2,\bar{3}4}^{1}\oplus q_{2,3\bar{4}}$
where $q_{2,\bar{3}4}^{1},\ q_{2,3\bar{4}}$ are the packets at the
head of line of queues $Q_{2,\bar{3}4}^{1},\ Q_{2,3\bar{4}}$ respectively.
If $q$ is erased at both nodes 3, 4, it is re-transmitted. Else, 
\begin{enumerate}
\item If $q$ is received by node 3, packet $q_{2,\bar{3}4}^{1}$ is removed
from $Q_{2,\bar{3}4}^{1}$ and $Q_{4,2\bar{3}}^{1}$. 
\item If $q$ is received by node 4, packet $q_{2,3\bar{4}}$ is removed
from $Q_{2,3\bar{4}}$ and $Q_{3,\bar{4}}^{2}$. 
\end{enumerate}
This process continues until at least one of the queues $Q_{2,\bar{3}4}^{1},\ Q_{2,3\bar{4}}$
empties; Then, the remaining packets (if any) of the queue that is
nonempty, are sent by Transmitter 2 until they are received by their
destination. 
\end{enumerate}

\subsection{Description of Algorithm 2}

The full description of Algorithm 2 is given in detail in subsection
\ref{subsec:Detailed-description-1}. Below we provide the rationale
for the steps taken by Algorithm 2, in addition to those taken by
Algorithm 1. In algorithm 2, we introduce three parameters, $g,$
s, $u$, corresponding to, and motivated by, the operational interpretation
of the parameters $G,$ $S$, $U$ appearing in the capacity outer
bound in Theorem \ref{thm:final}. This interpretation can be seen
from the proof of the theorem.

In Step \ref{enu:Alg1-st2} of Algorithm 1, packets from $Q_{2,\bar{3}\bar{4}}^{1}$
are always re-transmitted by node 2 until they are received by at
least one of the nodes 3, 4. While it can be shown that this option
is optimal (i.e., the algorithm is capacity achieving) if relation
(\ref{eq:conda}) holds, it may be sub-optimal in other cases. Specifically
if (\ref{eq:conda}) does not hold, two possibilities for improving
the performance of the Algorithm 1 arise. 
\begin{enumerate}
\item If $\epsilon_{24}^{1}<\epsilon_{34}^{2}$, it may be beneficial for
node 1 to re-transmit a portion $g$ of packets that are received
by node 2 and not received by nodes 3 and 4 (i.e the packets in $Q_{2,\bar{3}\bar{4}}^{1}$).
This can be done by selecting each packet in $Q_{2,\bar{3}\bar{4}}^{1}$
to be re-transmitted by node $1$ with probability $g$. We place
the selected packets in queue $G_{1,2\bar{3}\bar{4}}$ (Step 1e of
Algorithm 2). Packets in this queue are re-transmitted by node 1 until
they are received by either of the nodes 3, 4 (Step \ref{enu:Alg2St2}
of Algorithm 2). 
\item If $\epsilon_{4}^{1}<\epsilon_{4}^{2}$, it may be beneficial for
node 1 (instead of node 2) to transmit packets that, if received by
node 4, permit this node to reconstruct packets destined to it. However,
for this to be possible, since node 1 never receives packets transmitted
by node 2, node 4 must be able to discover a packet destined to it
(i.e., a packet that was originally in $Q_{2}$) by receiving a packet
transmitted by node 1 (that was originally in $Q_{1}$). This can
be accomplished as follows. Suppose that node 2, instead of transmitting
packet $q_{2,\bar{3}\bar{4}}^{1}$, transmits 
\begin{equation}
q=q_{2,\bar{3}\bar{4}}^{1}\oplus q_{2,3\bar{4}}.\label{eq:netcoded1}
\end{equation}
 Consider the following cases. 
\begin{enumerate}
\item \label{enu:case2a}\emph{$q$ is received by node 3 and erased at
node 4:} Then node 3 recovers $q_{2,\bar{3}\bar{4}}^{1}$ hence this
packet is removed from $Q_{2,\bar{3}\bar{4}}^{1}.$ 
\item \label{enu:case2b}\emph{$q$ is received by node 4 and erased at
node 3:} Then node 4 \emph{cannot }recover packet $q_{2,3\bar{4}}$
since it has not received $q_{2,\bar{3}\bar{4}}^{1}.$ However, now
the following flexibility regarding future transmissions is obtained:
Node 1 has the ability to re-transmit $q_{2,\bar{3}\bar{4}}^{1}$.
Upon such a re-transmission, we observe the following cases.
\begin{enumerate}
\item \label{enu:case2bi}\emph{$q_{2,\bar{3}\bar{4}}^{1}$ is received
by both nodes 3 and 4}: Then both nodes recover the constituent packet
of $q$ that is destined to them (node 3 packet $q_{2,\bar{3}\bar{4}}^{1}$
and node 4 packet $q_{2,3\bar{4}}=q\oplus q_{2,\bar{3}\bar{4}}^{1}$).
\item \label{enu:case2bii}\emph{$q_{2,\bar{3}\bar{4}}^{1}$ is received
by node 4 and erased at node 3}: Then node 4 recovers the constituent
packet of $q$ destined to it; in addition, it also knows $q_{2,\bar{3}\bar{4}}^{1}$,
hence this packet can be placed in $Q_{2,\bar{3}4}^{1}$ and $Q_{4,2\bar{3}}^{1}$.
\item \label{enu:case2biii}\emph{$q_{2,\bar{3}\bar{4}}^{1}$ is received
by node 3 and erased at node 4:} Then $q_{2,\bar{3}\bar{4}}^{1}$
is removed from $Q_{2,\bar{3}\bar{4}}^{1}.$ Even though received
by node 3, packet $q_{2,\bar{3}\bar{4}}^{1}$ is still useful, since
node 4 can recover $q_{2,3\bar{4}}$ if node 1 re-transmits $q_{2,\bar{3}\bar{4}}^{1}$. 
\end{enumerate}
\item \label{enu:case2c}\emph{$q$ is received by both nodes 3 and 4}:
Again, even though received by node 3, packet $q_{2,\bar{3}\bar{4}}^{1}$
is still useful, since node 4 can recover $q_{2,3\bar{4}}$ if node
1 re-transmits $q_{2,\bar{3}\bar{4}}^{1}$. 
\end{enumerate}
Motivated by this reasoning, in Step 1e of Algorithm 2, we select
a portion $s$ of the packets in $Q_{2,\bar{3}\bar{4}}^{1}$ to be
transmitted coded by node 2 in the form (\ref{eq:netcoded1}). These
packets are placed in queues \emph{$S_{1,2\bar{3}\bar{4}}$ }and\emph{
$S_{2,\bar{3}\bar{4}}^{1}$. }

In Step \ref{enu:Alg2St5} of Algorithm 2, node 2 transmits packets
of the form $q=s_{2,\bar{3}\bar{4}}^{1}\oplus q_{2,3\bar{4}}$. At
this step, based on channel feedback, queues are formed that contain
coded packets; to emphasize this fact and with a slight abuse of notation,
these queues are denoted by the capital bold letter $\boldsymbol{A}.$
A queue at node $1$ that contains the constituent packets of queue
$\boldsymbol{A}$ with origin node 1, will be denoted by $A$. For
example, \emph{$\boldsymbol{A}_{2,\bar{3}4}$ }is a queue containing
\emph{coded} packets of the form $q=s_{2,\bar{3}\bar{4}}^{1}\oplus q_{2,3\bar{4}}$,
transmitted by node 2, erased at node 3 and received by node 4,\emph{
$\boldsymbol{A}_{4,\bar{3}}^{2}$ }is located at node 4 and contains
the same packets as $\boldsymbol{A}_{2,\bar{3}4}$ , and \emph{$A_{1,2\bar{3}4}$
}is located at node 1 and contains all of the constituent packets
of the coded packets in $\boldsymbol{A}_{2,\bar{3}4}$\emph{ }that
have origin node 1. The placement of packets in these queues for each
possible feedback is based on the corresponding cases \ref{enu:case2a},
\ref{enu:case2b}, \ref{enu:case2c} described in the previous paragraph.

In Step \ref{enu:Alg2St6} of Algorithm 2, node 1 transmits packets
from queue \emph{$A_{1,2\bar{3}4}$} . The placement of packets in
these queues for each possible feedback is based on the corresponding
cases \ref{enu:case2bi}, \ref{enu:case2bii}, \ref{enu:case2biii}
described in the penultimate paragraph. 

At the end of Step \ref{enu:Alg2St6} of Algorithm 2, queues \emph{$A_{1,234}$,
$\boldsymbol{A}_{2,34}$, $\boldsymbol{A}_{4,3}^{2}$} may be nonempty.
Note that all packets with destination node 3 that are constituents
of packets in \emph{$\boldsymbol{A}_{2,34}$, $\boldsymbol{A}_{4,3}^{2}$,
}have already been received by node 3, hence node 4 has to recover
only the packets with destination node 4 that are constituents of\emph{
}packets in these queues. Two options for recovering these packets
are the following: a) by transmitting the uncoded packets by node
2 or b) by having node 1 transmit packets from \emph{$A_{1,234}$
}and doing the appropriate decoding at node 4\emph{. }The latter option
may seem preferable since the channel from 1 to 4 is better than the
channel from 2 to 4, i.e., $\epsilon_{4}^{1}<\epsilon_{4}^{2}$. However,
there is a third option: node 2 may have the opportunity to transmit
these packets network coded while attempting to deliver packets from
$Q_{2,\bar{3}4}^{1}$ to node 3, hence in effect at no transmission
cost (Step \ref{enu:Alg2St8} of Algorithm 2). To address this trade-off,
in Step \ref{enu:Alg2St7} of Algorithm 2 we select a portion $u$
of the packets from \emph{$A_{1,234}$ }to be transmitted by node
1; the rest are transmitted by node 2 in Step \ref{enu:Alg2St8} of
Algorithm 2.
\end{enumerate}

\subsubsection{Detailed description\label{subsec:Detailed-description-1}}

$ $

\textbf{Algorithm 2}
\begin{enumerate}
\item \label{enu:Alg2St1}If $Q_{1}$ is nonempty, transmitter 1 sends packet
$q_{1}$ from the head of line of $Q_{1}$ until it is received by
at least one of the nodes 2,3. Steps 1a, 1b, 1c, 1d are the same as
steps \ref{enu:Alg1St1a}, \ref{enu:Alg1St1b},\ref{enu:Alg1St1c},
\ref{enu:Alg1St1d} of Algorithm 1 respectively. 
\begin{enumerate}
\item[e)] \label{enu:Alg2st1e}If $q_{1}$ is received by node 2, erased at
nodes 3, 4 and $q_{1}$ is not in $B_{4,\bar{2}\bar{3}}^{1}$ (i.e.,
the packet has not been received earlier by node 4), \textbf{$q_{1}$
}is removed from $Q_{1}$.\textbf{ }\emph{With probability $g$ the
packet is placed in queues $G_{1,2\bar{3}\bar{4}}$ and $G_{2,\bar{3}\bar{4}}^{1}$,
with probability $s$, where $g+s\leq1$, it is placed in queues $S_{1,2\bar{3}\bar{4}}$
and $S_{2,\bar{3}\bar{4}}^{1}$, and with probability $1-s-g$ the
packet is placed in $Q_{2,\bar{3}\bar{4}}^{1}$. } 
\end{enumerate}
\textbf{Possible nonempty queues at this point: }\emph{$G_{1,2\bar{3}\bar{4}}$,
$G_{2,\bar{3}\bar{4}}^{1}$, $Q_{2,\bar{3}\bar{4}}^{1}$, $Q_{2}$,
$S_{1,2\bar{3}\bar{4}}$, $S_{2,\bar{3}\bar{4}}^{1}$, }$Q_{2,\bar{3}4}^{1}$,
\emph{,}$Q_{4,\bar{23}}^{1}$\emph{. } 
\item \emph{\label{enu:Alg2St2}If $G_{1,2\bar{3}\bar{4}}$ is nonempty,
transmitter 1 sends packet $g_{1,2\bar{3}\bar{4}}$ from the head
of line of $G_{1,2\bar{3}\bar{4}}$ until it is received by} \emph{at
least one of} \emph{the nodes 3,4. } 
\begin{enumerate}
\item \emph{If $g_{1,2\bar{3}\bar{4}}$ is received by node 3, it is removed
from $G_{1,\bar{23}\bar{4}}$ and $G_{2,\bar{3}\bar{4}}^{1}$. } 
\item \emph{If $g_{1,2\bar{3}\bar{4}}$ is erased at by node 3 and received
by node 4, it is removed from $G_{1,\bar{23}\bar{4}}$ and $G_{2,\bar{3}\bar{4}}^{1}$,
and placed in $Q_{2,\bar{3}4}^{1}$ and $Q_{4,2\bar{3}}^{1}$}. 
\end{enumerate}
\textbf{Possible nonempty queues at this point:}\emph{ $Q_{2,\bar{3}\bar{4}}^{1}$,~$Q_{2}$,$S_{1,2\bar{3}\bar{4}}$,
$S_{2,\bar{3}\bar{4}}^{1}$, }$Q_{2,\bar{3}4}^{1}$, \emph{,}$Q_{4,2\bar{3}}^{1}$\emph{.
} 
\item \label{enu:Alg2St3}If $Q_{2,\bar{3}\bar{4}}^{1}$ is nonempty, transmitter
2 sends packet $q_{2,\bar{3}\bar{4}}^{1}$ from from the head of line
of $Q_{2,\bar{3}\bar{4}}^{1}$ until it is received by at least one
of the nodes 3, 4. The same actions as in Step \ref{enu:Alg1-st2}
of Algorithm 1 are taken.

\textbf{Possible nonempty queues at this point:}\emph{ $Q_{2}$, $S_{1,2\bar{3}\bar{4}}$,
$S_{2,\bar{3}\bar{4}}^{1}$, }$Q_{2,\bar{3}4}^{1}$, \emph{,}$Q_{4,2\bar{3}}^{1}$\emph{.
} 
\item \label{enu:Alg2St4}If $Q_{2}$ is nonempty, transmitter 2 sends packet
$q_{2}$ from the head of line of $Q_{2}$ until it is received by
at least one of the nodes 3, 4. The same actions as in Step \ref{enu:Alg1-st3}
of Algorithm 1 are taken.

\textbf{Possible nonempty queues at this point:}\emph{ $S_{1,2\bar{3}\bar{4}}$,
$S_{2,\bar{3}\bar{4}}^{1}$, }$Q_{2,\bar{3}4}^{1}$, \emph{,}$Q_{4,2\bar{3}}^{1}$,
$Q_{2,3\bar{4}}$, $Q_{3,\bar{4}}^{2}$. 
\item \emph{\label{enu:Alg2St5}If }\textbf{\emph{$S_{2,\bar{3}\bar{4}}^{1}$
}}\emph{(hence also $S_{1,2\bar{3}\bar{4}}$) and $Q_{2,3\bar{4}}$}\textbf{\emph{
}}\emph{are nonempty, transmitter 2 sends packet $q=s_{2,\bar{3}\bar{4}}^{1}\oplus q_{2,3\bar{4}}$
where $s_{2,\bar{3}4}^{1},\ q_{2,3\bar{4}}$ are the packets at the
head of line of queues $S_{2,\bar{3}\bar{4}}^{1},\ Q_{2,3\bar{4}}$
respectively, until $q$ is received by at least one of the nodes
3, 4. } 
\begin{enumerate}
\item \emph{If $q$ is received by node 3 and erased at node 4, packet $s_{2,\bar{3}\bar{4}}^{1}$
is removed from queues }\textbf{\emph{$S_{1,2\bar{3}\bar{4}}$ }}\emph{and}\textbf{\emph{
$S_{2,\bar{3}\bar{4}}^{1}$. }}\emph{The reason for this action is
that node 3 can recover $s_{2,\bar{3}\bar{4}}^{1}$ as $s_{2,\bar{3}\bar{4}}^{1}=q\oplus q_{2,3\bar{4}}$.} 
\item \emph{If $q$ is received by node 4 and erased at node 3, $q$ is
removed from }\textbf{\emph{$S_{2,\bar{3}\bar{4}}^{1}$ }}\emph{and
placed in $\boldsymbol{A}_{2,\bar{3}4}$ and $\boldsymbol{A}_{4,\bar{3}}^{2}$
Moreover, its constituent packet $s_{2,\bar{3}\bar{4}}^{1}$ is removed
from $S_{1,2\bar{3}\bar{4}}$ and placed in $A_{1,2\bar{3}4}$.} 
\item \emph{If $q$ is received by both nodes 3, 4, $q$ is removed from
}\textbf{\emph{$S_{2,\bar{3}\bar{4}}^{1}$ }}\emph{and placed in queue
$\boldsymbol{A}_{2,34}$ and $\boldsymbol{A}_{4,3}^{2}$. Moreover,
its constituent packet $s_{2,\bar{3}\bar{4}}^{1}$ is removed from
$S_{1,2\bar{3}\bar{4}}$ and placed in $A_{1,234}$.} 
\end{enumerate}
\textbf{Possible nonempty queues at this point: }\emph{$A_{1,2\bar{3}4}$,
$\boldsymbol{A}_{2,\bar{3}4}$, $\boldsymbol{A}_{4,\bar{3}}^{2}$,
$A_{1,234}$, $\boldsymbol{A}_{2,34}$, $\boldsymbol{A}_{4,3}^{2}$,
}$Q_{2,\bar{3}4}^{1}$\emph{,}$Q_{4,\bar{23}}^{1}$, $Q_{2,3\bar{4}}$,
$Q_{3,\bar{4}}^{2}$. 
\item \emph{\label{enu:Alg2St6}If $A_{1,2\bar{3}4}$ is nonempty, transmitter
1 transmits packet $a_{1,2\bar{3}4}$ from the head of line of $A_{1,2\bar{3}4}$
until the packet is received by at least one of the nodes 3, 4. } 
\begin{enumerate}
\item \emph{If $a_{1,2\bar{3}4}$ is received by both nodes 3, 4, $a_{1,2\bar{3}4}$
is removed from $A_{1,2\bar{3}\bar{4}}$. Moreover the packet whose
constituent is $h_{1,2\bar{3}4}$ in $\boldsymbol{A}_{2,\bar{3}4}$
and $\boldsymbol{A}_{4,\bar{3}}^{2}$ is also removed.} 
\item \emph{If $a_{1,2\bar{3}4}$ is received by node 4 and erased at node
3, $a_{1,2\bar{3}4}$ is removed from $A_{1,2\bar{3}4}$ and added
to $Q_{2,\bar{3}4}^{1}$ and $Q_{4,2\bar{3}}^{1}$. Moreover, the
packet whose constituent is $a_{1,2\bar{3}4}$ in $\boldsymbol{A}_{2,\bar{3}4}$
and $\boldsymbol{A}_{4,\bar{3}}^{2}$ is removed. } 
\item \emph{If $a_{1,2\bar{3}4}$ is received by node 3 and erased at node
4, $a_{1,2\bar{3}4}$ is moved to $A_{1,234}$. Moreover, the packet
whose constituent is $a_{1,2\bar{3}4}$ in $\boldsymbol{A}_{2,\bar{3}4}$
is moved to $\boldsymbol{A}_{2,34}$.} 
\end{enumerate}
\textbf{Possible nonempty queues at this point:}\emph{ $A_{1,234}$,
$\boldsymbol{A}_{2,34}$, $\boldsymbol{A}_{4,3}^{2}$, }$Q_{2,\bar{3}4}^{1}$,
\emph{,}$Q_{4,\bar{3}}^{1}$, $Q_{2,3\bar{4}}$, $Q_{3,\bar{4}}^{2}$. 
\item \emph{\label{enu:Alg2St7}If $A_{1,234}$ is nonempty, with probability
$1-u$ each packet $a_{1,234}\in A_{1,234}$ is removed from $A_{1,234},$
and the packet $q$ whose constituent is $a_{1,234}$ in $\boldsymbol{A}_{2,34}$
and $\boldsymbol{A}_{4,3}^{2}$, is removed. Moreover, the other constituent
of packet $q$ is moved to $Q_{2,3\bar{4}}.$ Node 1 re-transmits
any remaining packet $a_{1,234}$ in $A_{1,234}$ until it is received
by node 4, at which point the packet is removed from $A_{1,234}$;
moreover, the packet $q$ in $\boldsymbol{A}_{2,34}$, $\boldsymbol{A}_{4,3}^{2}$
whose constituent is $a_{1,234}$ is also removed (note that reception
of $a_{1,234}$ by node 4 enables the recovery of the constituent
of $q$ with destination node 4). }

\textbf{Possible nonempty queues at this point:}\emph{ }$Q_{2,\bar{3}4}^{1}$
\emph{,}$Q_{4,\bar{23}}^{1}$, $Q_{2,3\bar{4}}$, $Q_{3,\bar{4}}^{2}$. 
\item \label{enu:Alg2St8}Transmitter 2 sends packet $q=q_{2,\bar{3}4}^{1}\oplus q_{2,3\bar{4}}$
where $q_{2,\bar{3}4}^{1},\ q_{2,3\bar{4}}$ are the packets at the
head of line of queues $Q_{2,\bar{3}4}^{1},\ Q_{2,3\bar{4}}$ respectively.
The same actions as in Step \ref{enu:Alg1-st4} of Algorithm 1 are
taken. 
\end{enumerate}

\subsection{Performance Analysis of Algorithms\label{subsec:Performance-Analysis}}

The performance analysis of Algorithm 1 is done in a similar way as
in \cite{gatzianas2013multiuser,papadopoulos_journal}. Let $\left\lceil n\boldsymbol{R}\right\rceil =(\left\lceil nR_{1}\right\rceil ,\left\lceil nR_{2}\right\rceil )$
be the vector consisting of the number of packets destined to each
of the receivers. Let $T(\left\lceil n\boldsymbol{R}\right\rceil )$
be the (random) time it takes for all packets to be delivered to their
destinations when Algorithm 1 is employed. 
\begin{prop}
\label{prop:Perf_Alg_1}It holds,
\end{prop}
\begin{eqnarray}
\lim_{n\rightarrow\infty}\frac{T(\left\lceil n\boldsymbol{R}\right\rceil )}{n} & = & \max\left\{ R_{1}\left(\frac{\epsilon_{3}^{1}-\epsilon_{23}^{1}}{\left(1-\epsilon_{3}^{2}\right)\left(1-\epsilon_{23}^{1}\right)}+\frac{1}{1-\epsilon_{23}^{1}}\right)+\frac{R_{2}}{1-\epsilon_{34}^{2}},\right.\nonumber \\
 &  & \left.\frac{R_{2}}{1-\epsilon_{4}^{2}}+R_{1}\left(\frac{1}{1-\epsilon_{23}^{1}}+\frac{\epsilon_{34}^{1}-\epsilon_{234}^{1}}{\left(1-\epsilon_{234}^{1}\right)\left(1-\epsilon_{34}^{2}\right)}\right)\right\} \nonumber \\
 & = & \hat{T}\left(\boldsymbol{R}\right).\label{eq:T_complete}
\end{eqnarray}

\begin{IEEEproof}
We provide an outline of the arguments; detailed description can be
found in \cite{gatzianas2013multiuser}. Let $T_{i}$ be the time
it takes for Step $i$ of Algorithm 1 to complete. According to the
description of the algorithm, and based on the Strong Law of Large
Numbers, we derive the following limiting quantities at the end of
each step.

\begin{enumerate}

\item The following limit holds for time $T_{1}$,
\begin{equation}
\lim_{n\rightarrow\infty}\frac{T_{1}}{n}=\frac{R_{1}}{1-\epsilon_{23}^{1}}.\label{eq:T_1}
\end{equation}
Furthermore at the end of this step, it also hold for the number packets
that are placed in queues $Q_{2,\bar{3}\bar{4}}^{1}$ and $Q_{2,\bar{3}4}^{1}$
- for simplicity we denote the number of packets in a queue $X$ with
the same letter. 
\begin{eqnarray}
\lim_{n\rightarrow\infty}\frac{Q_{2,\bar{3}\bar{4}}^{1}}{n} & = & R_{1}\frac{\epsilon_{34}^{1}-\epsilon_{234}^{1}}{1-\epsilon_{234}^{1}},\label{eq:step1_queue1}\\
\lim_{n\rightarrow\infty}\frac{Q_{2,\bar{3}4}^{1}}{n} & = & R_{1}\left(\frac{\epsilon_{3}^{1}-\epsilon_{23}^{1}}{1-\epsilon_{23}^{1}}-\frac{\epsilon_{34}^{1}-\epsilon_{234}^{1}}{1-\epsilon_{234}^{1}}\right).\label{eq:step1_queue2}
\end{eqnarray}

\item The following limit holds for the time needed in order for
transmitter 2 to send the $Q_{2,\bar{3}\bar{4}}^{1}$ packets to either
node 3 or 4.
\begin{align}
\lim_{n\rightarrow\infty}\frac{T_{2}}{n} & =\lim_{n\rightarrow\infty}\frac{1}{n}\frac{Q_{2,\bar{3}\bar{4}}^{1}}{1-\epsilon_{34}^{2}}=R_{1}\frac{\epsilon_{34}^{1}-\epsilon_{234}^{1}}{\left(1-\epsilon_{234}^{1}\right)\left(1-\epsilon_{34}^{2}\right)}.\label{eq:T_2}
\end{align}
Moreover, if $M$ if the number of packets from $Q_{2,\bar{3}\bar{4}}^{1}$
that are received by node 3 and erased at node 4, it holds, 
\begin{equation}
\lim_{n\rightarrow\infty}\frac{M}{n}=\frac{\epsilon_{3}^{2}-\epsilon_{34}^{2}}{1-\epsilon_{34}^{2}}\lim_{n\rightarrow\infty}\frac{Q_{2,\bar{3}\bar{4}}^{1}}{n}=R_{1}\frac{\left(\epsilon_{34}^{1}-\epsilon_{234}^{1}\right)\left(\epsilon_{3}^{2}-\epsilon_{34}^{2}\right)}{\left(1-\epsilon_{234}^{1}\right)\left(1-\epsilon_{34}^{2}\right)}.\label{eq:T2M}
\end{equation}
At the end of this step, only queue $Q_{2,\bar{3}4}^{1}$ exists and
the new number of the packets in this queue, denoted as $\hat{Q}_{2,\bar{3}4}^{1}$,
is: $\hat{Q}_{2,\bar{3}4}^{1}=Q_{2,\bar{3}4}^{1}+M$. According to
(\ref{eq:step1_queue2}) and (\ref{eq:T2M}) we have, 
\begin{align}
\lim_{n\rightarrow\infty}\frac{\hat{Q}_{2,\bar{3}4}^{1}}{n} & =R_{1}\left(\frac{\epsilon_{3}^{1}-\epsilon_{23}^{1}}{1-\epsilon_{23}^{1}}\right)+R_{1}\frac{\epsilon_{34}^{1}-\epsilon_{234}^{1}}{1-\epsilon_{234}^{1}}\left(\frac{\epsilon_{3}^{2}-\epsilon_{34}^{2}}{1-\epsilon_{34}^{2}}-1\right).\label{eq:step2_queue2}
\end{align}

\item The following limit holds for the time needed in order for
transmitter 2 to send $\left\lceil nR_{2}\right\rceil $ packets to
either node 3 or 4 is given by: 
\begin{equation}
\lim_{n\rightarrow\infty}\frac{T_{3}}{n}=\frac{R_{2}}{1-\epsilon_{34}^{2}}.\label{eq:T_3}
\end{equation}
Furthermore, at the end of this step, it holds for the number of packets
in queue\textbf{ $Q_{2,3\bar{4}}$}, 
\begin{equation}
\lim_{n\rightarrow\infty}\frac{Q_{2,3\bar{4}}}{n}=R_{2}\frac{\epsilon_{4}^{2}-\epsilon_{34}^{2}}{1-\epsilon_{34}^{2}}.\label{eq:Step3_queue3}
\end{equation}

\item Let $T_{4,3},\ T_{4,4}$, be the time needed for node 2 to
deliver packets in queue $\hat{Q}_{2,\bar{3}4}^{1}$, $Q_{2,3\bar{4}},$
to destinations 3 and 4 respectively, if packets in these queues were
sent uncoded. It then holds, 
\begin{align*}
\lim_{n\rightarrow\infty}\frac{T_{4,3}}{n} & =\frac{\hat{Q}_{2,\bar{3}4}^{1}}{1-\epsilon_{3}^{2}},\\
\lim_{n\rightarrow\infty}\frac{T_{4,4}}{n} & =\frac{Q_{2,3\bar{4}}}{1-\epsilon_{4}^{2}}.
\end{align*}
Since at this step packets are sent coded whenever both queues $\hat{Q}_{2,\bar{3}4}^{1}$,
$Q_{2,3\bar{4}},$ are nonempty, we have, $T_{4}=\max\left\{ T_{4,3},T_{4,4}\right\} $,
hence according to (\ref{eq:step2_queue2}) and (\ref{eq:Step3_queue3}),
\begin{equation}
\lim_{n\rightarrow\infty}\frac{T_{4}}{n}=\max\left\{ \frac{R_{1}\frac{\epsilon_{3}^{1}-\epsilon_{23}^{1}}{1-\epsilon_{23}^{1}}+R_{1}\frac{\epsilon_{34}^{1}-\epsilon_{234}^{1}}{1-\epsilon_{234}^{1}}\left(\frac{\epsilon_{3}^{2}-\epsilon_{34}^{2}}{1-\epsilon_{34}^{2}}-1\right)}{1-\epsilon_{3}^{2}},\frac{R_{2}\frac{\epsilon_{4}^{2}-\epsilon_{34}^{2}}{1-\epsilon_{34}^{2}}}{1-\epsilon_{4}^{2}}\right\} .\label{eq:T_4-1}
\end{equation}

\end{enumerate}

Since the total time for the algorithm to complete is given by 
\[
T\left(\left\lceil n\boldsymbol{R}\right\rceil \right)=T_{1}+T_{2}+T_{3}+T_{4},
\]
taking into account (\ref{eq:T_1}), (\ref{eq:T_2}), (\ref{eq:T_3}),
and (\ref{eq:T_4-1}), (\ref{eq:T_complete}) arises.
\end{IEEEproof}
The next proposition provides a sufficient condition for achievability
and follows easily from Proposition \ref{prop:Perf_Alg_1}. 
\begin{prop}
\label{prop:Achievability-1}If the rate vector $\boldsymbol{R}=\left(R_{1},R_{2}\right)$
, $R_{i}\geq0,\ i=1,2,$satisfies,
\[
\max\left\{ R_{1}\left(\frac{\epsilon_{3}^{1}-\epsilon_{23}^{1}}{\left(1-\epsilon_{3}^{2}\right)\left(1-\epsilon_{23}^{1}\right)}+\frac{1}{1-\epsilon_{23}^{1}}\right)+\frac{R_{2}}{1-\epsilon_{34}^{2}},\frac{R_{2}}{1-\epsilon_{4}^{2}}+R_{1}\left(\frac{1}{1-\epsilon_{23}^{1}}+\frac{\epsilon_{34}^{1}-\epsilon_{234}^{1}}{\left(1-\epsilon_{234}^{1}\right)\left(1-\epsilon_{34}^{2}\right)}\right)\right\} <1
\]

then $\boldsymbol{R}$ is achievable. 
\end{prop}
\begin{IEEEproof}
The proof is identical to the proof in \cite[Appendix C-B]{gatzianas2013multiuser}.
We present it here for completeness. Consider the following code:\begin{enumerate}

\item Use Algorithm 1 to transmit $\left\lceil n\boldsymbol{R}\right\rceil $
packets. 

\item If $T(\left\lceil n\boldsymbol{R}\right\rceil )\le n$ then
transmit $n-T(\left\lceil n\boldsymbol{R}\right\rceil )$ arbitrary
packets and stop. In this case, both receivers receive correctly their
packets. 

\item Else declare error. 

The probability of error of this code is computed as follows.
\begin{eqnarray*}
\lim_{n\rightarrow\infty}p_{e}\left(n\right) & = & \lim_{n\rightarrow\infty}\Pr\left(T\left(\left\lceil n\boldsymbol{R}\right\rceil \right)>n\right)\\
 & = & \lim_{n\rightarrow\infty}\Pr\left(\frac{T\left(\left\lceil n\boldsymbol{R}\right\rceil \right)}{n}>1\right)\\
 & = & \lim_{n\rightarrow\infty}\Pr\left(\frac{T\left(\left\lceil n\boldsymbol{R}\right\rceil \right)}{n}-\hat{T}\left(\boldsymbol{R}\right)>1-\hat{T}\left(\boldsymbol{R}\right)\right)\\
 & = & 0\text{ by (\ref{eq:T_complete}).}
\end{eqnarray*}
\end{enumerate} 
\end{IEEEproof}
The performance analysis of Algorithm 2, although more complicated,
is similar. From this analysis it follows that the rate region of
Algorithm 2 is the set of pairs $\boldsymbol{R}=\left(R_{1},R_{2}\right)$
that satisfy the following relations.

\[
\left(\frac{\epsilon_{3}^{1}-\epsilon_{23}^{1}}{\left(1-\epsilon_{3}^{2}\right)\left(1-\epsilon_{23}^{1}\right)}+\frac{1}{1-\epsilon_{23}^{1}}\right)R_{1}+\frac{R_{2}}{1-\epsilon_{34}^{2}}\leq1-G-S-U+\left(\frac{1-\epsilon_{3}^{1}}{1-\epsilon_{3}^{2}}\right)\left(G+S\right),
\]
\[
\left(\frac{\epsilon_{34}^{1}-\epsilon_{234}^{1}}{\left(1-\epsilon_{234}^{1}\right)\left(1-\epsilon_{34}^{2}\right)}+\frac{1}{1-\epsilon_{23}^{1}}\right)R_{1}+\frac{1}{1-\epsilon_{4}^{2}}R_{2}\leq1-G-S-U+\frac{1-\epsilon_{34}^{1}}{1-\epsilon_{34}^{2}}G+\frac{1-\epsilon_{4}^{1}}{1-\epsilon_{4}^{2}}\left(S+U\right),
\]
\[
G=g\frac{R_{1}\left(\epsilon_{34}^{1}-\epsilon_{234}^{1}\right)}{\left(1-\epsilon_{34}^{1}\right)\left(1-\epsilon_{234}^{1}\right)},
\]
\[
S=s\frac{R_{1}\left(\epsilon_{3}^{2}-\epsilon_{34}^{2}\right)\left(\epsilon_{34}^{1}-\epsilon_{234}^{1}\right)}{\left(1-\epsilon_{34}^{1}\right)\left(1-\epsilon_{234}^{1}\right)\left(1-\epsilon_{34}^{2}\right)},
\]
\begin{align*}
U & =us\left(\frac{R_{1}\left(\epsilon_{34}^{1}-\epsilon_{234}^{1}\right)\left(1-\epsilon_{4}^{2}\right)}{\left(1-\epsilon_{4}^{1}\right)\left(1-\epsilon_{34}^{2}\right)\left(1-\epsilon_{234}^{1}\right)}-\frac{R_{1}\left(\epsilon_{3}^{2}-\epsilon_{34}^{2}\right)\left(\epsilon_{34}^{1}-\epsilon_{234}^{1}\right)}{\left(1-\epsilon_{34}^{1}\right)\left(1-\epsilon_{234}^{1}\right)\left(1-\epsilon_{34}^{2}\right)}\right),
\end{align*}
\[
g+s\leq1,\ u\leq1,\ g\geq0,\ s\geq0,u\geq0,S\geq0,R_{i}\geq0,i\in\left\{ 1,2\right\} .
\]
It can be seen that the region of pairs $\boldsymbol{R}=\left(R_{1},R_{2}\right)$
satisfying these relations is the same as the region defining the
inner bound in part \ref{enu:Th4-part3} of Theorem \ref{thm:4}.
Note that when $s=u=g=0$ ($u=s=0)$, the region is identical to the
outer bound in part \ref{enu:cor1} (\ref{enu:cor2}) of Corollary
\ref{cor:FinCor}, hence the algorithm is capacity achieving in theses
cases.

\section{\label{sec:Conclusion}Conclusion \& Further Work}

In this paper we developed an outer bound for the capacity of a fundamental
cooperative cognitive network. We distinguished three cases based
on the statistical parameters of the channel. Through the design of
appropriate algorithm, we showed that in the first two cases the outer
bound is indeed tight. For the third case, the rate region of the
developed algorithm is close to the outer bound for a wide range of
channel statistics.

Directions for future work include the investigation of benefits of
cooperation in the cases of multiple secondary user and/or primary
user pairs.

\appendices{}

\section{\label{sec:Proof-of-Theorem}Proof of Theorem \ref{lem:BasicLemma}}

In the following, for convenience in notation, we write $nR_{i}$
instead of $\left\lceil nR_{i}\right\rceil $, and we omit the index
$n$ whenever there is no possibility for confusion, e.g. we write
$\boldsymbol{W}_{i}$ and $k_{i}$ instead of $\boldsymbol{W}_{i,n}$
and $k_{i,n}$ respectively. Also we use base $\left|\mathcal{F}\right|$
for logarithms concerning information measures. Hence, since packets
are uniformly selected elements of $\mathcal{F},$ $H\left(W_{i,k}\right)=1,\ 1\leq k\leq k_{i}$
and since packets are independent, $H\left(\boldsymbol{W}_{i}\right)=k_{i}.$

\subsection{Preliminary Results}

In this subsection we present preliminary results that are used in
the development of the outer bound to system capacity in Section \ref{sec:Main-Results}.
The next lemma relates achievable rates to mutual information measures. 
\begin{lem}
\label{lem:achievable}Let $\left(R_{1},R_{2}\right)$ be achievable. 
\begin{itemize}
\item If $\left\{ 2,3\right\} \subseteq{\cal S}_{1}$ then 
\begin{equation}
0\leq nR_{1}-I(\boldsymbol{W}_{1};\boldsymbol{Y}_{1{\cal S}_{1}}^{n},\boldsymbol{Z}^{n}|\boldsymbol{W}_{2})\leq o\left(n\right).\label{eq:Lem2a}
\end{equation}
\item If $4\in{\cal S}_{2}$ then 
\begin{equation}
0\leq nR_{2}-I\left(\boldsymbol{W}_{2};\boldsymbol{Y}_{2{\cal S}_{2}}^{n},\boldsymbol{Z}^{n}|\boldsymbol{W}_{1}\right)\leq o(n).\label{eq:Lem2a1}
\end{equation}
\item If $3\in{\cal S}_{1}\cap{\cal S}_{2}$ then 
\begin{equation}
0\leq nR_{1}-I(\boldsymbol{W}_{1};\boldsymbol{Y}_{1{\cal S}_{1}}^{n},\boldsymbol{Y}_{2S_{2}}^{n},\boldsymbol{Z}^{n})\leq o\left(n\right),\label{eq:Lem2b}
\end{equation}
\begin{equation}
0\leq nR_{1}-I(\boldsymbol{W}_{1};\boldsymbol{Y}_{1{\cal S}_{1}}^{n},\boldsymbol{Y}_{2S_{2}}^{n},\boldsymbol{Z}^{n}|\boldsymbol{W}_{2})\leq o\left(n\right).\label{eq:Lem2b1}
\end{equation}
\item If $4\in{\cal S}_{1}\cap{\cal S}_{2}$ then 
\begin{equation}
0\leq nR_{2}-I(\boldsymbol{W}_{2};\boldsymbol{Y}_{1{\cal S}_{1}}^{n},\boldsymbol{Y}_{2S_{2}}^{n},\boldsymbol{Z}^{n})\leq o(n).\label{eq:Lem2c}
\end{equation}
\end{itemize}
\end{lem}
\begin{IEEEproof}
We prove (\ref{eq:Lem2a}). The rest of relations follow by similar
arguments.

\begin{align*}
nR_{1} & =H\left(\boldsymbol{W}_{1}\right)\mbox{ "indeq. uniformly distributed packets}"\\
 & =H\left(\boldsymbol{W}_{1}|\boldsymbol{W}_{2}\right)\text{ \ensuremath{\text{"}\boldsymbol{W}_{1}\perp\boldsymbol{W}_{2}\text{ "}}}\\
 & =I(\boldsymbol{W}_{1};\boldsymbol{Y}_{1{\cal S}_{1}}^{n},\boldsymbol{Z}^{n}|\boldsymbol{W}_{2})+H(\boldsymbol{W}_{1}|\boldsymbol{Y}_{1{\cal S}_{1}}^{n},\boldsymbol{Z}^{n},\boldsymbol{W}_{2})\\
 & =I(\boldsymbol{W}_{1};\boldsymbol{Y}_{1{\cal S}_{1}}^{n},\boldsymbol{Z}^{n}|\boldsymbol{W}_{2})+H\left(\boldsymbol{W}_{1}|\boldsymbol{Y}_{1{\cal S}_{1}}^{n},\boldsymbol{Z}^{n},\boldsymbol{W}_{2},\boldsymbol{Y}_{2\{3\}}^{n}\right)\ \text{"}\ensuremath{\boldsymbol{Y}_{2\{3\}}^{n}=f_{2}\left(\boldsymbol{W}_{2},\boldsymbol{Y}_{1\{2\}}^{n},\boldsymbol{Z}^{n}\right)\text{"}}\\
 & =I(\boldsymbol{W}_{1};\boldsymbol{Y}_{1{\cal S}_{1}}^{n},\boldsymbol{Z}^{n}|\boldsymbol{W}_{2})+H\left(\boldsymbol{W}_{1}|\boldsymbol{Y}_{1{\cal S}_{1}}^{n},\boldsymbol{Z}^{n},\boldsymbol{W}_{2},\boldsymbol{Y}_{2\{3\}}^{n},\hat{\boldsymbol{W}}_{1}\right)\ \text{"}\ensuremath{\hat{\boldsymbol{W}_{1}}=g_{3}(\boldsymbol{Y}_{1\{3\}}^{n},\boldsymbol{Y}_{2\left\{ 3\right\} }^{n},\boldsymbol{Z}^{n})\text{"}}\\
 & \leq I(\boldsymbol{W}_{1};\boldsymbol{Y}_{1{\cal S}_{1}}^{n},\boldsymbol{Z}^{n}|\boldsymbol{W}_{2})+H\left(\boldsymbol{W}_{1}|\hat{\boldsymbol{W}}_{1}\right)\\
 & \leq I(\boldsymbol{W}_{1};\boldsymbol{Y}_{1{\cal S}_{1}}^{n},\boldsymbol{Z}^{n}|\boldsymbol{W}_{2})+o\left(n\right)\ \text{"Fano inequality"}
\end{align*}
Also, 
\begin{align*}
nR_{1} & =H\left(\boldsymbol{W}_{1}|\boldsymbol{W}_{2}\right)\\
 & \geq I\left(\boldsymbol{W}_{1};\boldsymbol{Y}_{1{\cal S}_{1}}^{n},\boldsymbol{\boldsymbol{Z}}^{n}|\boldsymbol{W}_{2}\right).
\end{align*}
\end{IEEEproof}
The next lemma expresses mutual information measures appearing in
Lemma \ref{lem:achievable} in terms of more elementary ones, which
will be useful in the \textcolor{black}{development of the outer bound
in Section \ref{sec:Capacity-Outer-Bound}.} 
\begin{lem}
\label{lem:Erase}Let $\mathcal{S}_{i}\subseteq\mathcal{N}_{i},\ i\in\left\{ 1,2\right\} .$
For random variables $\boldsymbol{A},\ \boldsymbol{B},$ it holds,
\begin{multline}
I\left(\boldsymbol{A};\boldsymbol{Y}_{1{\cal S}_{1}}(t),\boldsymbol{Y}_{2S_{2}}(t)\mid\boldsymbol{B},\boldsymbol{Z}(t),\sigma(t)\right)=\sum_{\boldsymbol{z}\in{\cal Z}_{{\cal S}_{1}}^{1}}I\left(\boldsymbol{A};X_{1}(t)\mid\boldsymbol{B},\boldsymbol{Z}_{1}(t)=\boldsymbol{z},\sigma(t)=1\right)\Pr\left(\boldsymbol{Z}_{1}(t)=\boldsymbol{\boldsymbol{z}}\right)\Pr\left(\sigma(t)=1\right)\\
+\sum_{\boldsymbol{z}\in{\cal Z}_{{\cal S}_{2}}^{2}}I\left(\boldsymbol{A};X_{2}(t)\mid\boldsymbol{B},\boldsymbol{Z}_{2}(t)=\boldsymbol{z},\sigma(t)=2\right)\Pr\left(\boldsymbol{Z}_{2}(t)=\boldsymbol{\boldsymbol{z}}\right)\Pr\left(\sigma(t)=2\right).\label{eq:Erase1}
\end{multline}
For random variables $\boldsymbol{Q}$ and $\boldsymbol{P}$, if $\boldsymbol{Z}(t)$
is independent of $\left(\boldsymbol{Q},X_{1}(t),X_{2}(t),\boldsymbol{P},\boldsymbol{Y}_{1{\cal S}_{1}}^{t-1},\boldsymbol{Y}_{2S_{2}}^{t-1},\boldsymbol{Z}^{t-1}\right)$
then 
\begin{align}
I\left(\boldsymbol{Q};\boldsymbol{Y}_{1{\cal S}_{1}}^{n},\boldsymbol{Y}_{2S_{2}}^{n},\boldsymbol{Z}^{n}\mid\boldsymbol{P}\right) & =\left(1-\epsilon_{{\cal S}_{1}}^{1}\right)\sum_{t=1}^{n}I\left(\boldsymbol{Q};X_{1}(t)\mid\boldsymbol{P},\boldsymbol{Y}_{1{\cal S}_{1}}^{t-1},\boldsymbol{Y}_{2S_{2}}^{t-1},\boldsymbol{Z}^{t-1},\sigma(t)=1\right)\Pr\left(\sigma(t)=1\right)\nonumber \\
+ & \left(1-\epsilon_{{\cal S}_{2}}^{2}\right)\sum_{t=1}^{n}I\left(\boldsymbol{Q};X_{2}(t)\mid\boldsymbol{P},\boldsymbol{Y}_{1{\cal S}_{1}}^{t-1},\boldsymbol{Y}_{2S_{2}}^{t-1},\boldsymbol{Z}^{t-1},\sigma(t)=2\right)\Pr\left(\sigma(t)=2\right).\label{eq:Erase4}
\end{align}
\end{lem}
\begin{IEEEproof}
To show (\ref{eq:Erase1}), using the definition of conditional mutual
information we have, 
\begin{align}
 & I\left(\boldsymbol{A};\boldsymbol{Y}_{1{\cal S}_{1}}(t),\boldsymbol{Y}_{2S_{2}}(t)\mid\boldsymbol{B},\boldsymbol{Z}(t),\sigma(t)\right)\nonumber \\
 & =\sum_{z\in\mathcal{Z}_{1}}I\left(\boldsymbol{A};\boldsymbol{Y}_{1{\cal S}_{1}}(t),\boldsymbol{Y}_{2S_{2}}(t)\mid\boldsymbol{B},\boldsymbol{Z}_{1}(t)=\boldsymbol{z},\sigma(t)=1\right)\Pr\left(\boldsymbol{Z}_{1}(t)=\boldsymbol{z},\sigma(t)=1\right)\nonumber \\
 & +\sum_{\boldsymbol{z}\in\mathcal{Z}_{2}}I\left(\boldsymbol{A};\boldsymbol{Y}_{1{\cal S}_{1}}(t),\boldsymbol{Y}_{2S_{2}}(t)\mid\boldsymbol{B},\boldsymbol{Z}(t)=\boldsymbol{z},\sigma(t)=2\right)\Pr\left(\boldsymbol{Z}_{2}(t)=\boldsymbol{z},\sigma(t)=2\right)\nonumber \\
 & =\sum_{z\in\mathcal{Z}_{1}}I\left(\boldsymbol{A};\boldsymbol{Y}_{1{\cal S}_{1}}(t)\mid\boldsymbol{B},\boldsymbol{Z}_{1}(t)=\boldsymbol{z},\sigma(t)=1\right)\Pr\left(\boldsymbol{Z}_{1}(t)=\boldsymbol{z}\right)\Pr\left(\sigma(t)=1\right)\nonumber \\
 & +\sum_{\boldsymbol{z}\in\mathcal{Z}_{2}}I\left(\boldsymbol{A};\boldsymbol{Y}_{2{\cal S}_{2}}(t)\mid\boldsymbol{B},\boldsymbol{Z}_{2}(t)=\boldsymbol{z},\sigma(t)=2\right)\Pr\left(\boldsymbol{Z}_{2}(t)=\boldsymbol{z}\right)\Pr\left(\sigma(t)=2\right)\nonumber \\
 & \ \text{\text{ "\ensuremath{\boldsymbol{Z}\left(t\right)\perp\sigma\left(t\right)} \ensuremath{\text{and }}}\mbox{if \ensuremath{\sigma(t)=i} \ensuremath{\mbox{then}} \ensuremath{\boldsymbol{Y}_{i^{c}{\cal S}_{i^{c}}}(t)}=\ensuremath{\boldsymbol{\eta}}." }}\nonumber \\
 & =\sum_{\boldsymbol{z}\in{\cal Z}_{{\cal S}_{1}}^{1}}I\left(\boldsymbol{A};\boldsymbol{Y}_{1{\cal S}_{1}}(t)\mid\boldsymbol{B},\boldsymbol{Z}_{1}(t)=\boldsymbol{z},\sigma(t)=1\right)\Pr\left(\boldsymbol{Z}_{1}(t)=\boldsymbol{z}\right)\Pr\left(\sigma(t)=1\right)\nonumber \\
 & +\sum_{\boldsymbol{z}\in{\cal Z}_{{\cal S}_{2}}^{2}}I\left(\boldsymbol{A};\boldsymbol{Y}_{2{\cal S}_{2}}(t)\mid\boldsymbol{B},\boldsymbol{Z}_{2}(t)=\boldsymbol{z},\sigma(t)=2\right)\Pr\left(\boldsymbol{Z}_{2}(t)=\boldsymbol{z}\right)\Pr\left(\sigma(t)=2\right)\nonumber \\
 & \ \text{"if \ensuremath{z\notin\mathcal{Z}_{\mathcal{S}_{1}}^{i}} then \ensuremath{\boldsymbol{Y}_{i\mathcal{S}_{i}}(t)=\boldsymbol{\varepsilon}}." }\nonumber \\
 & =\sum_{\boldsymbol{z}\in{\cal Z}_{{\cal S}_{1}}^{1}}I\left(\boldsymbol{A};X_{1}(t)\mid\boldsymbol{B},\boldsymbol{Z}_{1}(t)=\boldsymbol{z},\sigma(t)=1\right)\Pr\left(\boldsymbol{Z}_{1}(t)=\boldsymbol{z}\right)\Pr\left(\sigma(t)=1\right)\nonumber \\
 & +\sum_{\boldsymbol{z}\in{\cal Z}_{{\cal S}_{2}}^{2}}I\left(\boldsymbol{A};X_{2}(t)\mid\boldsymbol{B},\boldsymbol{Z}_{2}(t)=\boldsymbol{z},\sigma(t)=2\right)\Pr\left(\boldsymbol{Z}_{2}(t)=\boldsymbol{z}\right)\Pr\left(\sigma(t)=2\right)\label{eq:hh1}\\
 & \text{ \text{ \mbox{"Fact \ref{fact:Def}, item }\ref{enu:one-to-one}}."}\nonumber 
\end{align}

To show equality (\ref{eq:Erase4}) we apply first the chain rule
to obtain, 
\[
I\left(\boldsymbol{Q};\boldsymbol{Y}_{1{\cal S}_{1}}^{n},\boldsymbol{Y}_{2S_{2}}^{n},\boldsymbol{Z}^{n}\mid\boldsymbol{P}\right)=\sum_{t=1}^{n}I\left(\boldsymbol{Q};\boldsymbol{Y}_{1{\cal S}_{1}}(t),\boldsymbol{Y}_{2S_{2}}(t),\boldsymbol{Z}(t)\mid\boldsymbol{P},\boldsymbol{Y}_{1{\cal S}_{1}}^{t-1},\boldsymbol{Y}_{2S_{2}}^{t-1},\boldsymbol{Z}^{t-1}\right).
\]

Next, we write, 
\begin{align}
 & I\left(\boldsymbol{Q};\boldsymbol{Y}_{1{\cal S}_{1}}(t),\boldsymbol{Y}_{2S_{2}}(t),\boldsymbol{Z}(t)\mid\boldsymbol{P},\boldsymbol{Y}_{1{\cal S}_{1}}^{t-1},\boldsymbol{Y}_{2S_{2}}^{t-1},\boldsymbol{Z}^{t-1}\right)\nonumber \\
 & =I\left(\boldsymbol{Q};\boldsymbol{Z}(t)\mid\boldsymbol{P},\boldsymbol{Y}_{1{\cal S}_{1}}^{t-1},\boldsymbol{Y}_{2S_{2}}^{t-1},\boldsymbol{Z}^{t-1}\right)+I\left(\boldsymbol{Q};\boldsymbol{Y}_{1{\cal S}_{1}}(t),\boldsymbol{Y}_{2S_{2}}(t)\mid\boldsymbol{P},\boldsymbol{Y}_{1{\cal S}_{1}}^{t-1},\boldsymbol{Y}_{2S_{2}}^{t-1},\boldsymbol{Z}^{t-1},\boldsymbol{Z}(t)\right)\nonumber \\
 & =I\left(\boldsymbol{Q};\boldsymbol{Y}_{1{\cal S}_{1}}(t),\boldsymbol{Y}_{2S_{2}}(t)\mid\boldsymbol{P},\boldsymbol{Y}_{1{\cal S}_{1}}^{t-1},\boldsymbol{Y}_{2S_{2}}^{t-1},\boldsymbol{Z}^{t-1},\boldsymbol{Z}(t)\right)\mbox{ \ensuremath{\text{ "}\boldsymbol{Z}(t)\perp\left(\boldsymbol{Q},\boldsymbol{P},\boldsymbol{Y}_{1{\cal S}_{1}}^{t-1},\boldsymbol{Y}_{2S_{2}}^{t-1},\boldsymbol{Z}^{t-1}\right)\text{"}}}\nonumber \\
 & =I\left(\boldsymbol{Q};\boldsymbol{Y}_{1{\cal S}_{1}}(t),\boldsymbol{Y}_{2S_{2}}(t)\mid\boldsymbol{P},\boldsymbol{Y}_{1{\cal S}_{1}}^{t-1},\boldsymbol{Y}_{2S_{2}}^{t-1},\boldsymbol{Z}^{t-1},\boldsymbol{Z}(t),\sigma(t)\right)\ \mbox{\ensuremath{\text{"}\sigma(t)=\sigma(Z^{t-1})}"}\nonumber \\
 & =\sum_{\boldsymbol{z}\in{\cal Z}_{{\cal S}_{1}}^{1}}I\left(\boldsymbol{Q};X_{1}(t)\mid\boldsymbol{P},\boldsymbol{Y}_{1{\cal S}_{1}}^{t-1},\boldsymbol{Y}_{2S_{2}}^{t-1},\boldsymbol{Z}^{t-1},\boldsymbol{Z}(t)=\boldsymbol{z},\sigma(t)=1\right)\Pr\left(\boldsymbol{Z}_{1}(t)=\boldsymbol{z}\right)\Pr\left(\sigma(t)=1\right)\nonumber \\
 & +\sum_{\boldsymbol{z}\in{\cal Z}_{{\cal S}_{2}}^{2}}I\left(\boldsymbol{Q};X_{2}(t)\mid\boldsymbol{P},\boldsymbol{Y}_{1{\cal S}_{1}}^{t-1},\boldsymbol{Y}_{2S_{2}}^{t-1},\boldsymbol{Z}^{t-1},\boldsymbol{Z}(t)=\boldsymbol{z},\sigma(t)=2\right)\Pr\left(\boldsymbol{Z}_{2}(t)=\boldsymbol{z}\right)\Pr\left(\sigma(t)=2\right)\text{ \text{"by }\ref{eq:Erase1}),"}\nonumber \\
 & =\sum_{\boldsymbol{z}\in{\cal Z}_{{\cal S}_{1}}^{1}}I\left(\boldsymbol{Q};X_{1}(t)\mid\boldsymbol{P},\boldsymbol{Y}_{1{\cal S}_{1}}^{t-1},\boldsymbol{Y}_{2S_{2}}^{t-1},\boldsymbol{Z}^{t-1},\sigma(t)=1\right)\Pr\left(\boldsymbol{Z}_{1}(t)=\boldsymbol{z}\right)\Pr\left(\sigma(t)=1\right)\nonumber \\
 & +\sum_{\boldsymbol{z}\in{\cal Z}_{{\cal S}_{2}}^{2}}I\left(\boldsymbol{Q};X_{2}(t)\mid\boldsymbol{P},\boldsymbol{Y}_{1{\cal S}_{1}}^{t-1},\boldsymbol{Y}_{2S_{2}}^{t-1},\boldsymbol{Z}^{t-1},\sigma(t)=2\right)\Pr\left(\boldsymbol{Z}_{2}(t)=\boldsymbol{z}\right)\Pr\left(\sigma(t)=2\right)\nonumber \\
 & \text{ "}Z(t)\perp\left(\boldsymbol{Q},X_{1}(t),X_{2}(t),\boldsymbol{P},\boldsymbol{Y}_{1{\cal S}_{1}}^{t-1},\boldsymbol{Y}_{2S_{2}}^{t-1},\boldsymbol{Z}^{t-1}\right)\text{"}\nonumber \\
 & =\left(1-\epsilon_{{\cal S}_{1}}^{1}\right)I\left(\boldsymbol{Q};X_{1}(t)\mid\boldsymbol{P},\boldsymbol{Y}_{1{\cal S}_{1}}^{t-1},\boldsymbol{Y}_{2S_{2}}^{t-1},\boldsymbol{Z}^{t-1},\sigma(t)=1\right)\Pr\left(\sigma(t)=1\right)\nonumber \\
 & +\left(1-\epsilon_{{\cal S}_{2}}^{2}\right)I\left(\boldsymbol{Q};X_{2}(t)\mid\boldsymbol{P},\boldsymbol{Y}_{1{\cal S}_{1}}^{t-1},\boldsymbol{Y}_{2S_{2}}^{t-1},\boldsymbol{Z}^{t-1},\sigma(t)=2\right)\Pr\left(\sigma(t)=2\right).\label{eq:hh2}
\end{align}
Equality (\ref{eq:Erase4}) follows from (\ref{eq:hh1}) and (\ref{eq:hh2}). 
\end{IEEEproof}

\subsubsection{Node Scheduling Times}

In this subsection we define certain node scheduling times and relate
them to information theoretic quantities. These relations are needed
for the development of the outer bound. 
\begin{itemize}
\item $T_{i},\ i\in\left\{ 1,2\right\} :$ the number of times $t,$ $1\leq t\leq n$,
that node $i$ is scheduled to transmit, i.e., 
\[
T_{i}=\sum_{t=1}^{n}I_{\left\{ \sigma\left(t\right)=1\right\} },
\]
where $I_{\mathcal{A}}$ denotes the indicator function of event $\mathcal{A}$.
Clearly, $T_{1}+T_{2}=n$, hence, denoting $\tilde{\tau}_{i}=E\left[T_{i}\right]$
we have, 
\begin{equation}
\tilde{\tau}_{1}+\tilde{\tau}_{2}=n,\label{eq:sum-times}
\end{equation}
where 
\begin{equation}
\tilde{\tau_{i}}=\sum_{t=1}^{n}\Pr\left(\sigma(t)=i\right).\label{eq:avtimes}
\end{equation}
\item At time $t$ let $H_{1,i}$$\left(t\right)$ be the index of packet
transmitted by node 1 and received by node $i$. If node 1 does not
transmit at time $t,$ or if node 1 transmits but the packet is erased
at node $i,$ we set the packet index to null, $\eta.$ Similarly,
we define $H_{1,\bar{i}}$$\left(t\right)$ the index of packet transmitted
by node 1 and \emph{not }received by node $i$. If node 1 does not
transmit at time $t,$ or if node 1 transmits but the packet is \emph{received}
by node $i,$ we set the packet index to null, $\eta.$ We extend
this definition to packet indices determined by functions $\phi(i,j,k)$
of node indices involving logical AND $(\wedge)$ , OR ($\vee)$ and
NOT ($\bar{x})$ operations. For example, if $\phi(i,j)=i\wedge\bar{j}$,
then $H_{1,i\wedge\bar{j}}(t)$ is the index of packets in $\boldsymbol{W}_{1}$
, transmitted by node 1, received by node $i$ and not received by
node $j$ at time $t$ (and null in the rest of the cases). Similarly,
$H_{1,i\vee j}(t)$ is the index of packet in $\boldsymbol{W}_{1}$,
transmitted by node 1 and received either by node $i$ or by node
$j$ (or both) at time $t$. This is the index of the packet in the
vector $\boldsymbol{Y}_{1,\left\{ i,j\right\} }(t)$. We now define
the following scheduling times. 
\begin{itemize}
\item $T_{1,\phi\left(i,j,k\right)}:$ number of times $t,$ $1\leq t\leq n,$
that node 1 transmits one of the packets in $\boldsymbol{W}_{1}$
with index in $\boldsymbol{H}_{1,\phi\left(i,j,k\right)}^{t-1}$,
i.e., 
\begin{equation}
T_{1,\phi\left(i,j,k\right)}=\sum_{t=1}^{n}I\left(\sigma\left(t\right)=1,J\left(\boldsymbol{Z}^{t-1}\right)\in\boldsymbol{H}_{1,\phi\left(i,j,k\right)}^{t-1}\right).\label{eq:gentimes}
\end{equation}
From the definitions, setting $\tilde{\tau}_{1,\phi\left(i,j,k\right)}=E\left[T_{1,\phi\left(i,j,k\right)}\right]$
it follows that 
\begin{equation}
\tilde{\tau}_{1,\phi\left(i,j,k\right)}+\tilde{\tau}_{1,\overline{\phi\left(i,j,k\right)}}=\tilde{\tau}_{1},\label{eq:gensumtimes}
\end{equation}
\begin{equation}
\tilde{\tau}_{1,\phi\left(i,j,k\right)}=\sum_{t=1}^{n}\Pr\left(\sigma\left(t\right)=1,J\left(\boldsymbol{Z}^{t-1}\right)\in\boldsymbol{H}_{1,\phi\left(i,j,k\right)}^{t-1}\right).\label{eq:genavtimes}
\end{equation}
\end{itemize}
\end{itemize}
We now discuss some properties that follow from the fact that node
1 performs only packet scheduling operations based on channel feedback.
Note that for a given feedback $\boldsymbol{z}^{t-1},$ the packet
indices in $h_{i,\phi\left(\cdot\right)}^{t-1}$ are completely determined.
For the conditional probabilities below, the conditioning event is
assumed to be nonempty. 
\begin{lem}
\label{lem:independent}a)For any $\mathcal{S}\subseteq\left\{ 2,3,4\right\} $,
if $\boldsymbol{Z}^{t-1}=\boldsymbol{z}^{t-1},\sigma\left(\boldsymbol{z}^{t-1}\right)=1$
and $J\left(\boldsymbol{z}^{t-1}\right)\in h_{1,\land{}_{i\in\mathcal{S}}\bar{i}}^{t-1}$,
it holds for any $m\in\mathcal{F},$ $\boldsymbol{y}_{1\mathcal{S}}^{t-1}$,
\begin{equation}
\Pr\left\{ X_{1}\left(t\right)=m|\boldsymbol{Y}_{1\mathcal{S}}^{t-1}=\boldsymbol{y}_{1\mathcal{S}}^{t-1},\ \boldsymbol{Z}^{t-1}=\boldsymbol{z}^{t-1},\sigma\left(\boldsymbol{z}^{t-1}\right)=1\right\} =\frac{1}{\left|\mathcal{F}\right|},\label{eq:cond1}
\end{equation}
\begin{equation}
\Pr\left\{ X_{1}\left(t\right)=m|\boldsymbol{Y}_{1\mathcal{S}}^{t-1}=\boldsymbol{y}_{1\mathcal{S}}^{t-1},\ \boldsymbol{Z}^{t-1}=\boldsymbol{z}^{t-1},\boldsymbol{W}_{2}=\boldsymbol{w_{2}},\sigma\left(\boldsymbol{z}^{t-1}\right)=1\right\} =\frac{1}{\left|\mathcal{F}\right|}.\label{eq:cond2}
\end{equation}
b) For any $\mathcal{S}_{1}\subseteq\mathcal{N}_{1},\ \mathcal{S}_{2}\subseteq\mathcal{N}_{2}$,
if $\boldsymbol{Z}^{t-1}=\boldsymbol{z}^{t-1},\sigma\left(\boldsymbol{z}^{t-1}\right)=1$
and $J\left(\boldsymbol{z}^{t-1}\right)\in h_{1,\bar{2}\land{}_{i\in\mathcal{S}_{1}}\bar{i}}^{t-1}$,
it holds for any $m\in\mathcal{F},$ $\boldsymbol{y}_{1\mathcal{S}_{1}}^{t-1},\ \boldsymbol{y}_{2\mathcal{S}_{2}}^{t-1}$,
\begin{equation}
\Pr\left\{ X_{1}\left(t\right)=m|\boldsymbol{Y}_{1\mathcal{S}_{1}}^{t-1}=\boldsymbol{y}_{1\mathcal{S}_{1}}^{t-1},\boldsymbol{Y}_{2\mathcal{S}_{2}}^{t-1}=\boldsymbol{y}_{2\mathcal{S}_{2}}^{t-1}\ \boldsymbol{Z}^{t-1}=\boldsymbol{z}^{t-1},\boldsymbol{W}_{2}=\boldsymbol{w}_{2},\sigma\left(\boldsymbol{z}^{t-1}\right)=1\right\} =\frac{1}{\left|\mathcal{F}\right|},\label{eq:cond3}
\end{equation}
\begin{equation}
\Pr\left\{ X_{1}\left(t\right)=m|\boldsymbol{Y}_{1\mathcal{S}_{1}}^{t-1}=\boldsymbol{y}_{1\mathcal{S}_{1}}^{t-1},\boldsymbol{Y}_{2\mathcal{S}_{2}}^{t-1}=\boldsymbol{y}_{2\mathcal{S}_{2}}^{t-1}\ \boldsymbol{Z}^{t-1}=\boldsymbol{z}^{t-1},\sigma\left(\boldsymbol{z}^{t-1}\right)=1\right\} =\frac{1}{\left|\mathcal{F}\right|}.\label{eq:cond4}
\end{equation}
\end{lem}
\begin{IEEEproof}
To show (\ref{eq:cond1}) notice that since $J\left(\boldsymbol{z}^{t-1}\right)\in h_{1,\land{}_{i\in\mathcal{S}}\bar{i}}^{t-1},$
the indices of all packets in $\boldsymbol{Y}_{1,\mathcal{S}}^{t-1}$
are different from $J\left(\boldsymbol{z}^{t-1}\right)$, hence, since
the elements of $\boldsymbol{W}_{1}$ are independent, $X_{1}\left(t\right)=W_{J\left(\boldsymbol{z}^{t-1}\right)}$
is independent of $\boldsymbol{Y}_{1\mathcal{S}}^{t-1}.$ Moreover,
since $\boldsymbol{W}_{1}$ is independent of $\boldsymbol{Z}^{t-1},$
given $\boldsymbol{Z}^{t-1}=z^{t-1},$ $\left(X_{1}\left(t\right),\ \boldsymbol{Y}_{1\mathcal{S}}^{t-1}\right)$
are independent of $\boldsymbol{Z}^{t-1}.$ Hence, 
\[
\Pr\left\{ X_{1}\left(t\right)=m|\boldsymbol{Y}_{1\mathcal{S}}^{t-1}=\boldsymbol{y}_{1\mathcal{S}}^{t-1},\ \boldsymbol{Z}^{t-1}=\boldsymbol{z}^{t-1},\sigma\left(\boldsymbol{z}^{t-1}\right)=1\right\} 
\]
\begin{align*}
 & =\Pr\left\{ W_{J\left(\boldsymbol{z}^{t-1}\right)}=m|\boldsymbol{Y}_{1\mathcal{S}}^{t-1}=\boldsymbol{y}_{1\mathcal{S}}^{t-1}\right\} \\
 & =\Pr\left\{ W_{J\left(\boldsymbol{z}^{t-1}\right)}=m\right\} \\
 & =\frac{1}{\left|\mathcal{F}\right|}\text{ "packet are selected uniformly from \ensuremath{\mathcal{F}}".}
\end{align*}
Equality (\ref{eq:cond2}) follows from the fact that $\boldsymbol{W}_{2}$
is independent of the rest of the variables.

To show (\ref{eq:cond3}) assume first that $2\in\text{\ensuremath{\mathcal{S}_{1}}}$.
Then, since $\boldsymbol{Y}_{2\mathcal{S}_{2}}^{t-1}=f_{2}\left(\boldsymbol{W}_{2},\boldsymbol{Y}_{1\left\{ 2\right\} }^{t-1},\boldsymbol{Z}^{t-1}\right)$
we have 
\[
\Pr\left\{ X_{1}\left(t\right)=m|\boldsymbol{Y}_{1\mathcal{S}_{1}}^{t-1}=\boldsymbol{y}_{1\mathcal{S}_{1}}^{t-1},\boldsymbol{Y}_{2\mathcal{S}_{2}}^{t-1}=\boldsymbol{y}_{2\mathcal{S}_{2}}^{t-1}\ \boldsymbol{Z}^{t-1}=\boldsymbol{z}^{t-1},\boldsymbol{W}_{2}=\boldsymbol{w}_{2},\sigma\left(\boldsymbol{z}^{t-1}\right)=1\right\} 
\]
\begin{align}
 & =\Pr\left\{ X_{1}\left(t\right)=m|\boldsymbol{Y}_{1\mathcal{S}_{1}}^{t-1}=\boldsymbol{y}_{1\mathcal{S}_{1}}^{t-1},\ \boldsymbol{Z}^{t-1}=\boldsymbol{z}^{t-1},\boldsymbol{W}_{2}=\boldsymbol{w}_{2},\sigma\left(\boldsymbol{z}^{t-1}\right)=1\right\} \nonumber \\
 & =\frac{1}{\left|\mathcal{F}\right|}\text{ "by (\ref{eq:cond2})"}.\label{eq:cond3a}
\end{align}
If $2\notin\mathcal{S}_{1}$ then we write, 
\[
\Pr\left\{ X_{1}\left(t\right)=m|\boldsymbol{Y}_{1\mathcal{S}_{1}}^{t-1}=\boldsymbol{y}_{1\mathcal{S}_{1}}^{t-1},\boldsymbol{Y}_{2\mathcal{S}_{2}}^{t-1}=\boldsymbol{y}_{2\mathcal{S}_{2}}^{t-1}\ \boldsymbol{Z}^{t-1}=\boldsymbol{z}^{t-1},\boldsymbol{W}_{2}=\boldsymbol{w}_{2},\sigma\left(\boldsymbol{z}^{t-1}\right)=1\right\} 
\]
\begin{align*}
 & =\sum_{\boldsymbol{y}_{1\left\{ 2\right\} }^{t-1}}\left(\Pr\left\{ X_{1}\left(t\right)=m|\boldsymbol{Y}_{1\mathcal{S}_{1}\cup\left\{ 2\right\} }^{t-1}=\boldsymbol{y}_{1\mathcal{S}_{1}\cup\left\{ 2\right\} }^{t-1},\boldsymbol{Y}_{2\mathcal{S}_{2}}^{t-1}=\boldsymbol{y}_{2\mathcal{S}_{2}}^{t-1}\ \boldsymbol{Z}^{t-1}=\boldsymbol{z}^{t-1},\boldsymbol{W}_{2}=\boldsymbol{w}_{2},\sigma\left(\boldsymbol{z}^{t-1}\right)=1\right\} \right.\\
 & \times\left.\Pr\left\{ \boldsymbol{Y}_{1\left\{ 2\right\} }^{t-1}=\boldsymbol{y}_{1\left\{ 2\right\} }^{t-1}|\boldsymbol{Y}_{1\mathcal{S}_{1}}^{t-1}=\boldsymbol{y}_{1\mathcal{S}_{1}}^{t-1},\boldsymbol{Y}_{2\mathcal{S}_{2}}^{t-1}=\boldsymbol{y}_{2\mathcal{S}_{2}}^{t-1}\ \boldsymbol{Z}^{t-1}=\boldsymbol{z}^{t-1},\boldsymbol{W}_{2}=\boldsymbol{w}_{2},\sigma\left(\boldsymbol{z}^{t-1}\right)=1\right\} \right)\\
 & =\frac{1}{\left|\mathcal{F}\right|}\text{ "by (\ref{eq:cond3a})".}
\end{align*}
Equality (\ref{eq:cond4}) follows by similar arguments. 
\end{IEEEproof}
From Lemma \ref{lem:independent} and the definitions above we conclude
the following. 
\begin{lem}
The following hold for all $\mathcal{S}_{1}\subseteq\mathcal{N}_{1},\ \mathcal{S}_{2}\subseteq\mathcal{N}_{2}$. 
\begin{enumerate}
\item If $J\left(\boldsymbol{z}^{t-1}\right)\in h_{1,\vee_{i\in\mathcal{S}_{1}}}^{t-1},\ \sigma\left(\boldsymbol{z}^{t-1}\right)=1$,
then 
\begin{align}
H\left(X_{1}\left(t\right)|\boldsymbol{Y}_{1\mathcal{S}_{1}}^{t-1}=\boldsymbol{y}_{1\mathcal{S}_{1}}^{t-1},\boldsymbol{Z}^{t-1}=\boldsymbol{\boldsymbol{z}}^{t-1},\sigma(t)=1\right) & =0.\label{eq:fund1}
\end{align}
\item If $J\left(\boldsymbol{z}^{t-1}\right)\in h_{1,\wedge_{i\in\mathcal{S}_{1}}\bar{i}}^{t-1}$
then 
\begin{equation}
H\left(X_{1}\left(t\right)|\boldsymbol{Y}_{1\mathcal{S}_{1}}^{t-1}=\boldsymbol{y}_{1\mathcal{S}_{1}}^{t-1},\boldsymbol{Z}^{t-1}=\boldsymbol{\boldsymbol{z}}^{t-1},\boldsymbol{W}_{2}=\boldsymbol{w_{2}},\sigma(t)=1\right)=1.\label{eq:fund2}
\end{equation}
\item If $J\left(\boldsymbol{z}^{t-1}\right)\in h_{1,\bar{2}\wedge_{i\in\mathcal{S}_{1}}\bar{i}}^{t-1}$
then 
\begin{align}
H\left(X_{1}\left(t\right)|\boldsymbol{Y}_{1\mathcal{S}_{1}}^{t-1}=\boldsymbol{y}_{1\mathcal{S}_{1}}^{t-1},\boldsymbol{Y}_{2\mathcal{S}_{2}}^{t-1}=\boldsymbol{y}_{2\mathcal{S}_{2}}^{t-1},\boldsymbol{Z}^{t-1}=\boldsymbol{\boldsymbol{z}}^{t-1},\boldsymbol{W}_{2}=\boldsymbol{w}_{2},\sigma(t)=1\right) & =1.\label{eq:fund3}
\end{align}
\item For all $\mathcal{G}_{1}\subseteq\mathcal{N}_{1},\mathcal{S}_{1}\subseteq\mathcal{N}_{1},\mathcal{G}_{2}\subseteq\mathcal{N}_{2}$,
$\mathcal{S}_{1}\subseteq\mathcal{N}_{1}$, if $J\left(\boldsymbol{z}^{t-1}\right)\in h_{1,\bar{2}\wedge_{i\in\mathcal{S}_{1}\cup\mathcal{G}_{1}}\bar{i}}^{t-1},$
then, 
\begin{align}
I\left(\boldsymbol{W}_{2},\boldsymbol{Y}_{1\mathcal{G}_{1}}^{t-1},\boldsymbol{Y}_{2\mathcal{G}_{2}}^{t-1};X_{1}\left(t\right)|\boldsymbol{Y}_{1\mathcal{S}_{1}}^{t-1}=\boldsymbol{y}_{1\mathcal{S}_{1}}^{t-1},\boldsymbol{Y}_{2\mathcal{S}_{2}}^{t-1}=\boldsymbol{y}_{2\mathcal{S}_{2}}^{t-1},\boldsymbol{Z}^{t-1}=\boldsymbol{z}^{t-1},\sigma\left(t\right)=1\right) & =0.\label{eq:fund4}
\end{align}
\end{enumerate}
\end{lem}
\begin{IEEEproof}
Equality (\ref{eq:fund1}) follows from the fact that if $J\left(\boldsymbol{z}^{t-1}\right)\in h_{1,\vee_{i\in\mathcal{S}}}^{t-1}$
then $X_{1}\left(t\right)=W_{J\left(\boldsymbol{z}^{t-1}\right)}$
is one of the packets in $\boldsymbol{y}_{1\mathcal{S}_{1}}^{t-1}.$

Equalities (\ref{eq:fund2}) and (\ref{eq:fund3}) follow from (\ref{eq:cond2})
and (\ref{eq:cond3}) respectively.

For (\ref{eq:fund4}) we write, 
\[
0\leq I\left(\boldsymbol{W}_{2},\boldsymbol{Y}_{1\mathcal{G}_{1}}^{t-1},\boldsymbol{Y}_{\mathcal{G}_{2}}^{t-1};X_{1}\left(t\right)|\boldsymbol{Y}_{1\mathcal{S}_{1}}^{t-1}=\boldsymbol{y}_{1\mathcal{S}_{1}}^{t-1},\boldsymbol{Y}_{2\mathcal{S}_{2}}^{t-1}=\boldsymbol{y}_{2\mathcal{S}_{2}}^{t-1},\boldsymbol{Z}^{t-1}=\boldsymbol{z}^{t-1},\sigma\left(t\right)=1\right)
\]
\begin{align*}
 & =H\left(X_{1}\left(t\right)|\boldsymbol{Y}_{1\mathcal{S}_{1}}^{t-1}=\boldsymbol{y}_{1\mathcal{S}_{1}}^{t-1},\boldsymbol{Y}_{2\mathcal{S}_{2}}^{t-1}=\boldsymbol{y}_{2S_{2}}^{t-1},\boldsymbol{Z}^{t-1}=\boldsymbol{z}^{t-1},\sigma\left(t\right)=1\right)\\
 & -H\left(X_{1}\left(t\right)|\boldsymbol{Y}_{1\mathcal{S}_{1}}^{t-1}=\boldsymbol{y}_{1\mathcal{S}_{1}}^{t-1},\boldsymbol{Y}_{1\mathcal{G}_{1}}^{t-1},\boldsymbol{Y}_{2\mathcal{S}_{2}}^{t-1}=\boldsymbol{y}_{2S_{2}}^{t-1},\boldsymbol{Y}_{2\mathcal{G}_{2}}^{t-1},\boldsymbol{Z}^{t-1}=\boldsymbol{z}^{t-1},\boldsymbol{W}_{2},\sigma\left(t\right)=1\right)\\
 & =H\left(X_{1}\left(t\right)|\boldsymbol{Y}_{1\mathcal{S}_{1}}^{t-1}=\boldsymbol{y}_{1\mathcal{S}_{1}}^{t-1},\boldsymbol{Y}_{2\mathcal{S}_{2}}^{t-1}=\boldsymbol{y}_{2S_{2}}^{t-1},\boldsymbol{Z}^{t-1}=\boldsymbol{z}^{t-1},\sigma\left(t\right)=1\right)\\
 & -\sum_{\boldsymbol{y}_{1\mathcal{G}_{1}}^{t-1},\boldsymbol{y}_{2\mathcal{G}_{2}}^{t-1},\boldsymbol{w}_{2}}\left(H\left(X_{1}\left(t\right)|\boldsymbol{Y}_{1\mathcal{S}_{1}\cup\mathcal{G}_{1}}^{t-1}=\boldsymbol{y}_{1\mathcal{S}_{1}\cup\mathcal{G}_{1}}^{t-1},\boldsymbol{Y}_{2\mathcal{S}_{2}\cup\mathcal{\mathcal{G}}_{2}}^{t-1}=\boldsymbol{y}_{2\mathcal{S}_{2}\cup\mathcal{G}_{2}}^{t-1},\boldsymbol{Z}^{t-1}=\boldsymbol{z}^{t-1},\boldsymbol{W}_{2}=\boldsymbol{w}_{2},\sigma\left(t\right)=1\right)\right.\\
 & \times\left.\Pr\left\{ \boldsymbol{Y}_{1\mathcal{G}_{1}}^{t-1}=\boldsymbol{y}_{1\mathcal{G}_{1}}^{t-1},\boldsymbol{Y}_{2\mathcal{\mathcal{G}}_{2}}^{t-1}=\boldsymbol{y}_{2\mathcal{G}_{2}}^{t-1},\boldsymbol{W}_{2}=\boldsymbol{w}_{2}|\boldsymbol{Y}_{1\mathcal{S}_{1}}^{t-1}=\boldsymbol{y}_{1\mathcal{S}_{1}}^{t-1},\boldsymbol{Y}_{2\mathcal{S}_{2}}^{t-1}=\boldsymbol{y}_{2\mathcal{S}_{2}}^{t-1},\boldsymbol{Z}^{t-1}=\boldsymbol{z}^{t-1},\sigma\left(t\right)=1\right\} \right)\\
 & \leq1-1\text{ "by (\ref{eq:fund3})"}\\
 & =0.
\end{align*}
\end{IEEEproof}
We can now connect information theoretic measures to scheduling times.
This is done in the next lemma. 
\begin{lem}
\label{lem:NoCodingtime}The following hold: 
\begin{enumerate}
\item For any $\mathcal{S}_{1}\subseteq\mathcal{N}_{1},$ 
\begin{equation}
\sum_{t=1}^{n}H\left(X_{1}\left(t\right)|\boldsymbol{Y}_{1\mathcal{S}_{1}}^{t-1},\boldsymbol{Z}^{t-1},\boldsymbol{W}_{2},\sigma(t)=1\right)\Pr\left(\sigma\left(t\right)=1\right)=\tilde{\tau}_{1,\wedge_{i\in\mathcal{\mathcal{S}}_{1}}\bar{i}}.\label{eq:G1}
\end{equation}
\item For any $\mathcal{S}_{1}\subseteq\mathcal{N}_{1},\ \mathcal{S}_{2}\subseteq\mathcal{N}_{2}$,
\begin{equation}
\sum_{t=1}^{n}H\left(X_{1}\left(t\right)|\boldsymbol{Y}_{1\mathcal{S}_{1}}^{t-1},\boldsymbol{Y}_{2\mathcal{S}_{2}}^{t-1},\boldsymbol{Z}^{t-1},\sigma(t)=1\right)\Pr\left(\sigma\left(t\right)=1\right)\leq\tilde{\tau}_{1,\wedge_{i\in\mathcal{\mathcal{S}}_{1}}\bar{i}}.\label{eq:G1.1}
\end{equation}
\item For any $\mathcal{S}_{1}\subseteq\mathcal{N}_{1},\ \mathcal{S}_{2}\subseteq\mathcal{N}_{2}$,
if $2\in\mathcal{S}_{1},$ 
\begin{equation}
\sum_{t=1}^{n}H\left(X_{1}\left(t\right)|\boldsymbol{Y}_{1\mathcal{S}_{1}}^{t-1},\boldsymbol{Y}_{2\mathcal{S}_{2}}^{t-1},\boldsymbol{Z}^{t-1},\sigma(t)=1\right)\Pr\left(\sigma\left(t\right)=1\right)=\tilde{\tau}_{1,\wedge_{i\in\mathcal{\mathcal{S}}_{1}}\bar{i}}.\label{eq:G2}
\end{equation}
\item If $\ensuremath{2\notin\mathcal{S}_{1},}$ 
\begin{equation}
\sum_{t=1}^{n}I\left(\boldsymbol{Y}_{1\left\{ 2\right\} }^{t-1},\boldsymbol{W}_{2};X_{1}\left(t\right)|\boldsymbol{Y}_{1\mathcal{S}_{1}}^{t-1},\boldsymbol{Y}_{2\mathcal{S}_{2}}^{t-1},\boldsymbol{Z}^{t-1},\sigma\left(t\right)=1\right)\Pr\left(\sigma\left(t\right)=1\right)\leq\tilde{\tau}_{1,2\wedge_{i\in\mathcal{\mathcal{S}}_{1}}\bar{i}}.\label{eq:G3}
\end{equation}
\end{enumerate}
\end{lem}
\begin{IEEEproof}
To show (\ref{eq:G1}), using the definition of conditional entropy
we write, 
\[
H\left(X_{1}\left(t\right)|\boldsymbol{Y}_{1\mathcal{S}_{1}}^{t-1},\boldsymbol{Z}^{t-1},\boldsymbol{W}_{2},\sigma(t)=1\right)
\]
\begin{align*}
= & \sum_{\boldsymbol{y}_{1\mathcal{S}_{1}}^{t-1},\boldsymbol{\boldsymbol{z}}^{t-1},\boldsymbol{w}_{2}}\left(H\left(X_{1}\left(t\right)|\boldsymbol{Y}_{1\mathcal{S}_{1}}^{t-1}=\boldsymbol{y}_{1\mathcal{S}_{1}}^{t-1},\boldsymbol{Z}^{t-1}=\boldsymbol{\boldsymbol{z}}^{t-1},\boldsymbol{W}_{2}=\boldsymbol{w}_{2},\sigma(t)=1\right)\right.\\
\times & \left.\Pr\left(\boldsymbol{Y}_{1\mathcal{S}_{1}}^{t-1}=\boldsymbol{y}_{1\mathcal{S}_{1}}^{t-1},\boldsymbol{Z}^{t-1}=\boldsymbol{\boldsymbol{z}}^{t-1},\boldsymbol{W}_{2}=\boldsymbol{w}_{2}|\sigma(t)=1\right)\right)\\
= & \sum_{\begin{array}{c}
\boldsymbol{y}_{1\mathcal{S}}^{t-1},\boldsymbol{\boldsymbol{z}}^{t-1},\boldsymbol{w}_{2}\\
J\left(\boldsymbol{z}^{t-1}\right)\in h_{1,\vee_{i\in\mathcal{S}_{1}}}^{t-1}
\end{array}}\left(H\left(X_{1}\left(t\right)|\boldsymbol{Y}_{1\mathcal{S}_{1}}^{t-1}=\boldsymbol{y}_{1\mathcal{S}_{1}}^{t-1},\boldsymbol{Z}^{t-1}=\boldsymbol{\boldsymbol{z}}^{t-1},\boldsymbol{W}_{2}=\boldsymbol{w}_{2},\sigma(t)=1\right)\right.\\
\times & \left.\Pr\left(\boldsymbol{Y}_{1\mathcal{S}_{1}}^{t-1}=\boldsymbol{y}_{1\mathcal{S}_{1}}^{t-1},\boldsymbol{Z}^{t-1}=\boldsymbol{\boldsymbol{z}}^{t-1},\boldsymbol{W}_{2}=\boldsymbol{w}_{2}|\sigma(t)=1\right)\right)\\
+ & \sum_{\begin{array}{c}
\boldsymbol{y}_{1\mathcal{S}}^{t-1},\boldsymbol{\boldsymbol{z}}^{t-1},\boldsymbol{w}_{2}\\
J\left(\boldsymbol{z}^{t-1}\right)\in h_{1,\wedge_{i\in\mathcal{S}_{1}}\bar{i}}^{t-1}
\end{array}}\left(H\left(X_{1}\left(t\right)|\boldsymbol{Y}_{1\mathcal{S}_{1}}^{t-1}=\boldsymbol{y}_{1\mathcal{S}_{1}}^{t-1},\boldsymbol{Z}^{t-1}=\boldsymbol{\boldsymbol{z}}^{t-1},\boldsymbol{W}_{2}=\boldsymbol{w}_{2},\sigma(t)=1\right)\right.\\
\times & \left.\Pr\left(\boldsymbol{Y}_{1\mathcal{S}_{1}}^{t-1}=\boldsymbol{y}_{1\mathcal{S}_{1}}^{t-1},\boldsymbol{Z}^{t-1}=\boldsymbol{\boldsymbol{z}}^{t-1}\boldsymbol{W}_{2}=\boldsymbol{w}_{2},|\sigma(t)=1\right)\right)\text{ }\\
= & \sum_{\begin{array}{c}
\boldsymbol{y}_{1\mathcal{S}}^{t-1},\boldsymbol{\boldsymbol{z}}^{t-1},\boldsymbol{w}_{2}\\
J\left(\boldsymbol{z}^{t-1}\right)\in h_{1,\wedge_{i\in\mathcal{S}_{1}}\bar{i}}^{t-1}
\end{array}}\left(H\left(X_{1}\left(t\right)|\boldsymbol{Y}_{1\mathcal{S}_{1}}^{t-1}=\boldsymbol{y}_{1\mathcal{S}_{1}}^{t-1},\boldsymbol{Z}^{t-1}=\boldsymbol{\boldsymbol{z}}^{t-1},\boldsymbol{W}_{2}=\boldsymbol{w}_{2},\sigma(t)=1\right)\right.\\
\times & \left.\Pr\left(\boldsymbol{Y}_{1\mathcal{S}_{1}}^{t-1}=\boldsymbol{y}_{1\mathcal{S}_{1}}^{t-1},\boldsymbol{Z}^{t-1}=\boldsymbol{\boldsymbol{z}}^{t-1}\boldsymbol{W}_{2}=\boldsymbol{w}_{2},|\sigma(t)=1\right)\right)\text{ "by (\ref{eq:fund1})"}\\
= & \sum_{\begin{array}{c}
\boldsymbol{y}_{1\mathcal{S}_{1}}^{t-1},\boldsymbol{\boldsymbol{z}}^{t-1},\boldsymbol{w}_{2}\\
J\left(\boldsymbol{z}^{t-1}\right)\in h_{1,\wedge_{i\in\mathcal{S}_{1}}\bar{i}}^{t-1}
\end{array}}\Pr\left(\boldsymbol{Y}_{1\mathcal{S}_{1}}^{t-1}=\boldsymbol{y}_{1\mathcal{S}_{1}}^{t-1},\boldsymbol{Z}^{t-1}=\boldsymbol{\boldsymbol{z}}^{t-1}\boldsymbol{W}_{2}=\boldsymbol{w}_{2},|\sigma(t)=1\right)\ \ \text{"by}(\ref{eq:fund2})"\\
= & \Pr\left(J\left(\boldsymbol{Z}^{t-1}\right)\in H_{1,\wedge_{i\in\mathcal{\mathcal{S}}_{1}}\bar{i}}^{t-1}|\sigma\left(t\right)=1\right).
\end{align*}
Hence, 
\[
\sum_{t=1}^{n}H\left(X_{1}\left(t\right)|\boldsymbol{Y}_{1,\mathcal{S}}^{t-1},\boldsymbol{Z}^{t-1},\sigma(t)=1\right)\Pr\left(\sigma\left(t\right)=1\right)
\]
\begin{align*}
 & =\sum_{t=1}^{n}\Pr\left(J\left(\boldsymbol{Z}^{t-1}\right)\in H_{1,\wedge_{i\in\mathcal{\mathcal{S}}}\bar{i}}^{t-1}|\sigma\left(t\right)=1\right)\Pr\left(\sigma\left(t\right)=1\right)\\
 & =\tilde{\tau}_{1,\wedge_{i\in\mathcal{\mathcal{S}}}\bar{i}}\ \text{"by (\ref{eq:genavtimes})".}
\end{align*}

Inequality (\ref{eq:G1.1}) follows by a similar argument, using the
fact that $H\left(X_{1}\left(t\right)\right)\leq1.$

Equality (\ref{eq:G2}) follows by a similar argument, using (\ref{eq:fund3}).

Inequality (\ref{eq:G3}) follows similarly by observing also that
\[
I\left(\boldsymbol{Y}_{1\left\{ 2\right\} }^{t-1},\boldsymbol{W}_{2};X_{1}\left(t\right)|\boldsymbol{Y}_{1\mathcal{S}_{1}}^{t-1},\boldsymbol{Y}_{2\mathcal{S}_{2}}^{t-1},\boldsymbol{Z}^{t-1},\sigma\left(t\right)=1\right)=0,
\]
if either $J\left(\boldsymbol{z}^{t-1}\right)\in h_{1,\bar{2}\wedge_{i\in\mathcal{S}_{1}}\bar{i}}^{t-1}$
(according to (\ref{eq:fund4})), or $J\left(\boldsymbol{z}^{t-1}\right)\in h_{1,\vee_{i\in\mathcal{S}_{1}}i}^{t-1}$
(according to (\ref{eq:fund1})). 
\end{IEEEproof}

\subsection{Capacity Outer Bound\label{sec:Capacity-Outer-Bound}}

We can now proceed with the development of an outer bound to system
capacity. The next lemma relates relates achievable rates to node
scheduling times. 
\begin{lem}
\label{lem:BasicLemma}Let $\left(R_{1},R_{2}\right)$ be achievable.
Then,

\begin{equation}
0\leq nR_{1}-\left(1-\epsilon_{\mathcal{S}}^{1}\right)\tilde{\tau}_{1,\wedge_{i\in\mathcal{S}}\bar{i}}\leq o\left(n\right),\ \left\{ 2,3\right\} \subseteq\mathcal{S},\label{eq:R1Ineq1}
\end{equation}

\begin{equation}
\frac{nR_{1}}{1-\epsilon_{23}^{1}}+\frac{nR_{2}}{1-\epsilon_{4}^{2}}\leq n-\tilde{\tau}_{1,2\wedge\bar{3}}-\tilde{\tau}_{1,3}+o(n),\label{eq:R2Ineq}
\end{equation}
\begin{equation}
\left(\frac{\epsilon_{3}^{1}-\epsilon_{23}^{1}}{\left(1-\epsilon_{3}^{2}\right)\left(1-\epsilon_{23}^{1}\right)}+\frac{1}{1-\epsilon_{23}^{1}}\right)nR_{1}+\frac{nR_{2}}{1-\epsilon_{34}^{2}}\leq n-\tilde{\tau}_{1,2\wedge\bar{3}}-\tilde{\tau}_{1,3}+\left(\frac{1-\epsilon_{3}^{1}}{1-\epsilon_{3}^{2}}\right)\tilde{\tau}_{1,2\wedge\bar{3}}+o\left(n\right),\label{eq:third}
\end{equation}
\begin{align}
\left(\frac{\epsilon_{34}^{1}-\epsilon_{234}^{1}}{\left(1-\epsilon_{234}^{1}\right)\left(1-\epsilon_{34}^{2}\right)}+\frac{1}{1-\epsilon_{23}^{1}}\right)nR_{1}+\frac{1}{1-\epsilon_{4}^{2}}nR_{2} & \leq n-\tilde{\tau}_{1,2\wedge\bar{3}}-\tilde{\tau}_{1,3}+\frac{1-\epsilon_{34}^{1}}{1-\epsilon_{34}^{2}}\tilde{u}_{1,2\wedge\bar{3}\wedge\bar{4}}+\frac{1-\epsilon_{4}^{1}}{1-\epsilon_{4}^{2}}\tilde{v}_{1,2\wedge\bar{4}}+o\left(n\right),\label{eq:fourth}
\end{align}
where 
\begin{equation}
\tilde{u}_{1,2\wedge\bar{3}\wedge\bar{4}}\leq\tilde{\tau}_{1,2\wedge\bar{3}\wedge\bar{4}},\label{eq:Rfifth}
\end{equation}
\begin{equation}
\tilde{u}_{1,2\wedge\bar{3}\wedge\bar{4}}+\tilde{v}_{1,2\wedge\bar{4}}\leq\tilde{\tau}_{1,2\wedge\bar{3}\wedge\bar{4}}+\tilde{\tau}_{1,2\wedge3\wedge\bar{4}}.\label{eq:Rsixth}
\end{equation}
\end{lem}
\begin{IEEEproof}
To show (\ref{eq:R1Ineq1}) we write according to (\ref{eq:Erase4}),
\begin{align*}
I\left(\boldsymbol{W}_{1};\boldsymbol{Y}_{1\mathcal{S}}^{n},\boldsymbol{Z}^{n}|\boldsymbol{W}_{2}\right) & =\left(1-\epsilon_{\mathcal{S}}^{1}\right)\sum_{t=1}^{n}I\left(\boldsymbol{W}_{1};X_{1}\left(t\right)|\boldsymbol{Y}_{1\mathcal{S}}^{t-1},\boldsymbol{Z}^{t-1},\boldsymbol{W}_{2},\sigma\left(t\right)=1\right)\Pr\left(\sigma\left(t\right)=1\right)\\
 & =\left(1-\epsilon_{\mathcal{S}}^{1}\right)\sum_{t=1}^{n}H\left(X_{1}\left(t\right)|\boldsymbol{Y}_{1\mathcal{S}}^{t-1},\boldsymbol{Z}^{t-1},\boldsymbol{W}_{2},\sigma\left(t\right)=1\right)\Pr\left(\sigma\left(t\right)=1\right)\\
 & \text{"X\ensuremath{\left(t\right)}=\ensuremath{W_{J\left(\boldsymbol{Z}^{t-1}\right)}}=f\ensuremath{\left(\boldsymbol{W}_{1},\boldsymbol{Z}^{t-1}\right)}"}\\
 & =\left(1-\epsilon_{\mathcal{S}}^{1}\right)\tilde{\tau}_{1,\wedge_{i\in\mathcal{S}}\bar{i}}\ \text{"by (\ref{eq:G1})"}.
\end{align*}
Relation (\ref{eq:R1Ineq1}) follows now from (\ref{eq:Lem2a}).

To show (\ref{eq:R2Ineq}), notice first that, 
\begin{align}
\tilde{\tau}_{2} & =n-\tilde{\tau}_{1}\text{ "by(\ref{eq:sum-times})"}\nonumber \\
 & =n-\tilde{\tau}_{1,\bar{2}\wedge\bar{3}}-\tilde{\tau}_{1,2\wedge\bar{3}}-\tilde{\tau}_{1,3}\text{ "by definition"}\nonumber \\
 & \leq n-\frac{nR_{1}}{1-\epsilon_{23}^{1}}-\tilde{\tau}_{1,2\wedge\bar{3}}-\tilde{\tau}_{1,3}+o(n)\text{\text{ "by (\ref{eq:R1Ineq1})"}}.\label{eq:timeineq}
\end{align}
Next according to (\ref{eq:Lem2a1}), 
\begin{equation}
nR_{2}\leq I\left(\boldsymbol{W}_{2};\boldsymbol{Y}_{2\left\{ 4\right\} }^{t-1},\boldsymbol{Z}^{n}|\boldsymbol{W}_{1}\right)+o(n),\label{eq:hsecond}
\end{equation}
and according to (\ref{eq:Erase4}), 
\begin{align*}
I\left(\boldsymbol{W}_{2};\boldsymbol{Y}_{2\left\{ 4\right\} }^{n},\boldsymbol{Z}^{n}|\boldsymbol{W}_{1}\right) & =\left(1-\epsilon_{4}^{2}\right)\sum_{t=1}^{n}I\left(\boldsymbol{W}_{2};X_{2}(t)\mid\boldsymbol{W}_{1},\boldsymbol{Y}_{2\left\{ 4\right\} }^{t-1},\boldsymbol{Z}^{t-1},\sigma(t)=2\right)\Pr\left(\sigma(t)=2\right)\\
 & \leq\left(1-\epsilon_{4}^{2}\right)\sum_{t=1}^{n}\Pr\left(\sigma(t)=2\right)\text{ "since \ensuremath{H\left(X_{2}\left(t\right)\leq1\right)}"}\\
 & =\left(1-\epsilon_{4}^{2}\right)\tilde{\tau}_{2}\ \text{"by \ref{eq:avtimes}"}\\
 & \leq\left(1-\epsilon_{4}^{2}\right)\left(n-\frac{nR_{1}}{1-\epsilon_{23}^{1}}-\tilde{\tau}_{1,2\wedge\bar{3}}-\tilde{\tau}_{1,3}\right)\text{ "by (\ref{eq:timeineq})"}.
\end{align*}
The last inequality and (\ref{eq:hsecond}) imply inequality (\ref{eq:R2Ineq}).

Next we show inequality (\ref{eq:third}). According to (\ref{eq:Lem2b}),
\begin{equation}
nR_{1}\leq I(\boldsymbol{W}_{1};\boldsymbol{Y}_{1\left\{ 3\right\} }^{t-1},\boldsymbol{Y}_{2\left\{ 3\right\} }^{t-1},\boldsymbol{Z}^{n})+o(n),\label{eq:hthird}
\end{equation}
and according to (\ref{eq:Erase4}), 
\begin{align*}
I\left(\boldsymbol{W}_{1};\boldsymbol{Y}_{1\left\{ 3\right\} ,2\left\{ 3\right\} }^{n},\boldsymbol{Z}^{n}\right) & =\left(1-\epsilon_{3}^{1}\right)\sum_{t=1}^{n}I\left(\boldsymbol{W}_{1};X_{1}\left(t\right)|\boldsymbol{Y}_{1\left\{ 3\right\} }^{t-1},\boldsymbol{Y}_{2\left\{ 3\right\} }^{t-1},\boldsymbol{Z}^{t-1},\sigma\left(t\right)=1\right)\Pr\left(\sigma\left(t\right)=1\right)\\
 & +\left(1-\epsilon_{3}^{2}\right)\sum_{t=1}^{n}I\left(\boldsymbol{W}_{1};X_{2}\left(t\right)|\boldsymbol{Y}_{1\left\{ 3\right\} }^{t-1},\boldsymbol{Y}_{2\left\{ 3\right\} }^{t-1},\boldsymbol{Z}^{t-1},\sigma\left(t\right)=2\right)\Pr\left(\sigma\left(t\right)=2\right)\\
 & =\left(1-\epsilon_{3}^{1}\right)\sum_{t=1}^{n}H\left(X_{1}\left(t\right)|\boldsymbol{Y}_{1\left\{ 3\right\} }^{t-1},\boldsymbol{Y}_{2\left\{ 3\right\} }^{t-1},\boldsymbol{Z}^{t-1},\sigma\left(t\right)=1\right)\Pr\left(\sigma\left(t\right)=1\right)\ \ \text{"}X_{1}\left(t\right)=f\left(\boldsymbol{W}_{1},\boldsymbol{Z}^{t-1}\right)\text{"}\\
 & +\left(1-\epsilon_{3}^{2}\right)\sum_{t=1}^{n}I\left(\boldsymbol{W}_{1};X_{2}\left(t\right)|\boldsymbol{Y}_{1\left\{ 3\right\} }^{t-1},\boldsymbol{Y}_{2\left\{ 3\right\} }^{t-1},\boldsymbol{Z}^{t-1},\sigma\left(t\right)=2\right)\Pr\left(\sigma\left(t\right)=2\right)\\
 & \leq\left(1-\epsilon_{3}^{1}\right)\tilde{\tau}_{1,\bar{3}}\text{ "by (\ref{eq:G1.1})"}\\
 & +\left(1-\epsilon_{3}^{2}\right)\sum_{t=1}^{n}I\left(\boldsymbol{W}_{1};X_{2}\left(t\right)|\boldsymbol{Y}_{1\left\{ 3\right\} }^{t-1},\boldsymbol{Y}_{2\left\{ 3\right\} }^{t-1},\boldsymbol{Z}^{t-1},\sigma\left(t\right)=2\right)\Pr\left(\sigma\left(t\right)=2\right).
\end{align*}
Combining the last inequality with (\ref{eq:hthird}) and using the
fact that by definition, 
\[
\tilde{\tau}_{1,\bar{3}}=\tilde{\tau}_{1,\bar{2}\wedge\bar{3}}+\tilde{\tau}_{1,2\wedge\bar{3}}\leq\frac{nR_{1}}{1-\epsilon_{23}^{1}}+\tilde{\tau}_{1,2\wedge\bar{3}}\text{ "by (\ref{eq:R1Ineq1})",}
\]
we get by rearranging terms, 
\begin{align}
\left(\frac{1}{\left(1-\epsilon_{3}^{2}\right)}-\frac{1}{\left(1-\epsilon_{23}^{1}\right)}\right)nR_{1} & =\frac{\epsilon_{3}^{1}-\epsilon_{23}^{1}}{\left(1-\epsilon_{3}^{2}\right)\left(1-\epsilon_{23}^{1}\right)}nR_{1}\nonumber \\
 & \leq\left(\frac{1-\epsilon_{3}^{1}}{1-\epsilon_{3}^{2}}\right)\tilde{\tau}_{1,2\wedge\bar{3}}\\
 & +\sum_{t=1}^{n}I\left(\boldsymbol{W}_{1};X_{2}\left(t\right)|\boldsymbol{Y}_{1\left\{ 3\right\} }^{t-1},\boldsymbol{Y}_{2\left\{ 3\right\} }^{t-1},\boldsymbol{Z}^{t-1},\sigma\left(t\right)=2\right)\Pr\left(\sigma\left(t\right)=2\right)+o\left(n\right).\label{eq:h1}
\end{align}
Next, according to (\ref{eq:Lem2a1}), 
\begin{equation}
nR_{2}\leq I\left(\boldsymbol{W}_{2};\boldsymbol{Y}_{2\left\{ 34\right\} }^{n},\boldsymbol{Z}^{n}|\boldsymbol{W}_{1}\right)+o(n),\label{eq:hthird1}
\end{equation}
and according to (\ref{eq:Erase4}), 
\begin{align*}
I\left(\boldsymbol{W}_{2};\boldsymbol{Y}_{2\left\{ 34\right\} }^{n},\boldsymbol{Z}^{n}|\boldsymbol{W}_{1}\right) & =\left(1-\epsilon_{34}^{2}\right)\sum_{t=1}^{n}I\left(\boldsymbol{W}_{2};X_{2}\left(t\right)|\boldsymbol{Y}_{2\left\{ 34\right\} }^{t-1},\boldsymbol{Z}^{t-1},\boldsymbol{W}_{1},\sigma\left(t\right)=2\right)\Pr\left(\sigma(t)=2\right)\\
 & =\left(1-\epsilon_{34}^{2}\right)\sum_{t=1}^{n}I\left(\boldsymbol{W}_{2};X_{2}\left(t\right)|\boldsymbol{Y}_{1\{34\}}^{t-1},\boldsymbol{Y}_{2\left\{ 34\right\} }^{t-1},\boldsymbol{Z}^{t-1},\boldsymbol{W}_{1},\sigma\left(t\right)=2\right)\text{ \ensuremath{"\boldsymbol{Y}_{1\{34\}}^{t-1}=f\left(\boldsymbol{Z}^{t-1},\boldsymbol{W}_{1}\right)"}}\\
 & \leq\left(1-\epsilon_{34}^{2}\right)\sum_{t=1}^{n}I\left(\boldsymbol{W}_{2},\boldsymbol{Y}_{1\{4\}}^{t-1},\boldsymbol{Y}_{2\left\{ 4\right\} }^{t-1};X_{2}\left(t\right)|\boldsymbol{Y}_{1\{3\}}^{t-1},\boldsymbol{Y}_{2\left\{ 3\right\} }^{t-1},\boldsymbol{Z}^{t-1},\boldsymbol{W}_{1},\sigma\left(t\right)=2\right)\text{ "chain rule".}
\end{align*}
Combining with (\ref{eq:hthird1}) we get 
\begin{align}
\frac{nR_{2}}{1-\epsilon_{34}^{2}} & \leq\sum_{t=1}^{n}I\left(\boldsymbol{W}_{2},\boldsymbol{Y}_{1\{4\},2\{4\}}^{n};X_{2}\left(t\right)|\boldsymbol{Y}_{1\{3\}}^{t-1},\boldsymbol{Y}_{2\left\{ 3\right\} }^{t-1},\boldsymbol{Z}^{t-1},\boldsymbol{W}_{1},\sigma\left(t\right)=2\right)+o\left(n\right).\label{eq:h2}
\end{align}
Adding (\ref{eq:h1}) and (\ref{eq:h2}) we have, 
\begin{align*}
\frac{\epsilon_{3}^{1}-\epsilon_{23}^{1}}{\left(1-\epsilon_{3}^{2}\right)\left(1-\epsilon_{23}^{1}\right)}nR_{1}+\frac{nR_{2}}{1-\epsilon_{34}^{2}} & \leq\left(\frac{1-\epsilon_{3}^{1}}{1-\epsilon_{3}^{2}}\right)\tilde{\tau}_{1,2\wedge\bar{3}}\\
 & +\sum_{t=1}^{n}I\left(\boldsymbol{W}_{1};X_{2}\left(t\right)|\boldsymbol{Y}_{1\{3\}}^{t-1},\boldsymbol{Y}_{2\left\{ 3\right\} }^{t-1},\boldsymbol{Z}^{t-1},\sigma\left(t\right)=2\right)\Pr\left(\sigma\left(t\right)=2\right)\\
 & +\sum_{t=1}^{n}I\left(\boldsymbol{W}_{2},\boldsymbol{Y}_{1\{4\},2\{4\}}^{n};X_{2}\left(t\right)|\boldsymbol{Y}_{1\{3\}}^{t-1},\boldsymbol{Y}_{2\left\{ 3\right\} }^{t-1},\boldsymbol{Z}^{t-1},\boldsymbol{W}_{1},\sigma\left(t\right)=2\right)+o\left(n\right)\\
 & =\left(\frac{1-\epsilon_{3}^{1}}{1-\epsilon_{3}^{2}}\right)\tilde{\tau}_{1,2\wedge\bar{3}}\\
 & +\sum_{t=1}^{n}I\left(\boldsymbol{W}_{1},\boldsymbol{W}_{2},\boldsymbol{Y}_{1\{4\},2\{4\}}^{n};X_{2}\left(t\right)|\boldsymbol{Y}_{1\{3\}}^{t-1},\boldsymbol{Y}_{2\left\{ 3\right\} }^{t-1},\boldsymbol{Z}^{t-1},\sigma\left(t\right)=2\right)\Pr\left(\sigma\left(t\right)=2\right)+o\left(n\right)\\
 & \leq\left(\frac{1-\epsilon_{3}^{1}}{1-\epsilon_{3}^{2}}\right)\tilde{\tau}_{1,2\wedge\bar{3}}+\tilde{\tau}_{2}+o\left(n\right)\text{ "by (}\ref{eq:avtimes}\text{) and since \ensuremath{H\left(X_{2}\left(t\right)\right)\leq1\text{"}}}\\
 & \leq\left(\frac{1-\epsilon_{3}^{1}}{1-\epsilon_{3}^{2}}\right)\tilde{\tau}_{1,2\wedge\bar{3}}+n-\frac{nR_{1}}{1-\epsilon_{23}^{1}}-\tilde{\tau}_{1,2\wedge\bar{3}}-\tilde{\tau}_{1,3}+o\left(n\right)\text{ "by (\ref{eq:timeineq})".}
\end{align*}
By rearranging terms we get (\ref{eq:third}).

It remains to show (\ref{eq:fourth}), (\ref{eq:Rfifth}) and (\ref{eq:Rsixth}).
According to (\ref{eq:Lem2b1}), 
\begin{equation}
nR_{1}\leq I(\boldsymbol{W}_{1};\boldsymbol{Y}_{1\left\{ 34\right\} }^{t-1},\boldsymbol{Y}_{2\left\{ 34\right\} }^{t-1},\boldsymbol{Z}^{n}|\boldsymbol{W}_{2})+o(n),\label{eq:h4}
\end{equation}
and according to (\ref{eq:Erase4}), 
\[
I\left(\boldsymbol{W}_{1};\boldsymbol{Y}_{1\left\{ 34\right\} }^{n},\boldsymbol{Y}_{2\left\{ 34\right\} }^{n},\boldsymbol{Z}^{n}|\boldsymbol{W}_{2}\right)
\]
\begin{eqnarray}
 & = & \left(1-\epsilon_{34}^{1}\right)\sum_{t=1}^{n}I\left(\boldsymbol{W}_{1};X_{1}\left(t\right)|\boldsymbol{Y}_{1\left\{ 34\right\} }^{t-1},\boldsymbol{Y}_{2\left\{ 34\right\} }^{t-1},\boldsymbol{Z}^{t-1},\boldsymbol{W}_{2,}\sigma\left(t\right)=1\right)\Pr\left(\sigma\left(t\right)=1\right)\nonumber \\
 & + & \left(1-\epsilon_{34}^{2}\right)\sum_{t=1}^{n}I\left(\boldsymbol{W}_{1};X_{2}\left(t\right)|\boldsymbol{Y}_{1\left\{ 34\right\} }^{t-1},\boldsymbol{Y}_{2\left\{ 34\right\} }^{t-1},\boldsymbol{Z}^{t-1},\boldsymbol{W}_{2},\sigma\left(t\right)=2\right)\Pr\left(\sigma\left(t\right)=2\right).\label{eq:n2}
\end{eqnarray}
Now, 
\[
I\left(\boldsymbol{W}_{1};X_{1}\left(t\right)|\boldsymbol{Y}_{1\left\{ 34\right\} }^{t-1},\boldsymbol{Y}_{2\left\{ 34\right\} }^{t-1},\boldsymbol{Z}^{t-1},\boldsymbol{W}_{2},\sigma\left(t\right)=1\right)
\]
\begin{align}
 & =H\left(X_{1}\left(t\right)|\boldsymbol{Y}_{1\left\{ 34\right\} }^{t-1},\boldsymbol{Y}_{2\left\{ 34\right\} }^{t-1},\boldsymbol{Z}^{t-1},\boldsymbol{W}_{2},\sigma\left(t\right)=1\right)\ \text{"\ensuremath{X_{1}\left(t\right)=f\left(\boldsymbol{W}_{1},\boldsymbol{Z}^{t-1}\right)} "}\nonumber \\
 & =I\left(\boldsymbol{Y}_{1\left\{ 2\right\} }^{t-1};X_{1}\left(t\right)|\boldsymbol{Y}_{1\left\{ 34\right\} }^{t-1},\boldsymbol{Y}_{2\left\{ 34\right\} }^{t-1},\boldsymbol{Z}^{t-1},\boldsymbol{W}_{2},\sigma\left(t\right)=1\right)\nonumber \\
 & +H\left(X_{1}\left(t\right)|\boldsymbol{Y}_{1\left\{ 234\right\} }^{t-1},\boldsymbol{Y}_{2\left\{ 34\right\} }^{t-1},\boldsymbol{Z}^{t-1},\boldsymbol{W}_{2},\sigma\left(t\right)=1\right)\nonumber \\
 & =I\left(\boldsymbol{Y}_{1\left\{ 2\right\} }^{t-1};X_{1}\left(t\right)|\boldsymbol{Y}_{1\left\{ 34\right\} }^{t-1},\boldsymbol{Y}_{2\left\{ 34\right\} }^{t-1},\boldsymbol{Z}^{t-1},\boldsymbol{W}_{2},\sigma\left(t\right)=1\right)\nonumber \\
 & +H\left(X_{1}\left(t\right)|\boldsymbol{Y}_{1\left\{ 234\right\} }^{t-1},\boldsymbol{Z}^{t-1},\boldsymbol{W}_{2},\sigma\left(t\right)=1\right)\ "\boldsymbol{Y}_{2\left\{ 34\right\} }^{t-1}=f_{2}\left(\boldsymbol{W}_{2},\boldsymbol{Y}_{1\{2\}}^{t-1},\boldsymbol{Z}^{t-1}\right)"\nonumber \\
 & =I\left(\boldsymbol{Y}_{1\left\{ 2\right\} }^{t-1};X_{1}\left(t\right)|\boldsymbol{Y}_{1\left\{ 34\right\} }^{t-1},\boldsymbol{Y}_{2\left\{ 34\right\} }^{t-1},\boldsymbol{Z}^{t-1},\boldsymbol{W}_{2},\sigma\left(t\right)=1\right)\nonumber \\
 & +H\left(X_{1}\left(t\right)|\boldsymbol{Y}_{1\left\{ 234\right\} }^{t-1},\boldsymbol{Z}^{t-1},\sigma\left(t\right)=1\right)\ \text{"\ensuremath{\boldsymbol{W}_{2}} independent of rest".}\label{eq:n3}
\end{align}
According to (\ref{eq:G1}), 
\begin{align}
\sum_{t=1}^{n}H\left(X_{1}\left(t\right)|\boldsymbol{Y}_{1\left\{ 234\right\} }^{t-1},\boldsymbol{Z}^{t-1},\sigma\left(t\right)=1\right)\Pr\left(\sigma\left(t\right)=1\right) & =\tilde{\tau}_{1,\bar{2}\wedge\bar{3}\wedge\bar{4}}\nonumber \\
 & \leq\frac{nR_{1}}{1-\epsilon_{234}^{1}}\text{ "by (\ref{eq:R1Ineq1})"}\label{eq:n4}
\end{align}
Replacing (\ref{eq:n3}) and (\ref{eq:n4}) in (\ref{eq:n2}) we have
\begin{align*}
I\left(\boldsymbol{W}_{1};\boldsymbol{Y}_{1\left\{ 34\right\} }^{n},\boldsymbol{Y}_{2\left\{ 34\right\} }^{n},\boldsymbol{Z}^{n}|\boldsymbol{W}_{2}\right) & \leq\frac{1-\epsilon_{34}^{1}}{1-\epsilon_{234}^{1}}nR_{1}\\
 & +\left(1-\epsilon_{34}^{1}\right)\sum_{t=1}^{n}I\left(\boldsymbol{Y}_{1\left\{ 2\right\} }^{t-1};X_{1}\left(t\right)|\boldsymbol{Y}_{1\left\{ 34\right\} }^{t-1},\boldsymbol{Y}_{2\left\{ 34\right\} }^{t-1},\boldsymbol{Z}^{t-1},\boldsymbol{W}_{2},\sigma\left(t\right)=1\right)\Pr\left(\sigma\left(t\right)=1\right)\\
 & +\left(1-\epsilon_{34}^{2}\right)\sum_{t=1}^{n}I\left(\boldsymbol{W}_{1};X_{2}\left(t\right)|\boldsymbol{Y}_{1\left\{ 34\right\} }^{t-1},\boldsymbol{Y}_{2\left\{ 34\right\} }^{t-1},\boldsymbol{Z}^{t-1},\boldsymbol{W}_{2},\sigma\left(t\right)=2\right)\Pr\left(\sigma\left(t\right)=2\right).
\end{align*}
Combining the last inequality with (\ref{eq:h4}) and rearranging
terms we have, 
\begin{align}
\left(\frac{1}{1-\epsilon_{34}^{2}}-\frac{1-\epsilon_{34}^{1}}{\left(1-\epsilon_{234}^{1}\right)\left(1-\epsilon_{34}^{2}\right)}\right)nR_{1} & \leq\frac{1-\epsilon_{34}^{1}}{1-\epsilon_{34}^{2}}\sum_{t=1}^{n}I\left(\boldsymbol{Y}_{1\left\{ 2\right\} }^{t-1};X_{1}\left(t\right)|\boldsymbol{Y}_{1\left\{ 34\right\} }^{t-1},\boldsymbol{Y}_{2\left\{ 34\right\} }^{t-1},\boldsymbol{Z}^{t-1},\boldsymbol{W}_{2},\sigma\left(t\right)=1\right)\Pr\left(\sigma\left(t\right)=1\right)\nonumber \\
 & +\sum_{t=1}^{n}I\left(\boldsymbol{W}_{1};X_{2}\left(t\right)|\boldsymbol{Y}_{1\left\{ 34\right\} }^{t-1},\boldsymbol{Y}_{2\left\{ 34\right\} }^{t-1},\boldsymbol{Z}^{t-1},\boldsymbol{W}_{2},\sigma\left(t\right)=2\right)\Pr\left(\sigma\left(t\right)=2\right)+o\left(n\right).\label{eq:fin1}
\end{align}
Next we write similarly, 
\[
nR_{2}\leq I\left(\boldsymbol{W}_{2};\boldsymbol{Y}_{1\left\{ 4\right\} }^{n},\boldsymbol{Y}_{2\left\{ 4\right\} }^{n},\boldsymbol{Z}^{n}\right)+o\left(n\right),
\]
\begin{eqnarray}
 &  & I\left(\boldsymbol{W}_{2};\boldsymbol{Y}_{1\left\{ 4\right\} }^{n},\boldsymbol{Y}_{2\left\{ 4\right\} }^{n},\boldsymbol{Z}^{n}\right)\nonumber \\
 & = & \left(1-\epsilon_{4}^{1}\right)\sum_{t=1}^{n}I\left(\boldsymbol{W}_{2};X_{1}\left(t\right)|\boldsymbol{Y}_{1\left\{ 4\right\} }^{t-1},\boldsymbol{Y}_{2\left\{ 4\right\} }^{t-1},\boldsymbol{Z}^{t-1},\sigma\left(t\right)=1\right)\Pr\left(\sigma\left(t\right)=1\right)\nonumber \\
 & + & \left(1-\epsilon_{4}^{2}\right)\sum_{t=1}^{n}I\left(\boldsymbol{W}_{2};X_{2}\left(t\right)|\boldsymbol{Y}_{1\left\{ 4\right\} }^{t-1},\boldsymbol{Y}_{2\left\{ 4\right\} }^{t-1},\boldsymbol{Z}^{t-1},\sigma\left(t\right)=2\right)\Pr\left(\sigma\left(t\right)=2\right)\nonumber \\
 & \leq & \left(1-\epsilon_{4}^{1}\right)\sum_{t=1}^{n}I\left(\boldsymbol{W}_{2};X_{1}\left(t\right)|\boldsymbol{Y}_{1\left\{ 4\right\} }^{t-1},\boldsymbol{Y}_{2\left\{ 4\right\} }^{t-1},\boldsymbol{Z}^{t-1},\sigma\left(t\right)=1\right)\Pr\left(\sigma\left(t\right)=1\right)\nonumber \\
 &  & +\left(1-\epsilon_{4}^{2}\right)\sum_{t=1}^{n}I\left(\boldsymbol{W}_{2},\boldsymbol{Y}_{1\left\{ 3\right\} }^{t-1},\boldsymbol{Y}_{2\left\{ 3\right\} }^{t-1};X_{2}\left(t\right)|\boldsymbol{Y}_{1\left\{ 4\right\} }^{t-1},\boldsymbol{Y}_{2\left\{ 4\right\} }^{t-1},\boldsymbol{Z}^{t-1},\sigma\left(t\right)=2\right)\Pr\left(\sigma\left(t\right)=2\right).\label{eq:n7}
\end{eqnarray}
Combining the last two relations we conclude, 
\begin{align}
\frac{nR_{2}}{1-\epsilon_{4}^{2}} & \leq\frac{1-\epsilon_{4}^{1}}{1-\epsilon_{4}^{2}}\sum_{t=1}^{n}I\left(\boldsymbol{W}_{2};X_{1}\left(t\right)|\boldsymbol{Y}_{1\left\{ 4\right\} }^{t-1},\boldsymbol{Y}_{2\left\{ 4\right\} }^{t-1},\boldsymbol{Z}^{t-1},\sigma\left(t\right)=1\right)\Pr\left(\sigma\left(t\right)=1\right)\nonumber \\
 & +\sum_{t=1}^{n}I\left(\boldsymbol{W}_{2},\boldsymbol{Y}_{1\left\{ 3\right\} }^{t-1},\boldsymbol{Y}_{2\left\{ 3\right\} }^{t-1};X_{2}\left(t\right)|\boldsymbol{Y}_{1\left\{ 4\right\} }^{t-1},\boldsymbol{Y}_{2\left\{ 4\right\} }^{t-1},\boldsymbol{Z}^{t-1},\sigma\left(t\right)=2\right)\Pr\left(\sigma\left(t\right)=2\right)+o\left(n\right).\label{eq:fin2}
\end{align}
Adding (\ref{eq:fin1}), (\ref{eq:fin2}), observing that by the chain
rule, 
\begin{align*}
I\left(\boldsymbol{W}_{1};X_{2}\left(t\right)|\boldsymbol{Y}_{1\left\{ 34\right\} }^{t-1},\boldsymbol{Y}_{2\left\{ 34\right\} }^{t-1},\boldsymbol{Z}^{t-1},\boldsymbol{W}_{2},\sigma\left(t\right)=2\right) & +I\left(\boldsymbol{W}_{2},\boldsymbol{Y}_{1\left\{ 3\right\} }^{t-1},\boldsymbol{Y}_{2\left\{ 3\right\} }^{t-1};X_{2}\left(t\right)|\boldsymbol{Y}_{1\left\{ 4\right\} }^{t-1},\boldsymbol{Y}_{2\left\{ 4\right\} }^{t-1},\boldsymbol{Z}^{t-1},\sigma\left(t\right)=2\right)\\
 & =I\left(\boldsymbol{W}_{1},\boldsymbol{W}_{2},\boldsymbol{Y}_{1\left\{ 3\right\} }^{t-1},\boldsymbol{Y}_{2\left\{ 3\right\} }^{t-1};X_{2}\left(t\right)|\boldsymbol{Y}_{1\left\{ 4\right\} }^{t-1},\boldsymbol{Y}_{2\left\{ 4\right\} }^{t-1},\boldsymbol{Z}^{t-1},\sigma\left(t\right)=2\right)\\
 & \leq1,
\end{align*}
and using (\ref{eq:timeineq}) we obtain after rearranging terms,
\begin{align*}
\left(\frac{\epsilon_{34}^{1}-\epsilon_{234}^{1}}{\left(1-\epsilon_{234}^{1}\right)\left(1-\epsilon_{34}^{2}\right)}+\frac{1}{1-\epsilon_{23}^{1}}\right)nR_{1}+\frac{1}{1-\epsilon_{4}^{2}}nR_{2} & \leq n-\tilde{\tau}_{1,2\wedge\bar{3}}-\tilde{\tau}_{1,3}
\end{align*}
\begin{align*}
 & +\frac{1-\epsilon_{34}^{1}}{1-\epsilon_{34}^{2}}\sum_{t=1}^{n}I\left(\boldsymbol{Y}_{1\left\{ 2\right\} }^{t-1};X_{1}\left(t\right)|\boldsymbol{Y}_{1\left\{ 34\right\} }^{t-1},\boldsymbol{Y}_{2\left\{ 34\right\} }^{t-1},\boldsymbol{Z}^{t-1},\boldsymbol{W}_{2},\sigma\left(t\right)=1\right)\Pr\left(\sigma\left(t\right)=1\right)\\
 & +\frac{1-\epsilon_{4}^{1}}{1-\epsilon_{4}^{2}}\sum_{t=1}^{n}I\left(\boldsymbol{W}_{2};X_{1}\left(t\right)|\boldsymbol{Y}_{1\left\{ 4\right\} }^{t-1},\boldsymbol{Y}_{2\left\{ 4\right\} }^{t-1},\boldsymbol{Z}^{t-1},\sigma\left(t\right)=1\right)\Pr\left(\sigma\left(t\right)=1\right).
\end{align*}

Next, we write,

\[
I\left(\boldsymbol{Y}_{1\left\{ 2\right\} }^{t-1};X_{1}\left(t\right)|\boldsymbol{Y}_{1\left\{ 34\right\} }^{t-1},\boldsymbol{Y}_{2\left\{ 34\right\} }^{t-1},\boldsymbol{Z}^{t-1},\boldsymbol{W}_{2},\sigma\left(t\right)=1\right)+I\left(\boldsymbol{W}_{2};X_{1}\left(t\right)|\boldsymbol{Y}_{1\left\{ 4\right\} }^{t-1},\boldsymbol{Y}_{2\left\{ 4\right\} }^{t-1},\boldsymbol{Z}^{t-1},\sigma\left(t\right)=1\right)
\]
\begin{align}
 & =I\left(\boldsymbol{Y}_{1\left\{ 2\right\} }^{t-1};X_{1}\left(t\right)|\boldsymbol{Y}_{1\left\{ 34\right\} }^{t-1},\boldsymbol{Y}_{2\left\{ 34\right\} }^{t-1},\boldsymbol{Z}^{t-1},\boldsymbol{W}_{2},\sigma\left(t\right)=1\right)\nonumber \\
 & +I\left(\boldsymbol{Y}_{1\left\{ 3\right\} }^{t-1},\boldsymbol{Y}_{2\left\{ 3\right\} }^{t-1},\boldsymbol{W}_{2};X_{1}\left(t\right)|\boldsymbol{Y}_{1\left\{ 4\right\} }^{t-1},\boldsymbol{Y}_{2\left\{ 4\right\} }^{t-1},\boldsymbol{Z}^{t-1},\sigma\left(t\right)=1\right)\nonumber \\
 & -I\left(\boldsymbol{Y}_{1\left\{ 3\right\} }^{t-1},\boldsymbol{Y}_{2\left\{ 3\right\} }^{t-1};X_{1}\left(t\right)|\boldsymbol{Y}_{1\left\{ 4\right\} }^{t-1},\boldsymbol{Y}_{2\left\{ 4\right\} }^{t-1},\boldsymbol{Z}^{t-1},\boldsymbol{W}_{2},\sigma\left(t\right)=1\right)\nonumber \\
 & =I\left(\boldsymbol{Y}_{1\left\{ 2\right\} }^{t-1},\boldsymbol{Y}_{1\left\{ 3\right\} }^{t-1},\boldsymbol{Y}_{2\left\{ 3\right\} }^{t-1},\boldsymbol{W}_{2};X_{1}\left(t\right)|\boldsymbol{Y}_{1\left\{ 4\right\} }^{t-1},\boldsymbol{Y}_{2\left\{ 4\right\} }^{t-1},\boldsymbol{Z}^{t-1},\sigma\left(t\right)=1\right)\nonumber \\
 & -I\left(\boldsymbol{Y}_{1\left\{ 3\right\} }^{t-1},\boldsymbol{Y}_{2\left\{ 3\right\} }^{t-1};X_{1}\left(t\right)|\boldsymbol{Y}_{1\left\{ 4\right\} }^{t-1},\boldsymbol{Y}_{2\left\{ 4\right\} }^{t-1},\boldsymbol{Z}^{t-1},\boldsymbol{W}_{2},\sigma\left(t\right)=1\right)\nonumber \\
 & =I\left(\boldsymbol{Y}_{1\left\{ 2\right\} }^{t-1},\boldsymbol{W}_{2};X_{1}\left(t\right)|\boldsymbol{Y}_{1\left\{ 4\right\} }^{t-1},\boldsymbol{Y}_{2\left\{ 4\right\} }^{t-1},\boldsymbol{Z}^{t-1},\sigma\left(t\right)=1\right)\nonumber \\
 & +I\left(\boldsymbol{Y}_{1\left\{ 3\right\} }^{t-1},\boldsymbol{Y}_{2\left\{ 3\right\} }^{t-1};X_{1}\left(t\right)|\boldsymbol{Y}_{1\left\{ 24\right\} }^{t-1},\boldsymbol{Y}_{2\left\{ 4\right\} }^{t-1},\boldsymbol{Z}^{t-1},\boldsymbol{W}_{2},\sigma\left(t\right)=1\right)\nonumber \\
 & -I\left(\boldsymbol{Y}_{1\left\{ 3\right\} }^{t-1},\boldsymbol{Y}_{2\left\{ 3\right\} }^{t-1};X_{1}\left(t\right)|\boldsymbol{Y}_{1\left\{ 4\right\} }^{t-1},\boldsymbol{Y}_{2\left\{ 4\right\} }^{t-1},\boldsymbol{Z}^{t-1},\boldsymbol{W}_{2},\sigma\left(t\right)=1\right).\label{eq:ineq2}
\end{align}
We claim that, 
\begin{multline}
I\left(\boldsymbol{Y}_{1\left\{ 3\right\} }^{t-1},\boldsymbol{Y}_{2\left\{ 3\right\} }^{t-1};X_{1}\left(t\right)|\boldsymbol{Y}_{1\left\{ 24\right\} }^{t-1},\boldsymbol{Y}_{2\left\{ 4\right\} }^{t-1},\boldsymbol{Z}^{t-1},\boldsymbol{W}_{2},\sigma\left(t\right)=1\right)\\
\leq I\left(\boldsymbol{Y}_{1\left\{ 3\right\} }^{t-1},\boldsymbol{Y}_{2\left\{ 3\right\} }^{t-1};X_{1}\left(t\right)|\boldsymbol{Y}_{1\left\{ 4\right\} }^{t-1},\boldsymbol{Y}_{2\left\{ 4\right\} }^{t-1},\boldsymbol{Z}^{t-1},\boldsymbol{W}_{2},\sigma\left(t\right)=1\right).\label{eq:ineq1}
\end{multline}
To see this write,
\[
I\left(\boldsymbol{Y}_{1\left\{ 3\right\} }^{t-1},\boldsymbol{Y}_{2\left\{ 3\right\} }^{t-1};X_{1}\left(t\right)|\boldsymbol{Y}_{1\left\{ 24\right\} }^{t-1},\boldsymbol{Y}_{2\left\{ 4\right\} }^{t-1},\boldsymbol{Z}^{t-1},\boldsymbol{W}_{2},\sigma\left(t\right)=1\right)
\]
\begin{align}
 & =\sum_{\boldsymbol{y}_{1\left\{ 24\right\} }^{t-1}\boldsymbol{y}_{2\left\{ 4\right\} }^{t-1}\boldsymbol{\boldsymbol{z}}^{t-1},\boldsymbol{w}}I\left(\boldsymbol{Y}_{1\left\{ 3\right\} }^{t-1},\boldsymbol{Y}_{2\left\{ 3\right\} }^{t-1};X_{1}\left(t\right)|\boldsymbol{Y}_{1\left\{ 24\right\} }^{t-1}=\boldsymbol{y}_{1\left\{ 24\right\} }^{t-1},\boldsymbol{Y}_{2\left\{ 4\right\} }^{t-1}=\boldsymbol{y}_{2\left\{ 4\right\} }^{t-1}\boldsymbol{Z}^{t-1}=\boldsymbol{z}^{t-1},\boldsymbol{W}_{2}=\boldsymbol{w},\sigma\left(t\right)=1\right)\nonumber \\
 & \times\Pr\left(\boldsymbol{Y}=\boldsymbol{y}_{1\left\{ 24\right\} }^{t-1},\boldsymbol{Y}_{2\left\{ 4\right\} }^{t-1}=\boldsymbol{y}_{2\left\{ 4\right\} }^{t-1}\boldsymbol{Z}^{t-1}=\boldsymbol{z}^{t-1},\boldsymbol{W}_{2}=\boldsymbol{w}|\sigma\left(t\right)=1\right),\label{eq:ff1}
\end{align}
and 
\[
I\left(\boldsymbol{Y}_{1\left\{ 3\right\} }^{t-1},\boldsymbol{Y}_{2\left\{ 3\right\} }^{t-1};X_{1}\left(t\right)|\boldsymbol{Y}_{1\left\{ 4\right\} }^{t-1},\boldsymbol{Y}_{2\left\{ 4\right\} }^{t-1},\boldsymbol{Z}^{t-1},\boldsymbol{W}_{2},\sigma\left(t\right)=1\right)
\]
\begin{align}
 & =\sum_{\boldsymbol{y}_{1\left\{ 24\right\} }^{t-1}\boldsymbol{y}_{2\left\{ 4\right\} }^{t-1}\boldsymbol{\boldsymbol{z}}^{t-1},\boldsymbol{w}}I\left(\boldsymbol{Y}_{1\left\{ 3\right\} }^{t-1},\boldsymbol{Y}_{2\left\{ 3\right\} }^{t-1};X_{1}\left(t\right)|\boldsymbol{Y}_{1\left\{ 4\right\} }^{t-1}=\boldsymbol{y}_{1\left\{ 4\right\} }^{t-1},\boldsymbol{Y}_{2\left\{ 4\right\} }^{t-1}=\boldsymbol{y}_{2\left\{ 4\right\} }^{t-1}\boldsymbol{Z}^{t-1}=\boldsymbol{z}^{t-1},\boldsymbol{W}_{2}=\boldsymbol{w},\sigma\left(t\right)=1\right)\nonumber \\
 & \times\Pr\left(\boldsymbol{Y}_{1\left\{ 24\right\} }^{t-1}=\boldsymbol{y}_{1\left\{ 24\right\} }^{t-1},\boldsymbol{Y}_{2\left\{ 4\right\} }^{t-1}=\boldsymbol{y}_{2\left\{ 4\right\} }^{t-1}\boldsymbol{Z}^{t-1}=\boldsymbol{z}^{t-1},\boldsymbol{W}_{2}=\boldsymbol{w}|\sigma\left(t\right)=1\right).\label{eq:ff2}
\end{align}

We now consider the following (exhaustive) cases regarding the summation
terms in (\ref{eq:ff1}) and (\ref{eq:ff2}). 
\begin{enumerate}
\item $J\left(\boldsymbol{z}^{t-1}\right)\in h_{1,4}^{t-1}$: In this case
by (\ref{eq:fund1}), both mutual information terms in (\ref{eq:ff1})
and (\ref{eq:ff2}) are zero. 
\item $J\left(\boldsymbol{z}^{t-1}\right)\in h_{1,2\bar{4}}^{t-1}$: In
this case by (\ref{eq:fund1}), the mutual information term in (\ref{eq:ff1})
is zero. 
\item $J\left(\boldsymbol{z}^{t-1}\right)\in h_{1,\bar{2}\bar{3}\bar{4}}^{t-1}$:
In this case by (\ref{eq:fund4}), both mutual information terms in
(\ref{eq:ff1}) and (\ref{eq:ff2}) are zero. 
\item $J\left(\boldsymbol{z}^{t-1}\right)\in h_{1,\bar{2}3\bar{4}}^{t-1}:$
In this case, the mutual information term in (\ref{eq:ff2}) is one.
This is due to the fact that by (\ref{eq:fund3}), 
\[
H\left(X_{1}\left(t\right)|\boldsymbol{Y}_{1\left\{ 4\right\} }^{t-1}=\boldsymbol{y}_{1\left\{ 4\right\} }^{t-1},\boldsymbol{Y}_{2\left\{ 4\right\} }^{t-1}=\boldsymbol{y}_{2\left\{ 4\right\} }^{t-1}\boldsymbol{Z}^{t-1}=\boldsymbol{z}^{t-1},\boldsymbol{W}_{2}=\boldsymbol{w},\sigma\left(t\right)=1\right)=1
\]
while by (\ref{eq:fund1}), 
\[
H\left(X_{1}\left(t\right)|\boldsymbol{Y}_{1\left\{ 34\right\} }^{t-1}=\boldsymbol{y}_{1\left\{ 34\right\} }^{t-1},\boldsymbol{Y}_{2\left\{ 34\right\} }^{t-1}=\boldsymbol{y}_{2\left\{ 34\right\} }^{t-1}\boldsymbol{Z}^{t-1}=\boldsymbol{z}^{t-1},\boldsymbol{W}_{2}=\boldsymbol{w},\sigma\left(t\right)=1\right)=0.
\]
\end{enumerate}
Set now, 
\begin{align*}
\tilde{u}_{1,2\wedge\bar{3}\wedge\bar{4}} & =\sum_{t=1}^{n}I\left(\boldsymbol{Y}_{1\left\{ 2\right\} }^{t-1};X_{1}\left(t\right)|\boldsymbol{Y}_{1\left\{ 34\right\} }^{t-1},\boldsymbol{Y}_{2\left\{ 34\right\} }^{t-1},\boldsymbol{Z}^{t-1},\boldsymbol{W}_{2},\sigma\left(t\right)=1\right)\Pr\left(\sigma\left(t\right)=1\right)\\
 & \leq\tilde{\tau}_{1,2\wedge\bar{3}\wedge\bar{4}}\text{ "by (\ref{eq:G3})"},
\end{align*}
\[
\tilde{v}_{1,2\wedge\bar{4}}=\sum_{t=1}^{n}I\left(\boldsymbol{W}_{2};X_{1}\left(t\right)|\boldsymbol{Y}_{1\left\{ 4\right\} }^{t-1},\boldsymbol{Y}_{2\left\{ 4\right\} }^{t-1},\boldsymbol{Z}^{t-1},\sigma\left(t\right)=1\right)\Pr\left(\sigma\left(t\right)=1\right).
\]
Using (\ref{eq:ineq2}), (\ref{eq:ineq1}) we conclude, 
\begin{align*}
\tilde{u}_{1,2\wedge\bar{3}\wedge\bar{4}}+\tilde{v}_{1,2\wedge\bar{4}} & \leq\sum_{t=1}^{n}I\left(\boldsymbol{Y}_{1\left\{ 2\right\} }^{t-1},\boldsymbol{W}_{2};X_{1}\left(t\right)|\boldsymbol{Y}_{1\left\{ 4\right\} }^{t-1},\boldsymbol{Y}_{2\left\{ 4\right\} }^{t-1},\boldsymbol{Z}^{t-1},\sigma\left(t\right)=1\right)\\
 & \leq\tilde{\tau}_{1,2\wedge\bar{4}}\text{ "by (\ref{eq:G3})"}\\
 & =\tilde{\tau}_{1,2\wedge\bar{3}\wedge\bar{4}}+\tilde{\tau}_{1,2\wedge3\wedge\bar{4}}.
\end{align*}
\end{IEEEproof}
We can now provide the proof of Theorem \ref{thm:final}.

\subsection{Proof of Theorem \emph{\ref{thm:final}}}
\begin{IEEEproof}
\emph{(of Theorem \ref{thm:final}) }Let $\delta>0$ and assume $n$
large enough so that $o\left(n\right)/n\leq\delta$ for all inequalities
in Lemma \ref{lem:BasicLemma} and all $n\geq n(\delta).$ Let $\mathcal{\mathcal{R}}\left(\delta\right)$
be the region of $\left(R_{1},R_{2}\right)$ defined by, 
\begin{align}
\frac{R_{1}}{1-\epsilon_{23}^{1}}+\frac{R_{2}}{1-\epsilon_{4}^{2}} & \leq1-Q-S-U+\delta,\label{eq:finfin1-1}\\
\left(\frac{\epsilon_{3}^{1}-\epsilon_{23}^{1}}{\left(1-\epsilon_{3}^{2}\right)\left(1-\epsilon_{23}^{1}\right)}+\frac{1}{1-\epsilon_{23}^{1}}\right)R_{1}+\frac{R_{2}}{1-\epsilon_{34}^{2}} & \leq1-Q-S-U+\left(\frac{1-\epsilon_{3}^{1}}{1-\epsilon_{3}^{2}}\right)\left(Q+S\right)+\delta,\label{eq:finfin2-1}\\
\left(\frac{\epsilon_{34}^{1}-\epsilon_{234}^{1}}{\left(1-\epsilon_{234}^{1}\right)\left(1-\epsilon_{34}^{2}\right)}+\frac{1}{1-\epsilon_{23}^{1}}\right)R_{1}+\frac{1}{1-\epsilon_{4}^{2}}R_{2} & \leq1-Q-S-U+\frac{1-\epsilon_{34}^{1}}{1-\epsilon_{34}^{2}}Q+\frac{1-\epsilon_{4}^{1}}{1-\epsilon_{4}^{2}}\left(S+U\right)+\delta,\label{eq:finfin3-1}
\end{align}
\[
Q\geq0,\ S\geq0,\ U\geq0,\ R_{i}\geq0,i\in\left\{ 1,2\right\} .
\]
Taking into account that $\tilde{\tau}_{1,2\wedge\bar{3}\wedge\bar{4}}\leq\tilde{\tau}_{1,2\wedge\bar{3}}$
$\tilde{\tau}_{1,2\wedge3\wedge\bar{4}}\leq\tilde{\tau}_{1,3}$, we
conclude from (\ref{eq:R2Ineq}) that, 
\[
\frac{R_{1}}{1-\epsilon_{23}^{1}}+\frac{nR_{2}}{1-\epsilon_{4}^{2}}\leq1-\frac{\tilde{\tau}_{1,2\wedge\bar{3}\wedge\bar{4}}}{n}-\frac{\tilde{\tau}_{1,2\wedge3\wedge\bar{4}}}{n}+\delta.
\]
Taking also into account that $\left(1-\epsilon_{3}^{1}\right)/\left(1-\epsilon_{3}^{2}\right)\leq1$,
(\ref{eq:third}) implies that, 
\begin{align*}
\left(\frac{\epsilon_{3}^{1}-\epsilon_{23}^{1}}{\left(1-\epsilon_{3}^{2}\right)\left(1-\epsilon_{23}^{1}\right)}+\frac{1}{1-\epsilon_{23}^{1}}\right)R_{1}+\frac{R_{2}}{1-\epsilon_{34}^{2}} & \leq1-\left(1-\frac{1-\epsilon_{3}^{1}}{1-\epsilon_{3}^{2}}\right)\frac{\tilde{\tau}_{1,2\wedge\bar{3}}}{n}-\frac{\tilde{\tau}_{1,3}}{n}+\delta\\
 & \leq1-\frac{\tilde{\tau}_{1,2\wedge\bar{3}\wedge\bar{4}}}{n}-\frac{\tilde{\tau}_{1,2\wedge3\wedge\bar{4}}}{n}+\left(\frac{1-\epsilon_{3}^{1}}{1-\epsilon_{3}^{2}}\right)\frac{\tilde{\tau}_{1,2\wedge\bar{3}\wedge\bar{4}}}{n}+\delta.
\end{align*}
and similarly (\ref{eq:fourth}) implies that, 
\begin{align*}
\left(\frac{\epsilon_{34}^{1}-\epsilon_{234}^{1}}{\left(1-\epsilon_{234}^{1}\right)\left(1-\epsilon_{34}^{2}\right)}+\frac{1}{1-\epsilon_{23}^{1}}\right)R_{1}+\frac{1}{1-\epsilon_{4}^{2}}R_{2} & \leq1-\frac{\tilde{\tau}_{1,2\wedge\bar{3}\wedge\bar{4}}}{n}-\frac{\tilde{\tau}_{1,2\wedge3\wedge\bar{4}}}{n}+\frac{1-\epsilon_{34}^{1}}{1-\epsilon_{34}^{2}}\frac{\tilde{u}_{1,2\wedge\bar{3}\wedge\bar{4}}}{n}+\frac{1-\epsilon_{4}^{1}}{1-\epsilon_{4}^{2}}\frac{\tilde{v}_{1,2\wedge\bar{4}}}{n}+\delta,
\end{align*}
where, 
\begin{equation}
\frac{\tilde{u}_{1,2\wedge\bar{3}\wedge\bar{4}}}{n}\leq\frac{\tilde{\tau}_{1,2\wedge\bar{3}\wedge\bar{4}}}{n},\label{eq:w1}
\end{equation}
\begin{equation}
\frac{\tilde{u}_{1,2\wedge\bar{3}\wedge\bar{4}}}{n}+\frac{\tilde{v}_{1,2\wedge\bar{4}}}{n}\leq\frac{\tilde{\tau}_{1,2\wedge\bar{3}\wedge\bar{4}}}{n}+\frac{\tilde{\tau}_{1,2\wedge3\wedge\bar{4}}}{n}.\label{eq:w2}
\end{equation}
Consider now the change of variables, 
\begin{align*}
Q_{n} & =\frac{\tilde{u}_{1,2\wedge\bar{3}\wedge\bar{4}}}{n}\geq0,\\
S_{n} & =\frac{\tilde{\tau}_{1,2\wedge\bar{3}\wedge\bar{4}}}{n}-\frac{\tilde{u}_{1,2\wedge\bar{3}\wedge\bar{4}}}{n}\geq0\text{, "by (\ref{eq:w1})"}\\
U_{n} & =\frac{\tilde{\tau}_{1,2\wedge3\wedge\bar{4}}}{n}\geq0.
\end{align*}
With this change of variables and since by (\ref{eq:w2}) 
\[
\frac{\tilde{v}_{1,2\wedge\bar{4}}}{n}\leq S_{n}+U_{n},
\]
we see that $\left(R_{1},R_{2}\right)$ satisfies the inequalities,
\[
\frac{R_{1}}{1-\epsilon_{23}^{1}}+\frac{nR_{2}}{1-\epsilon_{4}^{2}}\leq1-Q_{n}-S_{n}+\delta,
\]
\[
\left(\frac{\epsilon_{3}^{1}-\epsilon_{23}^{1}}{\left(1-\epsilon_{3}^{2}\right)\left(1-\epsilon_{23}^{1}\right)}+\frac{1}{1-\epsilon_{23}^{1}}\right)R_{1}+\frac{R_{2}}{1-\epsilon_{34}^{2}}\leq1-Q_{n}-S_{n}+\left(\frac{1-\epsilon_{3}^{1}}{1-\epsilon_{3}^{2}}\right)\left(Q_{n}+S_{n}\right)+\delta,
\]
\[
\left(\frac{\epsilon_{34}^{1}-\epsilon_{234}^{1}}{\left(1-\epsilon_{234}^{1}\right)\left(1-\epsilon_{34}^{2}\right)}+\frac{1}{1-\epsilon_{23}^{1}}\right)R_{1}+\frac{1}{1-\epsilon_{4}^{2}}R_{2}\leq1-Q_{n}-S_{n}+\frac{1-\epsilon_{34}^{1}}{1-\epsilon_{34}^{2}}Q_{n}+\frac{1-\epsilon_{4}^{1}}{1-\epsilon_{4}^{2}}\left(S_{n}+U_{n}\right)+\delta,
\]
\[
Q_{n}\geq0,S_{n}\geq0,U_{n}\geq0,R_{i}\geq0,i\in\left\{ 1,2\right\} .
\]
From the last set of inequalities we conclude that for all $\delta>0,$
$\left(R_{1},R_{2}\right)\in\mathcal{\mathcal{\mathcal{R}}}\left(\delta\right),$
which implies that $\left(R_{1},R_{2}\right)\in\mathcal{\mathcal{\mathcal{R}}}\left(0\right)=\mathcal{\mathcal{R}}.$ 
\end{IEEEproof}

\subsection{\label{sec:ProofOfFinCor}Proof of Corollary \ref{cor:FinCor}}
\begin{enumerate}
\item Let $\left(R_{1},R_{2}\right)\in\mathcal{R}.$ Then, conditions $\epsilon_{3}^{1}\geq\epsilon_{3}^{2}$
and (\ref{eq:conda}) imply that the right hand sides of (\ref{eq:finfin2})
and (\ref{eq:finfin3}) are both at most 1, hence (\ref{eq:newfin2a})
and (\ref{eq:newfin3a}) are satisfied, i.e., $\left(R_{1},R_{2}\right)\in\mathcal{R}_{1}$.
Assume next that $\left(R_{1},R_{2}\right)\in\mathcal{R}_{1}.$ Selecting
$Q=S=U=0,$ we conclude from (\ref{eq:finfin2}), (\ref{eq:finfin3})
that (\ref{eq:newfin2a}) and (\ref{eq:newfin3a}) are satisfied;
also, (\ref{eq:finfin1}) is implied by (\ref{eq:newfin3a}), i.e.,
$\left(R_{1},R_{2}\right)\in\mathcal{R}$. 
\item Since 
\[
\frac{1-\epsilon_{34}^{1}}{1-\epsilon_{34}^{2}}\geq\frac{1-\epsilon_{4}^{1}}{1-\epsilon_{4}^{2}},
\]
it is easy to see that the region $\mathcal{R}$ is equal to the region
$\mathcal{R}_{a}$ defined by the following inequalities. 
\begin{align}
\frac{R_{1}}{1-\epsilon_{23}^{1}}+\frac{R_{2}}{1-\epsilon_{4}^{2}} & \leq1-Q,\label{eq:newfin1b}\\
\left(\frac{\epsilon_{3}^{1}-\epsilon_{23}^{1}}{\left(1-\epsilon_{3}^{2}\right)\left(1-\epsilon_{23}^{1}\right)}+\frac{1}{1-\epsilon_{23}^{1}}\right)R_{1}+\frac{R_{2}}{1-\epsilon_{34}^{2}} & \leq1-Q+\left(\frac{1-\epsilon_{3}^{1}}{1-\epsilon_{3}^{2}}\right)Q,\label{eq:newfin2b}\\
\left(\frac{\epsilon_{34}^{1}-\epsilon_{234}^{1}}{\left(1-\epsilon_{234}^{1}\right)\left(1-\epsilon_{34}^{2}\right)}+\frac{1}{1-\epsilon_{23}^{1}}\right)R_{1}+\frac{1}{1-\epsilon_{4}^{2}}R_{2} & \leq1-Q+\frac{1-\epsilon_{34}^{1}}{1-\epsilon_{34}^{2}}Q,\label{eq:newfin3b}
\end{align}
\[
Q\geq0,R_{i}\geq0,i\in\left\{ 1,2\right\} .
\]
To show that $\mathcal{R}_{a}$ is equal to the region $\mathcal{R}_{2},$
we only need to show that replacing (\ref{eq:newfin1b}) with (\ref{eq:r2a})
does not affect the region, since the rest of the inequalities are
the same. Assume first that $\left(R_{1},R_{2}\right)\in\mathcal{R}_{2}$.
Inequalities (\ref{eq:r2a}) and (\ref{eq:r2c}) imply that 
\begin{align*}
\left(\frac{\epsilon_{34}^{1}-\epsilon_{234}^{1}}{\left(1-\epsilon_{234}^{1}\right)\left(1-\epsilon_{34}^{2}\right)}+\frac{1}{1-\epsilon_{23}^{1}}\right)R_{1}+\frac{1}{1-\epsilon_{4}^{2}}R_{2} & \leq1-Q+\frac{1-\epsilon_{34}^{1}}{1-\epsilon_{34}^{2}}\frac{\epsilon_{34}^{1}-\epsilon_{234}^{1}}{\left(1-\epsilon_{234}^{1}\right)\left(1-\epsilon_{34}^{1}\right)}R_{1}\\
 & =1-Q+\frac{\epsilon_{34}^{1}-\epsilon_{234}^{1}}{\left(1-\epsilon_{234}^{1}\right)\left(1-\epsilon_{34}^{2}\right)}R_{1},
\end{align*}
hence (\ref{eq:newfin1b}) holds. We conclude that $\mathcal{R}_{2}\subseteq\mathcal{R}_{1}$.
Next let $\left(R_{1},R_{2}\right)\in\mathcal{R}_{a}.$ If for the
selected pair $\left(R_{1},R_{2}\right)$ and $Q$ inequality (\ref{eq:r2a})
is satisfied, then $\mathcal{R}_{1}\subseteq\mathcal{R}_{2}$. Let
us assume now that inequality (\ref{eq:r2a}) is not satisfied, i.e.,
\begin{equation}
\frac{\epsilon_{34}^{1}-\epsilon_{234}^{1}}{\left(1-\epsilon_{234}^{1}\right)\left(1-\epsilon_{34}^{1}\right)}R_{1}<Q.\label{eq:Qineq}
\end{equation}
We claim that for the same pair $\left(R_{1},R_{2}\right)$ we can
also select $Q_{0}>0$ so that inequalities (\ref{eq:r2a})-(\ref{eq:r2c})
are satisfied, which will imply that $\left(R_{1},R_{2}\right)\in\mathcal{R}_{2}$.
To see this let $Q_{0}$ be the infimum of all $Q\geq0$ satisfying
(\ref{eq:Qineq}) and $\left(\ref{eq:newfin1b}\right)$-$(\ref{eq:newfin3b})$.
From the definition it easily follows that $Q_{0}$ satisfies $\left(\ref{eq:newfin1b}\right)$-$(\ref{eq:newfin3b})$
and 
\begin{equation}
\frac{\epsilon_{34}^{1}-\epsilon_{234}^{1}}{\left(1-\epsilon_{234}^{1}\right)\left(1-\epsilon_{34}^{1}\right)}R_{1}\leq Q_{0}.\label{eq:Qeq}
\end{equation}
We claim that $Q_{0}$ satisfies (\ref{eq:Qeq}) with equality. Indeed
assume that 
\begin{equation}
\frac{\epsilon_{34}^{1}-\epsilon_{234}^{1}}{\left(1-\epsilon_{234}^{1}\right)\left(1-\epsilon_{34}^{1}\right)}R_{1}<Q_{0}.\label{eq:Qgeq}
\end{equation}
Multiplying both terms of (\ref{eq:Qgeq}) by $\left(1-\epsilon_{34}^{1}\right)/\left(1-\epsilon_{34}^{2}\right)$
and adding the terms of the resulting inequality with those of (\ref{eq:newfin1b})
we get, 
\[
\left(\frac{\epsilon_{34}^{1}-\epsilon_{234}^{1}}{\left(1-\epsilon_{234}^{1}\right)\left(1-\epsilon_{34}^{2}\right)}+\frac{1}{1-\epsilon_{23}^{1}}\right)R_{1}+\frac{1}{1-\epsilon_{4}^{2}}R_{2}<1-\hat{Q}+\frac{1-\epsilon_{34}^{1}}{1-\epsilon_{34}^{2}}\hat{Q}.
\]
Hence we can reduce $Q_{0}$ without violating $(\ref{eq:newfin3b})$.
Since $\left(\ref{eq:newfin1b}\right),$$(\ref{eq:newfin2b})$ are
actually strengthened by this reduction, we conclude that we can find
$Q>0$ smaller that $Q_{0}$ satisfying (\ref{eq:Qineq}) and also
$\left(\ref{eq:newfin1b}\right)$-$(\ref{eq:newfin3b})$, which contradicts
the definition of $Q_{0}.$ 
\item The arguments for item \ref{enu:cor3} are similar to those of item
\ref{enu:cor2}. 
\end{enumerate}

\section*{Acknowledgment}

The work of Athanasios Papadopoulos was funded by ELIDEK since August
2017 and by the Onassis Foundation (October 2015-August 2017).

\bibliographystyle{IEEEtran}
\bibliography{IEEEabrv,References}

\end{document}